\documentclass[atoms,article,accept,pdftex,moreauthors]{mdpi} 
\usepackage{longtable}
\usepackage{makecell} 
\usepackage{color}
\usepackage{framed}
\usepackage{etoolbox}
\usepackage{tensor}
\usepackage{graphicx}
\usepackage{amsmath}
\usepackage{amssymb}
\usepackage{amsfonts}
\usepackage{dcolumn}
\usepackage{tikz}
\usepackage{comment}
\usepackage{verbatim}
\usepackage{fancyvrb}
\usepackage{cancel}
\usepackage{multirow}
\usepackage{lscape}
\usepackage{mathtools}
\usepackage{soul}
\usepackage{url}
\usepackage{makecell}
\usepackage[pdftex]{pict2e}
\usepackage{microtype}
\makeatletter
\renewcommand{\maketag@@@}[1]{\hbox{\m@th\normalsize\normalfont#1}}
\makeatother

\def\redmem#1#2#3{\left\langle
#1 \left\Vert
#2 \right\Vert
#3 \right\rangle}
\firstpage{1} 
\pubvolume{14}
\issuenum{5}
\articlenumber{40}
\pubyear{2026}
\copyrightyear{2026}
\externaleditor{Firstname Lastname} 
\datereceived{7 April 2026} 
\daterevised{14 May 2026} 
\dateaccepted{19 May 2026} 
\datepublished{ } 
\hreflink{ https://doi.org/10.3390/atoms14050040} 

\Title{Second-Order Rayleigh–Schr\"odinger Perturbation Theory for the {\sc \textbf{Grasp}}
2018 Package}

\Author{Gediminas 
 Gaigalas *\orcidA{}, Pavel Rynkun \orcidB{} and Laima Kitovien\.e \orcidC{}}

\AuthorNames{Gediminas Gaigalas, Pavel Rynkun and Laima Kitovien\. e}

\address [1]{%
 Institute of Theoretical Physics and Astronomy, Faculty of Physics,
               Vilnius University, Saul\.{e}tekio Ave. 3, \linebreak  LT-10257 Vilnius, Lithuania;  pavel.rynkun@tfai.vu.lt (P.R.); laima.radziute@tfai.vu.lt (L.K.) 
\\
}

\corres{\hangafter=1 \hangindent=1.05em \hspace{-0.82em} Correspondence: gediminas.gaigalas@tfai.vu.lt}

\abstract{A developed method, based on the stationary second-order Rayleigh–Schrödinger many-body perturbation theory in an irreducible tensorial form, allows us to determine the most important core–valence, core, core–core, and valence–valence correlations for any atom or ion with an arbitrary number of valence and core electrons. This paper presents the Feynman diagrams that describe these correlations. Additionally, it provides the rules for obtaining algebraic expressions in an irreducible tensorial form for any Feynman diagram coming from second-order many-body perturbation theory. Whereas some types of the valence–valence and core–valence correlations are described by the three-particle Feynman diagrams, additional developments to calculate the spin-angular parts of these diagrams have been made to the program library \texttt{librang} of the {\sc Grasp}2018. 
As an example of the application of the developed method, the atomic calculations of the energy level structure and transition data for Ar II are presented.}


\begin{document}



\section{Introduction} 

The General Relativistic Atomic Structure Package ({\sc Grasp})~\cite{Graetal:80a,Dyaetal:89a,GRASP92,Jonetal:2007a,Jonetal:2013a,Fro:14a}, which has been under development for decades, is one of the most effective tools available today for the theoretical study of the various properties of the atom. It allows the study of the energy spectrum, various transition characteristics, hyperfine structure, isotope shift with very high precision. The~origin of this program lies with some of the most prominent and distinguished atomic theorists of recent times, Prof. Ian Philip Grant (1930--2025) and Prof. Charlotte Froese Fischer (1929--2024).

The {\sc Grasp} software package is based on the multiconfiguration Dirac–Hartree–Fock (MCDHF) method and relativistic configuration interaction (RCI) method~\cite{Fisetal:16a,Theory_GRASP_ATOM,Manual_GRASP}. They have provided very accurate atomic data for a wide range of atoms and ions. These data have been used in various fields of science and technology. One of the most important conditions for obtaining such accurate atomic characteristics is the accurate incorporation of correlation and relativistic effects into the calculations. Correlation effects are included in these methods when the atomic state function (ASF) for which the eigenvalue problem is being solved is constructed from a set of configuration state functions (CSFs) containing the starting configuration and the other configurations constructed from it. The~latter are obtained by performing single, double, or~even sometimes triple and~quadruple excitations on the initial configuration. This, if~high accuracy is to be achieved, usually leads to a large CSF base. This base, therefore, needs to be limited. There are a number of methods/recommendations~\cite{Fisetal:16a,Theory_GRASP_ATOM,Bunge_2006,Bunge_1997} for selecting the optimal CSF base to achieve the desired accuracy. But~these recommendations are approximate, and, moreover, the~choice of such a base is not trivial and requires much additional research, and,~at the same time, it is not always possible to find the most optimal set of~CSFs.

Recently, a~new approach~\cite{Gaigetal:2024CV,Gaigetal:2024C,Gaigetal:2024CC,Gaigetal:2025VV,Gaigetal:2025VVT,Gaigetal:2026CVT} has been proposed in which, based on the stationary second-order Rayleigh–Schr\"odinger many-body perturbation theory (RSMBPT) in irreducible tensorial form, it is theoretically possible, on~an ab~initio basis, to~find a set of CSFs that are optimal in terms of the size of the CSF basis and that lead to maximally accurate atomic characteristics. It is based on a combination of RCI and RSMBPT in irreducible tensorial form. It allows the inclusion of core–valence (CV), core (C), core–core (CC), and~valence–valence (VV) correlations using the second-order of perturbation theory for any atom and ion with any number of valence electrons. This newly developed method, which can be used in three ways (RCI + RSMBPT (see, for example, Section 4.1 in
~\cite{Gaigetal:2024CV}), RCI (RSMBPT) (see, for example, Section 4.2 in~\cite{Gaigetal:2024CV}), and~MCDHF (RSMBPT) (see, for example, Section 5.1.2 in~\cite{Gaigetal:2025VVT}), allows for the reduction of the space of configuration state functions for complex atoms and ions, which extends the capability of the {\sc Grasp}2018 software package~\cite{Fro:14a}.
 
In the RCI + RSMBPT method, the~correlations considered in the RSMBPT method are added to the correlations considered in the conventional RCI method by adding them to the corresponding matrix elements from which the matrix is constructed in the RCI method (see, for example, for CV correlations, Equation (20) in~\cite{Gaigetal:2024CV}). In~this method, the~order of the matrix to be constructed is considerably reduced because correlations considered by the RSMBPT method are not represented by their own matrix elements in the matrix constructed by the RCI~method.

In the RCI (RSMBPT) method, the~value of the correlation influence on RCI calculations is assessed by considering all the correlation types separately according to $\Delta E_{PT}$ (see, for example, for CV correlations, Equation (22) in~\cite{Gaigetal:2024CV}) from the RSMBPT method.
The most important correlations are then selected according to the criterion given, and~these, together with the remaining correlations (which are not considered in the RSMBPT method), are included in the RCI calculation in a regular way (via the CSF expansion). In~this case, the~matrix constructed is of a higher order than the RCI + RSMBPT, but~the correlations are included more consistently (not only in the second-order of perturbation theory). In~both methods, only Coulomb interactions are considered for correlations investigated by the RSMBPT, which makes the RCI (RSMBPT) method more attractive, as it can also take into account the Breit operator and quantum electrodynamic (QED) effects~\cite{McKetal:80a,Gra:2007a} in the final RCI calculations, as~the correlations considered in the RCI (RSMBPT) method also have their own matrix elements in the constructed RCI matrix, and therefore the Breit interaction and QED effects can be added in a regular~way. 

The MCDHF (RSMBPT) method is similar to the RCI (RSMBPT) method, but~it is applied to the MCDHF calculation, i.e.,~instead of calculating the RCI, a~matrix is constructed for the MCDHF~calculation.

The papers~\cite{Gaigetal:2024CV,Gaigetal:2024C,Gaigetal:2024CC,Gaigetal:2025VV,Gaigetal:2025VVT,Gaigetal:2026CVT} demonstrate how these methods work for a wide range of ions, both in terms of energy spectra and transition properties. They also provide a minimal but comprehensive set of Feynman diagrams for the {\sc Grasp} package, describing all the most important and necessary CV, C, CC, and VV correlations based on RSMBPT theory.
Each paper was devoted to a particular correlation type because~the theory is quite complex; therefore, much work is needed to develop it, both in terms of obtaining expressions for Feynman diagrams in an irreducible tensorial form and~in terms of reformulating them and, if~that is not enough, to~extend the library \texttt{librang}~\cite{Gaigalas:2022} in such a way that the spin-angular theory~\cite{Gaietal:97a,Gaigalas:2026a} remains effective and known atomic symmetries (including the quasispin) are fully~applied.

This paper extends the method, based on the stationary second-order Rayleigh–Schrödinger many-body perturbation theory in an irreducible tensorial form, developed by Gaigalas
~et~al.~\cite{Gaigetal:2024CV,Gaigetal:2024C,Gaigetal:2024CC,Gaigetal:2025VV,Gaigetal:2025VVT,Gaigetal:2026CVT}.
It provides all the Feynman diagrams that describe CV, C, CC, and~VV correlations for
which analytical expressions are derived in an irreducible tensorial form and~presents the general rules
for obtaining algebraic expressions in an irreducible tensorial form for any Feynman diagram from the
second-order perturbation theory. By~generalizing the method developed in~\cite{Gaigetal:2024CV,Gaigetal:2024C,Gaigetal:2024CC,Gaigetal:2025VV,Gaigetal:2025VVT,Gaigetal:2026CVT}, this extension of the
approach introduced in this paper reaches the next level of application for more general cases and
essentially completes the methodology. It can be successfully applied in a similar way not only to other
packages but also to the regular Rayleigh–Schrödinger many-body perturbation theory (the radial part of
regular perturbation theory is, however, beyond~the scope of the present paper) \cite{LinMor:82a,HubWil:10a,Shavitt_Bartlett_2009} 
and to the orthogonal operators~\cite{Uylings_2021}. This paper,
together with the previous {ones~\cite{Gaigetal:2024CV,Gaigetal:2024C,Gaigetal:2024CC,Gaigetal:2025VV,Gaigetal:2025VVT,Gaigetal:2026CVT}}, provides a
comprehensive representation of the newly developed version of the stationary second-order Rayleigh–Schr\"odinger many-body perturbation~theory.

The paper consists of an introduction, four sections, a~conclusion, and~two appendices.
{Section~\ref{sec:seconOrder} presents all the Feynman diagrams that describe correlations; 
these can be computed using the RSMBPT method developed in~\cite{Gaigetal:2024CV,Gaigetal:2024C,Gaigetal:2024CC,Gaigetal:2025VV,Gaigetal:2025VVT,Gaigetal:2026CVT}, and for these, analytical expressions are derived in an irreducible tensorial form.
Section~\ref{Sec:rules} presents the rules for getting algebraic expressions in an irreducible tensorial form for any Feynman diagram in $jj$-coupling of the stationary second-order Rayleigh–Schr\"odinger many-body perturbation theory~\cite{Gaigalas:89}.
Section~\ref{Sec:spin_angular} shows how the library \texttt{librang}~\cite{Gaigalas:2022} is extended to calculate spin-angular parts of three-particle Feynman diagrams VV$_3$ and CV$_7$~\cite{Gaigetal:2025VVT,Gaigetal:2026CVT}. 
Section~\ref{Sec:Calculation} provides a test case that is
directly derived from the paper’s focus---the generalization and finalization of the methodology; therefore,
it is intended to illustrate the RSMBPT approach’s application at different stages of the calculation process (MCDHF and RCI).}
Conclusions are presented in Section~\ref{Sec:Conclusions}.


\section{\textls[15]{Relativistic Second-Order Effective Hamiltonian of an Atom or Ion \linebreak  in Irreducible Tensorial Form for Including Correlations  \linebreak in the {\sc\textbf{Grasp}}2018}}
\label{sec:seconOrder}
\begin{figure}[H]
\begin{adjustwidth}{-\extralength}{-\extralength}
\centering 
\setlength{\unitlength}{1mm}
\begin{picture}(180,170)(8,0)
\thicklines
\put(10,165){\line(0,1){10}}
\put(10,173){\vector(0,1){2}}
\put(10,175){\vector(0,1){2}}
\put(12.5,173){\makebox(0,0)[t]{\small{$m$}}}
\multiput(10,165)(1,0){10}{\circle*{0.35}}
\put(10,155){\line(0,10){10}}
\put(10,159){\vector(0,1){2}}
\put(8,160){\makebox(0,0)[t]{\small{$r$}}}
\qbezier(20,155)(15,160)(20,165)
\put(17.6,160){\vector(0,1){2}}
\put(15.5,160){\makebox(0,0)[t]{\small{$s$}}}
\qbezier(20,155)(25,160)(20,165)
\put(22.4,160){\vector(0,-1){2}}
\put(24.5,160){\makebox(0,0)[t]{\small{$a$}}}
\multiput(10,155)(1,0){10}{\circle*{0.35}}
\put(10,145){\line(0,1){10}}
\put(10,145){\vector(0,1){3}}
\put(12.5,148){\makebox(0,0){\small{$m^{\prime}$}}}
\put(10,143){\vector(0,1){3}}
\put(15,139){\makebox(0,0){$\text{CV}_{1}$}}
\put(48,165){\line(0,1){10}}
\put(48,173){\vector(0,1){2}}
\put(48,175){\vector(0,1){2}}
\put(50.5,173){\makebox(0,0)[t]{\small{$m$}}}
\put(58,165){\line(-1,-1){10}}
\put(56,157){\vector(-1,1){2}}
\put(49,162){\makebox(0,0)[t]{\small{$r$}}}
\multiput(48,165)(1,0){10}{\circle*{0.35}}
\put(58,155){\line(0,10){10}}
\put(58,160){\vector(0,-1){2}}
\put(60.5,160){\makebox(0,0)[t]{\small{$a$}}}
\multiput(48,155)(1,0){10}{\circle*{0.35}}
\put(48,145){\line(0,1){10}}
\put(48,145){\vector(0,1){3}}
\put(50.5,148){\makebox(0,0){\small{$m^{\prime}$}}}
\put(48,143){\vector(0,1){3}}
\put(58,155){\line(-1,1){10}}
\put(54,161){\vector(1,1){2}}
\put(49,159){\makebox(0,0)[t]{\small{$s$}}}
\put(53,139){\makebox(0,0){$\text{CV}_{2}$}}
\put(86,165){\line(0,1){10}}
\put(86,173){\vector(0,1){2}}
\put(86,175){\vector(0,1){2}}
\put(88.5,173){\makebox(0,0)[t]{\small{$m$}}}
\put(96,165){\line(1,-1){10}}
\put(104,163){\vector(1,1){2.4}}
\put(103,162){\vector(1,1){2}}
\put(103,166){\makebox(0,0)[t]{\small{$n$}}}
\multiput(86,165)(1,0){10}{\circle*{0.35}}
\put(86,155){\line(0,10){10}}
\put(86,159){\vector(0,1){2}}
\put(84,160){\makebox(0,0)[t]{\small{$r$}}}
\put(96,155){\line(0,10){10}}
\put(96,160){\vector(0,-1){2}}
\put(94,160){\makebox(0,0)[t]{\small{$a$}}}
\multiput(86,155)(1,0){10}{\circle*{0.35}}
\put(86,145){\line(0,1){10}}
\put(86,145){\vector(0,1){3}}
\put(88.5,148){\makebox(0,0){\small{$m^{\prime}$}}}
\put(86,143){\vector(0,1){3}}
\put(96,155){\line(1,1){10}}
\put(105,156){\vector(-1,1){2.4}}
\put(102.2,155.4){\makebox(0,0){\small{$n'$}}}
\put(106,155){\vector(-1,1){2}}
\put(93,139){\makebox(0,0){$\text{CV}_{3}$}}
\put(124,165){\line(0,1){10}}
\put(124,173){\vector(0,1){2}}
\put(124,175){\vector(0,1){2}}
\put(121.5,173){\makebox(0,0)[t]{\small{$m$}}}
\put(134,165){\line(0.5,-1){5}}
\put(131.5,173){\makebox(0,0)[t]{\small{$n$}}}
\multiput(124,165)(1,0){10}{\circle*{0.35}}
\put(124,155){\line(0,10){10}}
\put(139,160){\makebox(0,0)[t]{\small{$r$}}}
\put(129,173){\vector(0,1){2}}
\put(129,175){\vector(0,1){2}}
\put(129,155){\line(0,10){20}}
\put(132.5,162){\vector(-0.5,-1){2}}
\put(133.5,160){\makebox(0,0)[t]{\small{$a$}}}
\multiput(129,155)(1,0){10}{\circle*{0.35}}
\put(124,145){\line(0,1){10}}
\put(124,145){\vector(0,1){3}}
\put(121.5,148){\makebox(0,0){\small{$m^{\prime}$}}}
\put(124,143){\vector(0,1){3}}
\put(129,155){\line(0.5,1){5}}
\put(139,145){\line(0,1){10}}
\put(139,145){\vector(0,1){3}}
\put(139,143){\vector(0,1){3}}
\put(136.5,148){\makebox(0,0){\small{$n^{\prime}$}}}
\put(138,157){\vector(-0.5,1){2}}
\put(129,139){\makebox(0,0){$\text{CV}_{4}$}}
\put(10,115){\line(0,1){10}}
\put(10,123){\vector(0,1){2}}
\put(10,125){\vector(0,1){2}}
\put(7.5,123){\makebox(0,0)[t]{\small{$m$}}}
\put(10,115){\line(0.5,-1){5}}
\put(28,123){\makebox(0,0)[t]{\small{$n$}}}
\multiput(10,115)(1,0){10}{\circle*{0.35}}
\put(20,105){\line(0,10){10}}
\put(10,111){\makebox(0,0)[t]{\small{$r$}}}
\put(25,123){\vector(0,1){2}}
\put(25,125){\vector(0,1){2}}
\put(25,105){\line(0,10){20}}
\put(18.5,112){\vector(-0.5,-1){2}}
\put(15.5,111){\makebox(0,0)[t]{\small{$a$}}}
\multiput(15,105)(1,0){10}{\circle*{0.35}}
\put(20,95){\line(0,1){10}}
\put(20,95){\vector(0,1){3}}
\put(17.5,98){\makebox(0,0){\small{$m^{\prime}$}}}
\put(20,93){\vector(0,1){3}}
\put(15,105){\line(0.5,1){5}}
\put(25,95){\line(0,1){10}}
\put(25,95){\vector(0,1){3}}
\put(25,93){\vector(0,1){3}}
\put(27.5,98){\makebox(0,0){\small{$n^{\prime}$}}}
\put(14,107){\vector(-0.5,1){2}}
\put(15,89){\makebox(0,0){$\text{CV}_{5}$}}
\put(48,115){\line(0,1){10}}
\put(48,123){\vector(0,1){2}}
\put(48,125){\vector(0,1){2}}
\put(50.5,123){\makebox(0,0)[t]{\small{$m$}}}
\multiput(48,115)(1,0){10}{\circle*{0.35}}
\put(48,95){\line(0,1){10}}
\put(48,95){\vector(0,1){3}}
\put(50.5,98){\makebox(0,0){\small{$m^{\prime}$}}}
\put(48,93){\vector(0,1){3}}
\put(48,105){\line(0,10){10}}
\qbezier(58,105)(53,110)(58,115)
\put(55.6,110){\vector(0,1){2}}
\put(53.5,110){\makebox(0,0)[t]{\small{$r$}}}
\qbezier(58,105)(63,110)(58,115)
\put(60.4,110){\vector(0,-1){2}}
\put(62.5,110){\makebox(0,0)[t]{\small{$a$}}}
\multiput(58,105)(1,0){10}{\circle*{0.35}}
\put(68,115){\line(0,1){10}}
\put(68,123){\vector(0,1){2}}
\put(68,125){\vector(0,1){2}}
\put(70.5,123){\makebox(0,0)[t]{\small{$n$}}}
\put(68,105){\line(0,10){10}}
\put(68,93){\vector(0,1){3}}
\put(68,95){\line(0,1){10}}
\put(68,95){\vector(0,1){3}}
\put(70.5,98){\makebox(0,0){\small{$n^{\prime}$}}}
\put(53,89){\makebox(0,0){$\text{CV}_{6}$}}
\put(86,115){\line(0,1){10}}
\put(86,123){\vector(0,1){2}}
\put(86,125){\vector(0,1){2}}
\put(83.5,123){\makebox(0,0)[t]{\small{$m$}}}
\put(87.9,111.1){\makebox(0,0)[t]{\small{$n$}}}
\multiput(86,115)(1,0){10}{\circle*{0.35}}
\put(86,105){\line(0,10){10}}
\put(98.5,109.5){\makebox(0,0)[r]{\small{$a$}}}
\put(88.5,107.5){\vector(1,1){2.4}}
\put(87.5,106.5){\vector(1,1){2}}
\put(96,111){\vector(0,-1){3}}
\put(96,115){\line(-1,-1){8.7}}
\put(96,105){\line(0,10){10}}
\put(96,105){\line(-1,1){8}}
\put(90.5,110.5){\vector(-1,1){2}}
\put(89.5,111.5){\vector(-1,1){2.4}}
\put(88.5,105.4){\makebox(0,0){\small{$n^{\prime}$}}}
\multiput(96,105)(1,0){10}{\circle*{0.35}}
\put(86,95){\line(0,1){10}}
\put(86,95){\vector(0,1){3}}
\put(83.5,98){\makebox(0,0){\small{$m^{\prime}$}}}
\put(86,93){\vector(0,1){3}}
\put(106,95){\line(0,1){20}}
\put(106,115){\vector(0,1){3}}
\put(106,113){\vector(0,1){3}}
\put(103.5,113){\makebox(0,0){\small{$p$}}}
\put(106,95){\vector(0,1){3}}
\put(106,93){\vector(0,1){3}}
\put(103.5,98){\makebox(0,0){\small{$p^{\prime}$}}}
\put(96,89){\makebox(0,0){$\text{CV}_{7}$}}
\multiput(124,115)(1,0){10}{\circle*{0.35}}
\qbezier(124,105)(119,110)(124,115)
\put(121.6,110){\vector(0,1){2}}
\put(119.5,110){\makebox(0,0)[t]{\small{$r$}}}
\qbezier(124,105)(129,110)(124,115)
\put(126.4,110){\vector(0,-1){2}}
\put(127.5,113){\makebox(0,0)[t]{\small{$a$}}}
\qbezier(134,105)(129,110)(134,115)
\put(131.6,110){\vector(0,1){2}}
\put(130.5,107.5){\makebox(0,0)[t]{\small{$s$}}}
\qbezier(134,105)(139,110)(134,115)
\put(136.4,110){\vector(0,-1){2}}
\put(138.5,110){\makebox(0,0)[t]{\small{$b$}}}
\multiput(124,105)(1,0){10}{\circle*{0.35}}
\put(129,89){\makebox(0,0){$\text{CC}_{1}$}}
\multiput(10,68)(1,0){10}{\circle*{0.35}}
\put(20,68){\line(-1,-1){10}}
\put(18,60){\vector(-1,1){2}}
\put(11.5,65){\makebox(0,0)[t]{\small{$r$}}}
\put(10,58){\line(0,10){10}}
\put(10,63){\vector(0,-1){2}}
\put(7.5,63){\makebox(0,0)[t]{\small{$a$}}}
\put(20,58){\line(0,10){10}}
\put(20,63){\vector(0,-1){2}}
\put(22.5,63){\makebox(0,0)[t]{\small{$b$}}}
\put(20,58){\line(-1,1){10}}
\put(16,64){\vector(1,1){2}}
\put(11.5,62){\makebox(0,0)[t]{\small{$s$}}}
\multiput(10,58)(1,0){10}{\circle*{0.35}}
\put(15,54){\makebox(0,0){$\text{CC}_{2}$}}
\multiput(48,68)(1,0){10}{\circle*{0.35}}
\qbezier(48,58)(43,63)(48,68)
\put(45.6,63){\vector(0,1){2}}
\put(43.5,63){\makebox(0,0)[t]{\small{$r$}}}
\qbezier(48,58)(53,63)(48,68)
\put(50.4,63){\vector(0,-1){2}}
\put(52,63){\makebox(0,0)[t]{\small{$a$}}}
\put(58,68){\line(1,-1){10}}
\put(66,66){\vector(1,1){2.4}}
\put(65,65){\vector(1,1){2}}
\put(65,69){\makebox(0,0)[t]{\small{$m$}}}
\put(58,58){\line(0,10){10}}
\put(58,63){\vector(0,-1){2}}
\put(56,63.6){\makebox(0,0)[t]{\small{$b$}}}
\put(58,58){\line(1,1){10}}
\put(67,59){\vector(-1,1){2.4}}
\put(64.2,58.4){\makebox(0,0){\small{$m'$}}}
\put(68,58){\vector(-1,1){2}}
\multiput(48,58)(1,0){10}{\circle*{0.35}}
\put(53,54){\makebox(0,0){$\text{CC}_{3}$}}
\multiput(86,68)(1,0){10}{\circle*{0.35}}
\put(96,68){\line(-1,-1){10}}
\put(93,61){\vector(1,-1){2}}
\put(95.5,65){\makebox(0,0)[t]{\small{$a$}}}
\put(86,58){\line(0,10){10}}
\put(86,63){\vector(0,1){2}}
\put(83.5,63){\makebox(0,0)[t]{\small{$r$}}}
\put(96,68){\line(1,-1){10}}
\put(104,66){\vector(1,1){2.4}}
\put(103,65){\vector(1,1){2}}
\put(103,69){\makebox(0,0)[t]{\small{$m$}}}
\put(96,58){\line(-1,1){10}}
\put(94,66){\vector(-1,-1){2}}
\put(95.5,62.5){\makebox(0,0)[t]{\small{$b$}}}
\put(96,58){\line(1,1){10}}
\put(105,59){\vector(-1,1){2.4}}
\put(102.2,58.4){\makebox(0,0){\small{$m'$}}}
\put(106,58){\vector(-1,1){2}}
\multiput(86,58)(1,0){10}{\circle*{0.35}}
\put(91,54){\makebox(0,0){$\text{CC}_{4}$}}
\multiput(134,68)(1,0){10}{\circle*{0.35}}
\put(144,68){\line(-1,-1){10}}
\put(140,62){\vector(1,-1){2}}
\put(135.5,65){\makebox(0,0)[t]{\small{$a$}}}
\put(144,58){\line(0,10){10}}
\put(144,61.5){\vector(0,1){2}}
\put(144,63.5){\vector(0,1){2}}
\put(146.5,63){\makebox(0,0)[t]{\small{$n$}}}
\put(124,68){\line(1,-1){10}}
\put(125,59){\vector(1,1){2.4}}
\put(124,58){\vector(1,1){2}}
\put(127.2,69){\makebox(0,0)[t]{\small{$m$}}}
\put(144,58){\line(-1,1){10}}
\put(142,66){\vector(-1,-1){2}}
\put(135.5,62.9){\makebox(0,0)[t]{\small{$b$}}}
\put(124,58){\line(1,1){10}}
\put(126,66){\vector(-1,1){2.4}}
\put(127.5,58.4){\makebox(0,0){\small{$m'$}}}
\put(127,65){\vector(-1,1){2}}
\multiput(134,58)(1,0){10}{\circle*{0.35}}
\put(139,54){\makebox(0,0){$\text{CC}_{5}$}}
\multiput(20,30)(1,0){10}{\circle*{0.35}}
\put(10,30){\line(1,-1){10}}
\put(11,21){\vector(1,1){2.4}}
\put(10,20){\vector(1,1){2}}
\put(13.2,31){\makebox(0,0)[t]{\small{$m$}}}
\put(10,20){\line(1,1){10}}
\put(12,28){\vector(-1,1){2.4}}
\put(13.5,20.4){\makebox(0,0){\small{$m'$}}}
\put(13,27){\vector(-1,1){2}}
\put(20,20){\line(0,10){10}}
\put(20,25){\vector(0,-1){2}}
\put(22.5,25){\makebox(0,0)[t]{\small{$a$}}}
\put(30,20){\line(0,10){10}}
\put(30,25){\vector(0,-1){2}}
\put(27.5,25.7){\makebox(0,0)[t]{\small{$b$}}}
\put(30,30){\line(1,-1){10}}
\put(38,28){\vector(1,1){2.4}}
\put(37,27){\vector(1,1){2}}
\put(37,31){\makebox(0,0)[t]{\small{$n$}}}
\put(30,20){\line(1,1){10}}
\put(39,21){\vector(-1,1){2.4}}
\put(36.2,20.4){\makebox(0,0){\small{$n'$}}}
\put(40,20){\vector(-1,1){2}}
\multiput(20,20)(1,0){10}{\circle*{0.35}}
\put(25,04){\makebox(0,0){$\text{CC}_{6}$}}
\put(48,30){\line(0,1){10}}
\put(48,38){\vector(0,1){2}}
\put(48,40){\vector(0,1){2}}
\put(50.5,38){\makebox(0,0)[t]{\small{$m$}}}
\put(58,30){\line(-1,-1){10}}
\put(56,22){\vector(-1,1){2}}
\put(49,27){\makebox(0,0)[t]{\small{$r$}}}
\multiput(48,30)(1,0){10}{\circle*{0.35}}
\put(58,20){\line(0,10){10}}
\put(58,26.5){\vector(0,-1){2}}
\put(58,24.5){\vector(0,-1){2}}
\put(60.5,25){\makebox(0,0)[t]{\small{$n$}}}
\multiput(48,20)(1,0){10}{\circle*{0.35}}
\put(48,10){\line(0,1){10}}
\put(48,10){\vector(0,1){3}}
\put(50.5,13){\makebox(0,0){\small{$m^{\prime}$}}}
\put(48,08){\vector(0,1){3}}
\put(58,20){\line(-1,1){10}}
\put(54,26){\vector(1,1){2}}
\put(49,24){\makebox(0,0)[t]{\small{$s$}}}
\put(53,04){\makebox(0,0){$\text{VV}_{1}$}}
\put(86,30){\vector(0,1){10}}
\put(86,40){\vector(0,1){2}}
\put(83.5,38){\makebox(0,0)[t]{\small{$m$}}}
\put(96,30){\vector(0,1){10}}
\put(96,40){\vector(0,1){2}}
\put(98,38){\makebox(0,0)[t]{\small{$n$}}}
\multiput(86,30)(1,0){10}{\circle*{0.35}}
\put(86,20){\line(0,10){10}}
\put(86,24){\vector(0,1){2}}
\put(83.5,28){\makebox(0,0)[t]{\small{$r$}}}
\put(96,20){\line(0,10){10}}
\put(96,24){\vector(0,1){2}}
\put(98,28){\makebox(0,0)[t]{\small{$s$}}}
\multiput(86,20)(1,0){10}{\circle*{0.35}}
\put(86,10){\line(0,1){10}}
\put(86,10){\vector(0,1){3}}
\put(83.5,13){\makebox(0,0){\small{$m'$}}}
\put(86,08){\vector(0,1){3}}
\put(96,10){\line(0,1){10}}
\put(96,10){\vector(0,1){3}}
\put(91,04){\makebox(0,0){$\text{VV}_{2}$}}
\put(98,13){\makebox(0,0){\small{$n'$}}}
\put(96,08){\vector(0,1){3}}
\put(124,30){\line(0,1){10}}
\put(124,38){\vector(0,1){2}}
\put(124,40){\vector(0,1){2}}
\put(121.5,38){\makebox(0,0)[t]{\small{$m$}}}
\put(131.5,38){\makebox(0,0)[t]{\small{$n$}}}
\multiput(124,30)(1,0){10}{\circle*{0.35}}
\put(124,20){\line(0,10){10}}
\put(131.6,24){\makebox(0,0)[r]{\small{$r$}}}
\put(134,38){\vector(0,1){2}}
\put(134,40){\vector(0,1){2}}
\put(134,24){\vector(0,1){3}}
\put(134,10){\line(0,10){30}}
\put(134,10){\vector(0,1){3}}
\put(134,08){\vector(0,1){3}}
\put(131.5,13){\makebox(0,0){\small{$n^{\prime}$}}}
\multiput(134,20)(1,0){10}{\circle*{0.35}}
\put(124,10){\line(0,1){10}}
\put(124,10){\vector(0,1){3}}
\put(121.5,13){\makebox(0,0){\small{$m^{\prime}$}}}
\put(124,08){\vector(0,1){3}}
\put(144,10){\line(0,1){20}}
\put(144,30){\vector(0,1){3}}
\put(144,28){\vector(0,1){3}}
\put(141.5,28){\makebox(0,0){\small{$p$}}}
\put(144,10){\vector(0,1){3}}
\put(144,08){\vector(0,1){3}}
\put(141.5,13){\makebox(0,0){\small{$p^{\prime}$}}}
\put(134,04){\makebox(0,0){$\text{VV}_{3}$}}
\end{picture}
\end{adjustwidth}
\caption{The 
 Feynman diagrams of the second-order effective Hamiltonian that are included in the {\sc Grasp}2018{\sc\_PT} computer~package.}
\label{Feynman_Diagrams}
\end{figure}
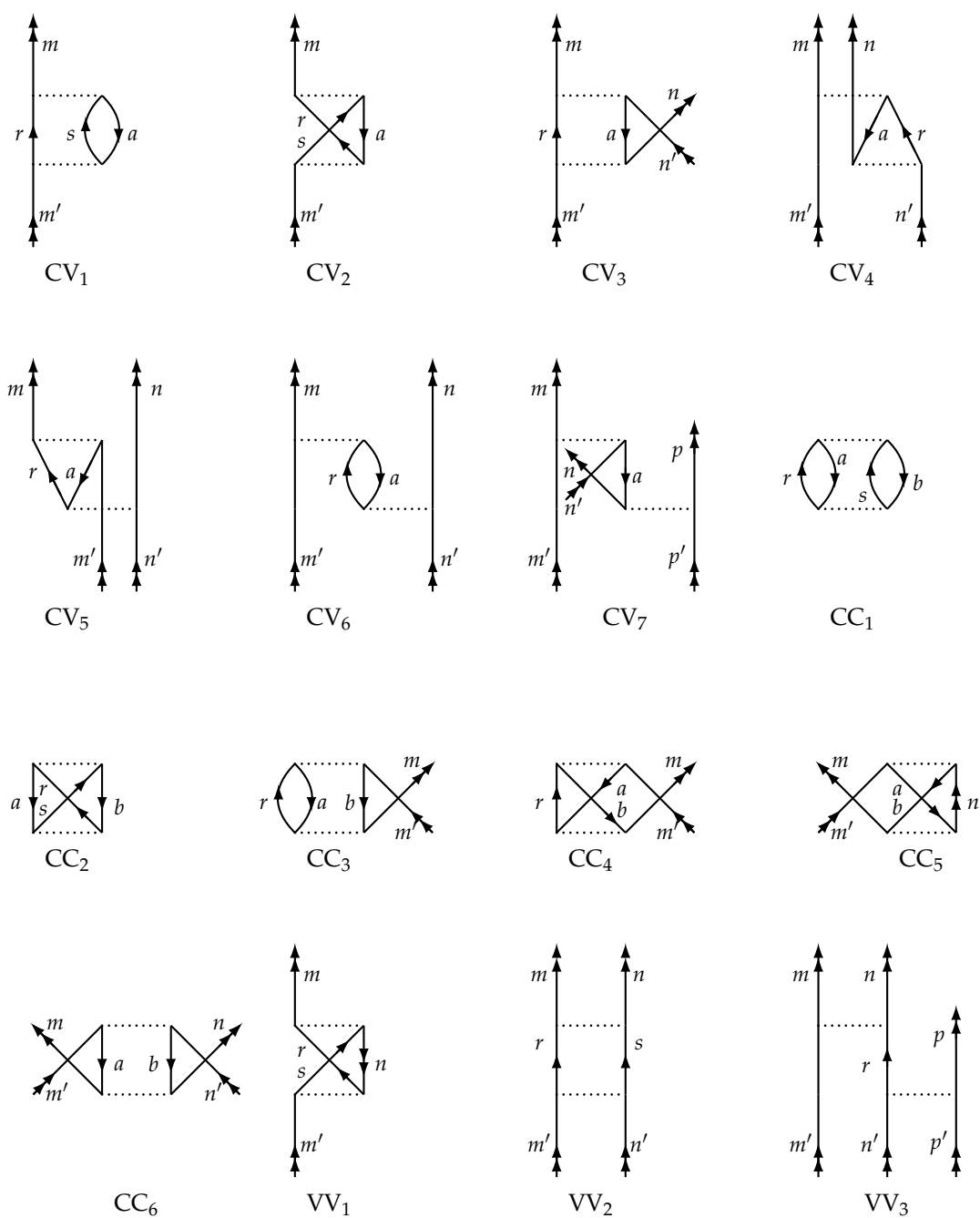

The relativistic second-order effective Hamiltonian for an atom or ion, formulated in an irreducible tensorial form for the inclusion of correlations, is implemented in the new extended {\sc Grasp} version, named {\sc Grasp}2018{\sc\_PT}. This version builds upon the latest release of the {\sc Grasp} package ({\sc Grasp}2018) and the {\sc Graspg} \cite{Li_2023_GRASPG,Si_2025_GRASPG}. {\sc Graspg} was applied because this program package allows the $F$, $F'$, and~$G$ orbital sets to be easily distinguished during atomic data computations. It should be mentioned that  extensions based on the theory detailed below can also be applied 
to older versions of the {\sc Grasp} packages~\cite{Dyaetal:89a,GRASP92,Jonetal:2007a,Jonetal:2013a}.
All the Feynman diagrams that describe a particular type of correlation are implemented in the {\sc Grasp}2018{\sc\_PT} packages, as shown in Figure~\ref{Feynman_Diagrams}. Analytical expressions in irreducible tensorial form have been derived for them, and~the appropriate software has been developed to calculate them. The~analytical expressions of the diagrams are general and suitable for any atom or ion with any number of subshells in the configurations (including open subshells) and with any number of electrons in the subshells. Therefore, the~developed methodology is general and can be applicable to any task and in any computational package.

These diagrams can thus be used to assess the importance of the correlation configurations, in~other words, to~determine their correlation influence on the calculations.
This includes single and double excitations attributed to CV, CC, C, and~VV correlations. In~Figure~\ref{Feynman_Diagrams}, all Feynman diagrams are grouped so that the most relevant and problematic correlations for RCI or MCDHF methods are mentioned before the others. The~details of how this method is used and the results it is expected to produce are published in papers~\cite{Gaigetal:2024CV,Gaigetal:2024C,Gaigetal:2024CC,Gaigetal:2025VV,Gaigetal:2025VVT,Gaigetal:2026CVT}. Analytical expressions for all these diagrams are also given in these~papers.

\subsection{Correlations Which Are Included in the {\sc\textit{Grasp}}2018{\sc\_\textit{PT}}}

In this section, we will present all the excitations (correlations) whose influence and importance the developed approach~\cite{Gaigetal:2024CV,Gaigetal:2024C,Gaigetal:2024CC,Gaigetal:2025VV,Gaigetal:2025VVT,Gaigetal:2026CVT} allows to be identified by the RSMBPT method in~advance of the RCI and MCDHF calculations. We will also list a few cases where correlations in the RCI and MCDHF methods need to be dealt with~routinely.

\subsubsection{Core–Valence~Correlations}

The first group of diagrams in Figure~\ref{Feynman_Diagrams} includes Feynman diagrams that describe CV correlations. They are characterized by the fact that there must be one internal downward-directed line (core (hole) line) and at least two free double arrow lines (particle lines acting on valence subshells). This group includes diagrams CV$_1$, CV$_2$, CV$_3$, CV$_4$, CV$_5$, CV$_6$, and~CV$_7$.
CV$_1$ and CV$_2$ are one-particle; CV$_3$, CV$_4$, CV$_5$, and~CV$_6$ are two-particle; and~CV$_7$ is a three-particle diagram. 
Feynman diagram CV$_1$ describing correlations has the same influence on the relativistic configuration.
Other diagrams affect both the configuration and its splitting. All of them can describe the four types of CV excitations (correlations) included in the {\sc Grasp}2018{\sc\_PT} package. They are as follows:

\begin{itemize}
\item The first type: 
\begin{equation}
\label{eq:CV-a}
    (n_{a} \ell_{a})\, j_{a}^{2j_a+1} \, (n_{m} \ell_{m})\, j_{m}^{w_m} 
   \rightarrow (n_{a} \ell_{a})\, j_{a}^{2j_a} \; (n_{m} \ell_{m})\, j_{m}^{w_m-1} \; (n_{r} \ell_{r})\, j_{r} \; (n_{s} \ell_{s})\, j_{s}
\end{equation}
\begin{equation}
\label{eq:CV-b}
    (n_{a} \ell_{a})\, j_{a}^{2j_a+1} \, (n_{m} \ell_{m})\, j_{m}^{w_m} 
   \rightarrow (n_{a} \ell_{a})\, j_{a}^{2j_a} \; (n_{m} \ell_{m})\, j_{m}^{w_m-1} \; (n_{s} \ell_{s})\, j_{s}^2
\end{equation}

\item {The second type:}
{\small
\begin{equation}
\label{eq:CV-c}
    (n_{a} \ell_{a})\, j_{a}^{2j_a+1} \, (n_{m} \ell_{m})\, j_{m}^{w_m} \, (n_{n} \ell_{n})\, j_{n}^{w_n} 
   \rightarrow (n_{a} \ell_{a})\, j_{a}^{2j_a} \; (n_{m} \ell_{m})\, j_{m}^{w_m-1} \; (n_{n} \ell_{n})\, j_{n}^{w_n+1}
	 \; (n_{r} \ell_{r})\, j_{r} 
\end{equation}}

\item {The third type:}
\begin{equation}
\label{eq:not-CV-a}
    (n_{a} \ell_{a})\, j_{a}^{2j_a+1} \, (n_{m} \ell_{m})\, j_{m}^{w_m} \, (n_{n} \ell_{n})\, j_{n}^{w_n} 
   \rightarrow (n_{a} \ell_{a})\, j_{a}^{2j_a} \; (n_{m} \ell_{m})\, j_{m}^{w_m-1} \, (n_{n} \ell_{n})\, j_{n}^{w_n+2} 
\end{equation}

\item {The fourth type:}
\begin{eqnarray}
\label{eq:not-CV-b}
&
\hspace{-7.0cm}
    (n_{a} \ell_{a})\, j_{a}^{2j_a+1} \, (n_{m} \ell_{m})\, j_{m}^{w_m} \, (n_{n} \ell_{n})\, j_{n}^{w_n} \, (n_{p} \ell_{p})\, j_{p}^{w_p} 
   \nonumber  \\[1ex]
&
\hspace{1.5cm}
   \rightarrow (n_{a} \ell_{a})\, j_{a}^{2j_a} \; (n_{m} \ell_{m})\, j_{m}^{w_m-1} \, (n_{n} \ell_{n})\,j_{n}^{w_n+1}  \; (n_{p} \ell_{p})\, j_{p}^{w_p+1} .
\end{eqnarray}
\end{itemize}

\subsubsection{Core~Correlations}

C correlations are described by the same diagrams, CV$_3$, CV$_4$, CV$_5$, and~CV$_6$, describing CV correlations. All open lines with double arrows act on the same valence subshell in this case. This leads to the type of C correlation that is included in this methodology:

\begin{itemize}
\item {the first type:}
\vspace{-6pt}
\begin{equation}
\label{eq:C-a}
    (n_{a} \ell_{a})\, j_{a}^{2j_a+1} \, (n_{m} \ell_{m})\, j_{m}^{w_m}
    \rightarrow (n_{a} \ell_{a})\, j_{a}^{2j_a} \; (n_{m} \ell_{m})\, j_{m}^{w_m} \; (n_{r} \ell_{r})\, j_{r} .
\end{equation}
\end{itemize}

\subsubsection{Core–Core~Correlations}

The second group of Feynman diagrams in Figure~\ref{Feynman_Diagrams} describes the CC correlations.
These diagrams are unique in that they have two internal downward-pointing lines (core (hole) lines). The~remaining lines are the inner upward-pointing lines and/or the free double arrow lines. CC$_1$ and CC$_2$ are vacuum; CC$_3$, CC$_4$, and~CC$_5$ are one-particle; and~CC$_6$ is a two-particle diagram. Feynman diagrams CC$_1$ and CC$_2$ describing correlations have the same influence on the relativistic configuration.
Other diagrams affect both the configuration and its splitting. All of them can describe the four types of CC excitations (correlations) included in the {\sc Grasp}2018{\sc\_PT} package. They are as follows:

\clearpage
\begin{itemize}
\item {The first type:}
\begin{eqnarray}
\label{eq:CC1-a}
&
\hspace{-7.0cm}
(n_{a} \ell_{a})\, j_{a}^{2j_a+1} \,  (n_{b} \ell_{b})\, j_{b}^{2j_b+1} \, 
(n_{m} \ell_{m})\, j_{m}^{w_m} \, (n_{n} \ell_{n})\, j_{n}^{w_n} 
   \nonumber  \\[1ex]
&
\hspace{1.5cm}	
   \rightarrow (n_{a} \ell_{a})\, j_{a}^{2j_a} \,  (n_{b} \ell_{b})\, j_{b}^{2j_b} \,
	(n_{m} \ell_{m})\, j_{m}^{w_m} \; (n_{n} \ell_{n})\, j_{n}^{w_n} \;
	(n_{r} \ell_{r})\, j_{r} \; (n_{s} \ell_{s})\, j_{s}
\end{eqnarray}
\vspace{-21pt}
\begin{eqnarray}
\label{eq:CC1-b}
&
\hspace{-7.0cm}
(n_{a} \ell_{a})\, j_{a}^{2j_a+1} \,  (n_{b} \ell_{b})\, j_{b}^{2j_b+1} \, 
(n_{m} \ell_{m})\, j_{m}^{w_m} \, (n_{n} \ell_{n})\, j_{n}^{w_n}
   \nonumber  \\
&
\hspace{0.3cm}
   \rightarrow (n_{a} \ell_{a})\, j_{a}^{2j_a} \,  (n_{b} \ell_{b})\, j_{b}^{2j_b} \,
	(n_{m} \ell_{m})\, j_{m}^{w_m} \; (n_{n} \ell_{n})\, j_{n}^{w_n} \;
	(n_{r} \ell_{r})\, j_{r}^2
\end{eqnarray}
\vspace{-21pt}
\begin{eqnarray}
\label{eq:CC1-c}
&
\hspace{-7.2cm}
(n_{a} \ell_{a})\, j_{a}^{2j_a+1} 
(n_{m} \ell_{m})\, j_{m}^{w_m} \, (n_{n} \ell_{n})\, j_{n}^{w_n} 
   \nonumber  \\
&
\hspace{1.6cm}
   \rightarrow (n_{a} \ell_{a})\, j_{a}^{2j_a-1} \,
	(n_{m} \ell_{m})\, j_{m}^{w_m} \; (n_{n} \ell_{n})\, j_{n}^{w_n} \;
	(n_{r} \ell_{r})\, j_{r} \;(n_{s} \ell_{s})\, j_{s}
\end{eqnarray}
\vspace{-21pt}
\begin{eqnarray}
\label{eq:CC1-d}
&
\hspace{-7.2cm}
(n_{a} \ell_{a})\, j_{a}^{2j_a+1} 
(n_{m} \ell_{m})\, j_{m}^{w_m} \, (n_{n} \ell_{n})\, j_{n}^{w_n} 
   \nonumber  \\
&
\hspace{1.6cm}
   \rightarrow (n_{a} \ell_{a})\, j_{a}^{2j_a-1} \,
	(n_{m} \ell_{m})\, j_{m}^{w_m} \; (n_{n} \ell_{n})\, j_{n}^{w_n} \;
	(n_{r} \ell_{r})\, j_{r}^2
\end{eqnarray}

\item {The second type:}
\begin{eqnarray}
\label{eq:CC2-a}
&
\hspace{-7.0cm}
(n_{a} \ell_{a})\, j_{a}^{2j_a+1} \,  (n_{b} \ell_{b})\, j_{b}^{2j_b+1} \, 
(n_{m} \ell_{m})\, j_{m}^{w_m} \, (n_{n} \ell_{n})\, j_{n}^{w_n} 
   \nonumber  \\
&
\hspace{1.5cm}	
   \rightarrow (n_{a} \ell_{a})\, j_{a}^{2j_a} \,  (n_{b} \ell_{b})\, j_{b}^{2j_b} \,
	(n_{m} \ell_{m})\, j_{m}^{w_m+1} \; (n_{n} \ell_{n})\, j_{n}^{w_n} \;
	(n_{r} \ell_{r})\, j_{r}
\end{eqnarray}
\vspace{-21pt}
\begin{eqnarray}
\label{eq:CC2-b}
&
\hspace{-7.2cm}
(n_{a} \ell_{a})\, j_{a}^{2j_a+1} \,
(n_{m} \ell_{m})\, j_{m}^{w_m} \, (n_{n} \ell_{n})\, j_{n}^{w_n} 
   \nonumber  \\
&
\hspace{1.6cm}
   \rightarrow (n_{a} \ell_{a})\, j_{a}^{2j_a-1} \,
	(n_{m} \ell_{m})\, j_{m}^{w_m+1} \; (n_{n} \ell_{n})\, j_{n}^{w_n} \;
	(n_{r} \ell_{r})\, j_{r}
\end{eqnarray}

\item {The third type:}
\begin{eqnarray}
\label{eq:CC3-a}
&
\hspace{-7.0cm}
(n_{a} \ell_{a})\, j_{a}^{2j_a+1} \,  (n_{b} \ell_{b})\, j_{b}^{2j_b+1} \, 
(n_{m} \ell_{m})\, j_{m}^{w_m} \, (n_{n} \ell_{n})\, j_{n}^{w_n} 
   \nonumber  \\
&
\hspace{1.5cm}	
   \rightarrow (n_{a} \ell_{a})\, j_{a}^{2j_a} \,  (n_{b} \ell_{b})\, j_{b}^{2j_b} \,
	(n_{m} \ell_{m})\, j_{m}^{w_m+2} \; (n_{n} \ell_{n})\, j_{n}^{w_n}
\end{eqnarray}
\vspace{-21pt}
\begin{eqnarray}
\label{eq:CC3-b}
&
\hspace{-7.2cm}
(n_{a} \ell_{a})\, j_{a}^{2j_a+1} \,
(n_{m} \ell_{m})\, j_{m}^{w_m} \, (n_{n} \ell_{n})\, j_{n}^{w_n} 
   \nonumber  \\
&
\hspace{1.6cm}
   \rightarrow (n_{a} \ell_{a})\, j_{a}^{2j_a-1} \,
	(n_{m} \ell_{m})\, j_{m}^{w_m+2} \; (n_{n} \ell_{n})\, j_{n}^{w_n}
\end{eqnarray}

\item {The fourth type:}
\begin{eqnarray}
\label{eq:CC4-a}
&
\hspace{-7.0cm}
(n_{a} \ell_{a})\, j_{a}^{2j_a+1} \,  (n_{b} \ell_{b})\, j_{b}^{2j_b+1} \, 
(n_{m} \ell_{m})\, j_{m}^{w_m} \, (n_{n} \ell_{n})\, j_{n}^{w_n} 
   \nonumber  \\
&
\hspace{1.5cm}	
   \rightarrow (n_{a} \ell_{a})\, j_{a}^{2j_a} \,  (n_{b} \ell_{b})\, j_{b}^{2j_b} \,
	(n_{m} \ell_{m})\, j_{m}^{w_m+1} \; (n_{n} \ell_{n})\, j_{n}^{w_n+1}
\end{eqnarray}
\vspace{-21pt}
\begin{eqnarray}
\label{eq:CC4-b}
&
\hspace{-7.0cm}
(n_{a} \ell_{a})\, j_{a}^{2j_a+1} \,  
(n_{m} \ell_{m})\, j_{m}^{w_m} \, (n_{n} \ell_{n})\, j_{n}^{w_n} 
   \nonumber  \\
&
\hspace{1.5cm}	
   \rightarrow (n_{a} \ell_{a})\, j_{a}^{2j_a-1} \,  
	(n_{m} \ell_{m})\, j_{m}^{w_m+1} \; (n_{n} \ell_{n})\, j_{n}^{w_n+1} .
\end{eqnarray}
\end{itemize}

\subsubsection{Valence–Valence~Correlations}

The third group of Feynman diagrams in Figure~\ref{Feynman_Diagrams} describes VV correlations.
These diagrams are unique in that they have at least one internal upward-directed line (particle line acting on virtual orbitals). The~remaining lines are the double arrow lines (particle lines acting on valence subshells). VV$_1$ is a one-particle diagram, VV$_2$ is a two-particle diagram, and~VV$_3$ is a three-particle diagram. All of them can describe the four types of VV excitations (correlations) included in the {\sc Grasp}2018{\sc\_PT} package. They are as follows:

\begin{itemize}
\item {The first type:}
\begin{equation}
\label{eq:VV-a} 
(n_{m} \ell_{m}) \; j_{m}^{w_m} \; (n_{n} \ell_{n}) \; j_{n}^{w_n}
   \rightarrow (n_{m} \ell_{m})\, j_{m}^{w_m-2} \; (n_{n} \ell_{n}) \; j_{n}^{w_n}
	 \; (n_{r} \ell_{r}) \; j_{r} \; (n_{s} \ell_{s}) \; j_{s}
\end{equation}
\vspace{-21pt}
\begin{equation}
\label{eq:VV-a2} 
(n_{m} \ell_{m}) \; j_{m}^{w_m} \; (n_{n} \ell_{n}) \; j_{n}^{w_n}
   \rightarrow (n_{m} \ell_{m})\, j_{m}^{w_m-2} \; (n_{n} \ell_{n}) \; j_{n}^{w_n}
	 \; (n_{s} \ell_{s})\, j_{s}^2
\end{equation}

\item {The second type:}
\begin{equation}
\label{eq:VV-b} 
(n_{m} \ell_{m}) \; j_{m}^{w_m} \; (n_{n} \ell_{n}) \; j_{n}^{w_n}
   \rightarrow (n_{m} \ell_{m})\, j_{m}^{w_m-1} \; (n_{n} \ell_{n}) \; j_{n}^{w_n-1}
	 \; (n_{r} \ell_{r}) \; j_{r} \; (n_{s} \ell_{s}) \; j_{s}
\end{equation}
\vspace{-21pt}
\begin{equation}
\label{eq:VV-b2} 
(n_{m} \ell_{m}) \; j_{m}^{w_m} \; (n_{n} \ell_{n}) \; j_{n}^{w_n}
   \rightarrow (n_{m} \ell_{m})\, j_{m}^{w_m-1} \; (n_{n} \ell_{n}) \; j_{n}^{w_n-1}
	 \; (n_{s} \ell_{s})\, j_{s}^2
\end{equation}

\item {The third type:}
\begin{equation}
\label{eq:not-VV-a}
    (n_{m} \ell_{m})\, j_{m}^{w_m} \; (n_{n} \ell_{n})\, j_{n}^{w_n} 
   \rightarrow (n_{m} \ell_{m})\, j_{m}^{w_m+1} \; (n_{n} \ell_{n})\, j_{n}^{w_n-2} \; (n_{r} \ell_{r})\, j_{r} 
\end{equation}

\item {The fourth type:}
{\small
\begin{equation}
\label{eq:not-VV-b}
    (n_{m} \ell_{m})\, j_{m}^{w_m} \; (n_{n} \ell_{n})\, j_{n}^{w_n} \; (n_{p} \ell_{p})\, j_{p}^{w_p} 
   \rightarrow (n_{m} \ell_{m})\, j_{m}^{w_m+1} \; (n_{n} \ell_{n})\, j_{n}^{w_n-1}  \; (n_{p} \ell_{p})\, j_{p}^{w_p-1} \; (n_{r} \ell_{r})\, j_{r} . 
\end{equation}}

\end{itemize}


\subsubsection{Contributions of RSMBPT to Off-Diagonal Matrix~Elements}

The following are contributions of RSMBPT to~off-diagonal 

$\redmem{(n_{m} \ell_{m}) \, j_{m}^{w_m} \, (n_{n} \ell_{n}) \, j_{n}^{w_n}}{\, \widehat{{\cal H}}^{(2)}_{Effective} \,}{(n_{m} \ell_{m})\, j_{m}^{w_m-2} \, (n_{n} \ell_{n})\, j_{n}^{w_n+2} }$  matrix elements:

\begin{itemize}
\item {The second type of core–valence correlations}
{\small
\begin{equation}
\label{eq:CV-off_Diagonal} 
(n_{a} \ell_{a}) \; j_{a}^{2j_a+1} \; (n_{m} \ell_{m}) \; j_{m}^{w_m} \; (n_{n} \ell_{n}) \; j_{n}^{w_n}
   \rightarrow (n_{a} \ell_{a})\, j_{a}^{2j_a} \; (n_{m} \ell_{m}) \; j_{m}^{w_m-1} \; (n_{n} \ell_{n}) \; j_{n}^{w_n+1}
	 \; (n_{r} \ell_{r}) \; j_{r} .
\end{equation}}

This type of correlation is described by two-particle Feynman diagrams CV$_3$, CV$_4$, CV$_5$, and~CV$_6$.

\item {The third type of core–core correlations}
\begin{eqnarray}
\label{eq:CC-off_Diagonal}
&
\hspace{-5.5cm}
(n_{a} \ell_{a}) \; j_{a}^{2j_a+1} \, (n_{b} \ell_{b}) \; j_{b}^{2j_b+1} \; (n_{m} \ell_{m})\; j_{m}^{w_m} \; (n_{n} \ell_{n})\; j_{n}^{w_n}
   \nonumber  \\[1ex]
&
\hspace{1.5cm}
   \rightarrow (n_{a} \ell_{a})\, j_{a}^{2j_a} \; (n_{b} \ell_{b}) \; j_{b}^{2j_b} \; (n_{m} \ell_{m}) \; j_{m}^{w_m} \; (n_{n} \ell_{n}) \; j_{n}^{w_n+2} .
\end{eqnarray}

This type of correlation is described by the two-particle Feynman diagram CC$_6$.

\item {The first type of valence–valence correlations}
\begin{equation}
\label{eq:VV-off_Diagonal} 
(n_{m} \ell_{m}) \; j_{m}^{w_m} \; (n_{n} \ell_{n}) \; j_{n}^{w_n}
   \rightarrow (n_{m} \ell_{m})\, j_{m}^{w_m-2} \; (n_{n} \ell_{n}) \; j_{n}^{w_n}
	 \; (n_{r} \ell_{r}) \; j_{r} \; (n_{s} \ell_{s}) \; j_{s} .
\end{equation}

This type of correlation is described by the two-particle Feynman diagram VV$_2$.

\end{itemize}

\subsection{Correlations That Are Not Included in the {\sc\textit{Grasp}}2018{\sc \_\textit{PT}}}

The following correlations are not included in the {\sc Grasp}2018{\sc\_PT} program:

\begin{itemize}
\item {Valence correlations}
\begin{equation}
\label{eq:not-V-a}
    (n_{m} \ell_{m})\, j_{m}^{w_m} 
   \rightarrow (n_{m} \ell_{m})\, j_{m}^{w_m-1} \; (n_{r} \ell_{r})\, j_{r} 
\end{equation}

\item {Core correlations}
\begin{equation}
\label{eq:not-C-a}
    (n_{a} \ell_{a})\, j_{a}^{2j_a+1} \, (n_{m} \ell_{m})\, j_{m}^{w_m} 
   \rightarrow (n_{a} \ell_{a})\, j_{a}^{2j_a} \; (n_{m} \ell_{m})\, j_{m}^{w_m+1} .
\end{equation}
\end{itemize}

The above excitations shall be calculated in the regular way as before in the {\sc Grasp}2018~package.

\subsection{Multireference~Space}

The choice of multireference (MR) space for the  Rayleigh–Schr\"odinger many-body perturbation theory ($P$-space according to many-body perturbation theory~\cite{LinMor:82a}) is the same as for the ordinary calculation of the {\sc Grasp}2018~\cite{Gaigetal:2024CV}.
The simplest way to construct an MR is to include in the MR only those CSFs for which the eigenvalue problem is solvable. But~in general, the~choice of the MR space is an important and responsible task in the search for accurate atomic characteristics. Often, the MR space is extended so that the CSFs in the MR are those that can be formed from nearly degenerate configurations~\cite{Fisetal:16a,CompAS-book} (see chapter~4 in~\cite{CompAS-book}). 
 Thus, although~in the RSMBPT theory we only consider single and double excitations, it can also be considered to include the most important excitations higher than double excitations when, in~the MR space, to~the CSFs for which eigenvalues are searched for, we add the CSFs for which eigenvalues are not searched for but are only included for correlations in single and double~excitations.

The wave function based on the CSFs in the MR is the first approximation, and~it is the starting point for further refinements. So the selection of the MR in advance or {\it a priori 
} is far from trivial, and~it often requires a number of exploratory calculations to find a good MR. For~details on how the MR space in RSMBPT theory is selected in the calculations, see the papers~\cite{Gaigetal:2024CV,Gaigetal:2024C,Gaigetal:2024CC,Gaigetal:2025VV,Gaigetal:2025VVT,Gaigetal:2026CVT}.

\section{The Rules for Getting Algebraic Expressions in an Irreducible Tensorial Form of Feynman~Diagrams}
\label{Sec:rules}


\textls[-15]{In this section, we give the rules for finding the algebraic expressions of the tensorial form of any linked-cluster Feynman diagrams. We analyze the Feynman diagrams of the second order of the effective operator of perturbation theory, describing the Coulomb interaction in the relativistic atomic theory.
These rules are similar to those in the non-relativistic theory of the atom~\cite{Gaigalas:89}. They are formulated on the basis of Wick's theorem in coupled tensorial form~\cite{Gaigalas:85,Gaigalas:89} and~the experience accumulated in previous research~\cite{Gaigetal:2024CV,Gaigetal:2024C,Gaigetal:2024CC,Gaigetal:2025VV,Gaigetal:2025VVT}.} Using these rules, we easily obtain all algebraic expressions for the Feynman diagrams coming into the second order of perturbation theory in an irreducible tensorial form. These rules are as follows:

\begin{enumerate}
\item[1.] The sum of the sets of quantum numbers $n$, $l$, $j$ describing all electron lines and the ranks $k$ describing all interaction (horizontal) lines is obtained. For~example, for~the diagram CV$_7$ from Figures~\ref{Feynman_Diagrams} and \ref{CV_7}, we have $\sum\limits_{m, m^{\prime}}~\sum\limits_{n, n^{\prime}}~\sum\limits_{p, p^{\prime}}~\sum\limits_{k, k^{\prime}}$.
\vspace{-6pt}

\begin{figure}[H]

\begin{adjustwidth}{-\extralength}{-\extralength}
\centering 
\setlength{\unitlength}{1mm}
\begin{picture}(180,43)(4,0)
\thicklines
\put(10,30){\line(0,1){10}}
\put(10,38){\vector(0,1){2}}
\put(10,40){\vector(0,1){2}}
\put(7.5,38){\makebox(0,0)[t]{\small{$m$}}}
\put(11.9,26.1){\makebox(0,0)[t]{\small{$n$}}}
\multiput(10,30)(1,0){10}{\circle*{0.35}}
\put(10,20){\line(0,10){10}}
\put(22.5,24.5){\makebox(0,0)[r]{\small{$a$}}}
\put(12.5,22.5){\vector(1,1){2.4}}
\put(11.5,21.5){\vector(1,1){2}}
\put(20,26){\vector(0,-1){3}}
\put(20,30){\line(-1,-1){8.7}}
\put(20,20){\line(0,10){10}}
\put(20,20){\line(-1,1){8}}
\put(14.5,25.5){\vector(-1,1){2}}
\put(13.5,26.5){\vector(-1,1){2.4}}
\put(12.5,20.4){\makebox(0,0){\small{$n^{\prime}$}}}
\multiput(20,20)(1,0){10}{\circle*{0.35}}
\put(10,10){\line(0,1){10}}
\put(10,10){\vector(0,1){3}}
\put(7.5,13){\makebox(0,0){\small{$m^{\prime}$}}}
\put(10,08){\vector(0,1){3}}
\put(30,10){\line(0,1){20}}
\put(30,30){\vector(0,1){3}}
\put(30,28){\vector(0,1){3}}
\put(27.5,28){\makebox(0,0){\small{$p$}}}
\put(30,10){\vector(0,1){3}}
\put(30,08){\vector(0,1){3}}
\put(27.5,13){\makebox(0,0){\small{$p^{\prime}$}}}
\put(20,04){\makebox(0,0){$\text{CV}_{7}$}}
\put(37,37){\makebox(0,0) [l] {$\displaystyle{ = -
	\sum_{k, k^{\prime}, x}~ \sqrt{\frac{\left[ x \right]}{\left[ k, k^{\prime} \right]}} ~\sum_{m, m^{\prime}}~\sum_{n, n^{\prime}}~\sum_{p, p^{\prime}}
	}$}}
\put(40,25){\makebox(0,0) [l] {$\displaystyle{ \times
		\left[\left[\left[\;  a^{\left( j_m \right) }  \times 
  \tilde a^{\left( j_{m^{\prime}} \right) } \; \right] ^{\left( k \right)} \times 
	\left[\; \tilde a^{\left( j_{n^{\prime}} \right) }  \times 
  a^{\left( j_{n} \right) } \; \right] ^{\left( x \right)} \right]^{\left( k^{\prime} \right)}	
  \times
	\left[\; a^{\left( j_{p} \right) }  \times 
  \tilde a^{\left( j_{p^{\prime}} \right) } \; \right] ^{\left( k^{\prime} \right)} \right]^{\left( 0 \right)}
	}$}}
\put(40,13){\makebox(0,0) [l] {$\displaystyle{ \times
\sum_{a}~\frac{1}
{\left( \varepsilon_{a}+\varepsilon_{p'}-\varepsilon_n-\varepsilon_p \right)}
	\left\{
    \begin{array}{ccc}
      j_{n} & j_{n'} & x \\
      k      & k'    & j_{a}  
     \end{array}
	\right\}  
	X_{k}(m a, m^{\prime} n^{\prime}) ~ X_{k'}(n p, a~p^{\prime})}$}}
\end{picture}
\end{adjustwidth}
\caption{The CV Feynman diagram of the second-order effective Hamiltonian for the third and fourth types of core–valence correlations 
$(n_{a} \ell_{a})\, j_{a}^{2j_a+1} \, (n_{m} \ell_{m})\, j_{m}^{w_m} \, (n_{n} \ell_{n})\, j_{n}^{w_n} 
   \rightarrow (n_{a} \ell_{a})\, j_{a}^{2j_a} \; (n_{m} \ell_{m})\, j_{m}^{w_m-1} \, (n_{n} \ell_{n})\, j_{n}^{w_n+2}$
and
$(n_{a} \ell_{a})\, j_{a}^{2j_a+1} \, (n_{m} \ell_{m})\, j_{m}^{w_m} \, (n_{n} \ell_{n})\, j_{n}^{w_n} \, (n_{p} \ell_{p})\, j_{p}^{w_p} 
   \rightarrow (n_{a} \ell_{a})\, j_{a}^{2j_a} \; (n_{m} \ell_{m})\, j_{m}^{w_m-1} \, (n_{n} \ell_{n})\,j_{n}^{w_n+1}  \; (n_{p} \ell_{p})\, j_{p}^{w_p+1}$.
}
\label{CV_7}
\end{figure}

\item[2.] A matrix element is assigned for each Coulomb interaction line (horizontal Feynman diagram line):
\begin{equation}
\label{eq:rule2}
   \nonumber
   \frac{1}{\sqrt{\left[ k \right]}} \; X_{k}(i j, i' j').
\end{equation}
For the diagram CV$_7$, we have $\frac{1}{\sqrt{\left[ k, k' \right]}} \; X_{k}(m a, m^{\prime} n^{\prime}) \; X_{k'}(n p, a~p^{\prime})$.

\item[3.] The phase multiplier
\begin{equation}
\label{eq:rule3}
   \nonumber
   (-1)^{a+c+h}
\end{equation}
 is assigned to the whole diagram, where $a$ is the number of $\left[ \tilde a^{\left( j \right)} \times 
  a^{\left( j' \right)} \right]^{(k)}$ pairs, where the annihilation operator of the secondary quantization is first and the creation operator is second in~the diagram (to find out how, see rule 6.2.2. below); $c$ is the number of closed loops of the electron line in the diagram; and~$h$ is the number of internal hole lines and internal lines with double arrows pointing downward. For~example, for~the diagram CV$_7$, we have $a=1$, $c=1$, and~$h=0$.

\item[4.] The whole chart corresponds to a weighted multiplier of
\begin{equation}
\label{eq:rule4}
   \nonumber
\frac{1}{\lambda}.
\end{equation}
For diagrams without internal double arrow lines, $\lambda = 2$ if the diagram is symmetrical (replacing all its vertices with respect to the line of interaction results in a typologically identical diagram), and~if not, $\lambda = 1$. For~diagrams with internal double arrows, $\frac{1}{\lambda}$ is equal to the weighted multiplier of the diagram obtained by disconnecting the double arrows. For~the diagram CV$_7$, we have $\lambda = 1$.

\item[5.] \textls[-15]{The lower horizontal interaction line of the diagram corresponds to the energy multiplier}
\begin{equation}
\label{eq:rule5}
   \frac{1}{D},
\end{equation}
 where 
$D = \sum \left( \varepsilon_{\text{down}} - \varepsilon_{\text{up}} \right)$ is an energy denominator and
$\varepsilon_{\text{down}}$ ($\varepsilon_{\text{up}}$) is the single-particle eigenvalue associated with the down- (up-)
orbital lines to (from) the lowest interaction line of the diagram.
For the diagram CV$_7$, we have $D = \left( \varepsilon_{a}+\varepsilon_{p'}-\varepsilon_n-\varepsilon_p \right)$.
This rule is the same as Goldstone evaluation rules~\cite{Goldstone_1957} (see, for example, rule c on page 265~\cite{LinMor:82a}).

\item[6.] The Feynman diagram has an irreducible tensorial form of operators of second quantization, which can be found using the graphical representation of the angular momentum technique~\cite{JucBan:77a,Gaietal:85a,Gaigalas:89} by applying the rules below:

\begin{enumerate}

\item[6.1] The diagram of angular momentum typologically equivalent to the initial Feynman one is drawn (for example, the~diagram A$_1$ of angular momentum for Feynman diagram CV$_7$ from Figures~\ref{Feynman_Diagrams} and \ref{CV_7} is shown in Figure~\ref{CV_7_angular}), where

\begin{enumerate}
\item[6.1.1] Free lines A$_2$ or A$_4$ (see Figure~\ref{rule_1}) from the Feynman diagram are replaced by graphic elements A$_3$ or A$_5$ in the angular momentum diagram, shown in Figure~\ref{rule_1}. The~diagram A$_3$ represents the creation operator in the angular momentum diagram, and~A$_5$ represents the annihilation operator.
\item[6.1.2] Internal electronic lines are carried over unchanged.

\item[6.1.3] The horizontal lines A$_6$ (see Figure~\ref{rule_2}) corresponding to the interactions in the Feynman diagram are replaced by the graphical elements A$_7$ presented in Figure~\ref{rule_2}. The~diagram A$_7$ shows the angular momentum line corresponding to the rank ($k$ or $k'$) of the Coulomb interaction in the angular momentum diagram.

\vspace{-10pt}
\begin{figure}[H]
\setlength{\unitlength}{1mm}
\resizebox{.3\textwidth}{!}{%
\begin{picture}(33,43)(4,0)
\thicklines
\put(10,30){\line(0,1){10}}
\put(10,39){\vector(0,1){2}}
\put(7,30){\makebox(0,0){\small{$+$}}}
\put(22,32){\makebox(0,0){\small{$-$}}}
\put(10,40){\oval(5,5)[b]}
\put(10,30.6){\line(1,0){10}}
\put(10,30.5){\line(1,0){10}}
\put(10,30.4){\line(1,0){10}}
\put(10,30.3){\line(1,0){10}}
\put(10,30.2){\line(1,0){10}}
\put(10,30.1){\line(1,0){10}}
\put(10,30){\line(1,0){10}}
\put(10,29.9){\line(1,0){10}}
\put(10,29.8){\line(1,0){10}}
\put(10,29.7){\line(1,0){10}}
\put(10,29.6){\line(1,0){10}}
\put(10,29.5){\line(1,0){10}}
\put(10,29.4){\line(1,0){10}}
\put(15,28){\line(0,4){4}}
\put(18,30.6){\vector(-1,0){3}}
\put(18,29.4){\vector(-1,0){3}}
\put(10,20){\line(0,10){10}}
\put(18,18){\makebox(0,0){\small{$-$}}}
\put(12.6,22.2){\vector(1,1){1.5}}
\put(12.6,22.6){\oval(3.5,3.5)[r]}
\put(20,26){\vector(0,-1){3}}
\put(20.4,30){\line(-1,-1){8}}
\put(20,20){\line(0,10){10}}
\put(33,20){\makebox(0,0){\small{$-$}}}
\put(20,19.6){\line(-1,1){7.6}}
\put(13.5,26.1){\vector(-1,1){1.5}}
\put(12.6,26.7){\oval(3.5,3.5)[r]}
\put(20,20.6){\line(1,0){10}}
\put(20,20.5){\line(1,0){10}}
\put(20,20.4){\line(1,0){10}}
\put(20,20.3){\line(1,0){10}}
\put(20,20.2){\line(1,0){10}}
\put(20,20.1){\line(1,0){10}}
\put(20,20){\line(1,0){10}}
\put(20,19.9){\line(1,0){10}}
\put(20,19.8){\line(1,0){10}}
\put(20,19.7){\line(1,0){10}}
\put(20,19.6){\line(1,0){10}}
\put(20,19.5){\line(1,0){10}}
\put(20,19.4){\line(1,0){10}}
\put(25,18){\line(0,4){4}}
\put(28,20.6){\vector(-1,0){3}}
\put(28,19.4){\vector(-1,0){3}}
\put(10,09){\line(0,1){11}}
\put(10,10){\vector(0,1){2}}
\put(10,10){\oval(5,5)[t]}
\put(30,09){\line(0,1){21}}
\put(30,29){\vector(0,1){2}}
\put(30,30){\oval(5,5)[b]}
\put(30,10){\vector(0,1){2}}
\put(30,10){\oval(5,5)[t]}
\put(20,04){\makebox(0,0){$\text{A}_{1}$}}
\put(10,30){\color{white}\circle*{1.7}} 
\put(10,30){\color{black}\circle{1.7}}  
\put(20,30){\color{white}\circle*{1.7}} 
\put(20,30){\color{black}\circle{1.7}}  
\put(20,20){\color{white}\circle*{1.7}} 
\put(20,20){\color{black}\circle{1.7}}  
\put(30,20){\color{white}\circle*{1.7}} 
\put(30,20){\color{black}\circle{1.7}}  
\end{picture}%
}
\caption{The diagram of angular momentum for Feynman diagram $\text{CV}_{7}$ from Figure~\ref{CV_7}.}
\label{CV_7_angular}
\end{figure}
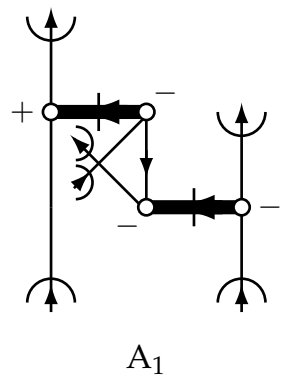

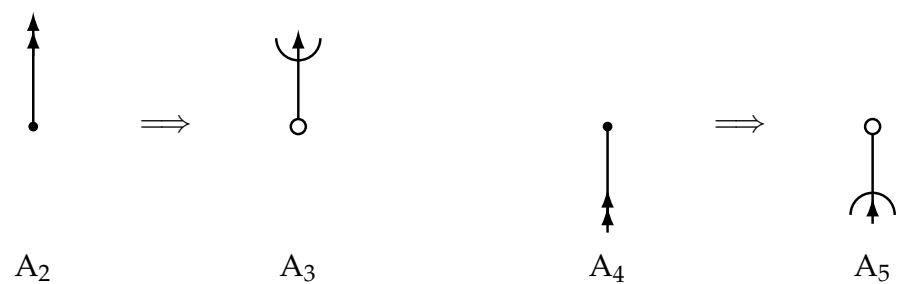
\begin{figure}[H]
\setlength{\unitlength}{1mm}
\resizebox{.9\textwidth}{!}{%
\hspace{5pt}\begin{picture}(105,31)(4,0)
\thicklines
\put(5,20){\line(0,1){10}}
\put(5,30){\vector(0,1){3}}
\put(5,28){\vector(0,1){3}}
\put(5,20){\circle*{1,1}}
\put(5,04){\makebox(0,0){$\text{A}_{2}$}}
\put(20,20){\makebox(0,0){$\Longrightarrow$}}
\put(35,20){\line(0,1){10}}
\put(35,29){\vector(0,1){2}}
\put(35,30){\oval(5,5)[b]}
\put(35,04){\makebox(0,0){$\text{A}_{3}$}}
\put(70,10){\line(0,1){10}}
\put(70,10){\vector(0,1){3}}
\put(70,08){\vector(0,1){3}}
\put(70,20){\circle*{1,1}}
\put(70,04){\makebox(0,0){$\text{A}_{4}$}}
\put(85,20){\makebox(0,0){$\Longrightarrow$}}
\put(100,09){\line(0,1){11}}
\put(100,10){\vector(0,1){2}}
\put(100,10){\oval(5,5)[t]}
\put(100,04){\makebox(0,0){$\text{A}_{5}$}}
\put(35,20){\color{white}\circle*{1.7}} 
\put(35,20){\color{black}\circle{1.7}}  
\put(100,20){\color{white}\circle*{1.7}} 
\put(100,20){\color{black}\circle{1.7}}  
\end{picture}%
}
\caption{The free lines $A_2$ or $A_4$ from the Feynman diagram are replaced by graphic elements $A_3$ or $A_5$ in angular momentum diagram, respectively.}
\label{rule_1}
\end{figure}
\vspace{-9pt}

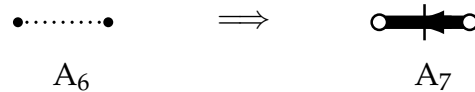
\begin{figure}[H]
\setlength{\unitlength}{1mm}
\resizebox{.5\textwidth}{!}{%
\begin{picture}(58,13)(4,0)
\thicklines
\put(5,10){\circle*{1,1}}
\put(15,10){\circle*{1,1}}
\multiput(5,10)(1,0){10}{\circle*{0.35}}
\put(30,10){\makebox(0,0){$\Longrightarrow$}}
\put(45,10.6){\line(1,0){10}}
\put(45,10.5){\line(1,0){10}}
\put(45,10.4){\line(1,0){10}}
\put(45,10.3){\line(1,0){10}}
\put(45,10.2){\line(1,0){10}}
\put(45,10.1){\line(1,0){10}}
\put(45,10){\line(1,0){10}}
\put(45,9.9){\line(1,0){10}}
\put(45,9.8){\line(1,0){10}}
\put(45,9.7){\line(1,0){10}}
\put(45,9.6){\line(1,0){10}}
\put(45,9.5){\line(1,0){10}}
\put(45,9.4){\line(1,0){10}}
\put(50,8){\line(0,4){4}}
\put(53,10.6){\vector(-1,0){3}}
\put(53,9.4){\vector(-1,0){3}}
\put(11,04){\makebox(0,0){$\text{A}_{6}$}}
\put(51,04){\makebox(0,0){$\text{A}_{7}$}}
\put(45,10){\color{white}\circle*{1.7}} 
\put(45,10){\color{black}\circle{1.7}}  
\put(55,10){\color{white}\circle*{1.7}} 
\put(55,10){\color{black}\circle{1.7}}  
\end{picture}%
}
\caption{The horizontal lines $A_6$ corresponding to the interactions of the Feynman diagram are replaced by the graphical elements $A_7$ in the angular momentum~diagram.}
\label{rule_2}
\end{figure}
\item[6.1.4] Nodes without signs are provided with a “$+$” sign if the line with an arrow coming out of the node can be turned counterclockwise to align it with the line having an arrow entering the node without crossing the thickened line (see A$_8$ in Figure~\ref{CV_7_sign}); otherwise, the~node is given the sign “$-$” (see A$_9$ in Figure~\ref{CV_7_sign}).
\end{enumerate}

\vspace{-12pt}
\begin{figure}[H]
\setlength{\unitlength}{1mm}
\resizebox{.7\textwidth}{!}{%
\begin{picture}(70,33)(4,0)
\thicklines
\linethickness{0.6pt}
\qbezier[20](9,25)(0,20)(9,15)
\put(9.5,14.64){\vector(1,-0.72){0}} 
\thicklines
\put(10,20){\line(0,1){10}}
\put(10,29){\vector(0,1){2}}
\put(7,20){\makebox(0,0){\small{$+$}}}
\put(10,30){\oval(5,5)[b]}
\put(10,20.6){\line(1,0){10}}
\put(10,20.5){\line(1,0){10}}
\put(10,20.4){\line(1,0){10}}
\put(10,20.3){\line(1,0){10}}
\put(10,20.2){\line(1,0){10}}
\put(10,20.1){\line(1,0){10}}
\put(10,20){\line(1,0){10}}
\put(10,19.9){\line(1,0){10}}
\put(10,19.8){\line(1,0){10}}
\put(10,19.7){\line(1,0){10}}
\put(10,19.6){\line(1,0){10}}
\put(10,19.5){\line(1,0){10}}
\put(10,19.4){\line(1,0){10}}
\put(15,18){\line(0,4){4}}
\put(18,20.6){\vector(-1,0){3}}
\put(18,19.4){\vector(-1,0){3}}
\linethickness{0.6pt}
\qbezier[20](61,25)(70,20)(61,15)
\put(60.5,14.64){\vector(-1,-0.72){0}} 
\thicklines
\put(63,20){\makebox(0,0){\small{$-$}}}
\put(50,20.6){\line(1,0){10}}
\put(50,20.5){\line(1,0){10}}
\put(50,20.4){\line(1,0){10}}
\put(50,20.3){\line(1,0){10}}
\put(50,20.2){\line(1,0){10}}
\put(50,20.1){\line(1,0){10}}
\put(50,20){\line(1,0){10}}
\put(50,19.9){\line(1,0){10}}
\put(50,19.8){\line(1,0){10}}
\put(50,19.7){\line(1,0){10}}
\put(50,19.6){\line(1,0){10}}
\put(50,19.5){\line(1,0){10}}
\put(50,19.4){\line(1,0){10}}
\put(55,18){\line(0,4){4}}
\put(58,20.6){\vector(-1,0){3}}
\put(58,19.4){\vector(-1,0){3}}
\put(10,09){\line(0,1){11}}
\put(10,10){\vector(0,1){2}}
\put(10,10){\oval(5,5)[t]}
\put(60,09){\line(0,1){21}}
\put(60,29){\vector(0,1){2}}
\put(60,30){\oval(5,5)[b]}
\put(60,10){\vector(0,1){2}}
\put(60,10){\oval(5,5)[t]}
\put(15,04){\makebox(0,0){$\text{A}_{8}$}}
\put(55,04){\makebox(0,0){$\text{A}_{9}$}}
\put(10,20){\color{white}\circle*{1.7}} 
\put(10,20){\color{black}\circle{1.7}}  
\put(60,20){\circle*{1,7}}
\put(60,20){\color{white}\circle*{1.7}} 
\put(60,20){\color{black}\circle{1.7}}  
\end{picture}%
}
\caption{An illustration of the 6.1.4 rule for determining the signs at the vertex of the angular momentum~diagram.}
\label{CV_7_sign}
\end{figure}
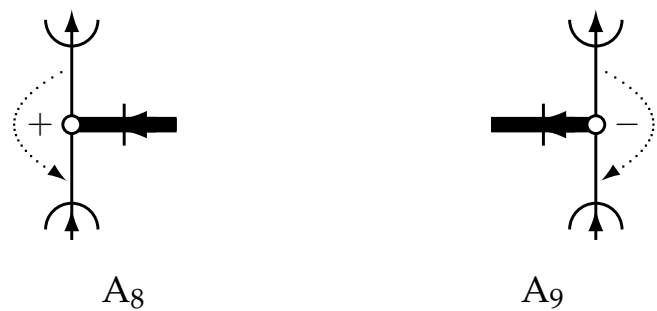

\item[6.2] The obtained recoupling diagram and tensorial product of operators of second quantization from the diagram of angular momentum (for example, from the A$_1$, see Figure~\ref{CV_7_angular}):
\begin{enumerate}
\item[6.2.1] The lines related to the secondary quantization operators need to be cut.

\item[6.2.2] Then, through the diagram depicting the generalized Clebsch–Gordan coefficient, the~graphical elements representing the secondary quantization operators are connected into pairs, and~the pairs---successively from left to right---into the resulting rank~\cite{Gaigalas:85,Gaigalas:89}. In~a pair, the first is the creation operator, and~if the pair consists of operators connected in the Feynman diagram by internal electron lines, then the first is the operator belonging to the leftmost upper node. This yields a graphical representation of the irreducible tensorial product of operators of the second quantization.
For example, we obtain diagram $A_{10}$ (see \mbox{Figure~\ref{A_10}}), representing the tensorial product of the second quantization operators for Feynman diagram $A_1$ 
(see Figure~\ref{CV_7_angular}), with~the additional summations $\sum\limits_{J_1, J_2, J}$ according to~\cite{JucBan:77a,Gaietal:85a,Gaigalas:89}.
\begin{figure}[H]
\setlength{\unitlength}{1mm}
\hspace{7pt}\begin{picture}(90,39)(4,0)
\thicklines
%
%
\put(15,36){\line(-1,-1){10}}
\put(5,21){\line(0,1){5}}
\put(6,31){\makebox(0,0){$ j_{m}$}}
\put(5,24){\vector(0,-1){3}}
\put(5,23){\oval(5,5)[t]}
\put(30,39){\makebox(0,0){$J_1$}}
\put(15,36){\line(1,-1){10}}
\put(25,21){\line(0,1){5}}
\put(25,31){\makebox(0,0){$ j_{m'}$}}
\put(25,21){\vector(0,1){3}}
\put(25,23){\oval(5,5)[t]}
\put(15,33){\makebox(0,0){\small{$+$}}}
\put(15,36.6){\line(1,0){15}}
\put(15,36.5){\line(1,0){15}}
\put(15,36.4){\line(1,0){15}}
\put(15,36.3){\line(1,0){15}}
\put(15,36.2){\line(1,0){15}}
\put(15,36.1){\line(1,0){15}}
\put(15,36){\line(1,0){30}}
\put(15,35.9){\line(1,0){15}}
\put(15,35.8){\line(1,0){15}}
\put(15,35.7){\line(1,0){15}}
\put(15,35.6){\line(1,0){15}}
\put(15,35.5){\line(1,0){15}}
\put(15,35.4){\line(1,0){15}}
%
%
\put(42,33){\makebox(0,0){${+}$}}
\put(45,23){\makebox(0,0){${+}$}}
\put(45.6,26){\line(0,1){5}}
\put(45.5,26){\line(0,1){5}}
\put(45.4,26){\line(0,1){5}}
\put(45.3,26){\line(0,1){5}}
\put(45.2,26){\line(0,1){5}}
\put(45.1,26){\line(0,1){5}}
\put(45,26){\line(0,1){10}}
\put(44.9,26){\line(0,1){5}}
\put(44.8,26){\line(0,1){5}}
\put(44.7,26){\line(0,1){5}}
\put(44.6,26){\line(0,1){5}}
\put(44.5,26){\line(0,1){5}}
\put(44.4,26){\line(0,1){5}}
\put(48,29){\makebox(0,0){$J_2$}}
\put(35,16){\line(1,1){10}}
\put(35,11){\line(0,1){5}}
\put(36,21){\makebox(0,0){$ j_{n'}$}}
\put(35,11){\vector(0,1){3}}
\put(35,13){\oval(5,5)[t]}
\put(55,16){\line(-1,1){10}}
\put(55,11){\line(0,1){5}}
\put(55,21){\makebox(0,0){$ j_{n}$}}
\put(55,14){\vector(0,-1){3}}
\put(55,13){\oval(5,5)[t]}
\put(45,4){\makebox(0,0){\large{$\text{A}_{10}$}}}
%
%
\put(75,36){\line(-1,-1){10}}
\put(65,21){\line(0,1){5}}
\put(65,24){\vector(0,-1){3}}
\put(65,23){\oval(5,5)[t]}
\put(66,31){\makebox(0,0){$ j_{p}$}}
\put(75,36){\line(1,-1){10}}
\put(85,21){\line(0,1){5}}
\put(60,39){\makebox(0,0){$J$}}
\put(85,31){\makebox(0,0){$ j_{p'}$}}
\put(85,21){\vector(0,1){3}}
\put(85,23){\oval(5,5)[t]}
\put(75,33){\makebox(0,0){\small{$+$}}}
\put(45,36.6){\line(1,0){30}}
\put(45,36.5){\line(1,0){30}}
\put(45,36.4){\line(1,0){30}}
\put(45,36.3){\line(1,0){30}}
\put(45,36.2){\line(1,0){30}}
\put(45,36.1){\line(1,0){30}}
\put(45,36){\line(1,0){30}}
\put(45,35.9){\line(1,0){30}}
\put(45,35.8){\line(1,0){30}}
\put(45,35.7){\line(1,0){30}}
\put(45,35.6){\line(1,0){30}}
\put(45,35.5){\line(1,0){30}}
\put(45,35.4){\line(1,0){30}}
\put(58,34){\line(0,4){4}}
\put(63,36.6){\vector(-1,0){3}}
\put(63,35.4){\vector(-1,0){3}}
\put(15,36){\color{white}\circle*{1.7}} 
\put(15,36){\color{black}\circle{1.7}}  
\put(45,36){\color{white}\circle*{1.7}} 
\put(45,36){\color{black}\circle{1.7}}  
\put(45,26){\color{white}\circle*{1.7}} 
\put(45,26){\color{black}\circle{1.7}}  
\put(75,36){\color{white}\circle*{1.7}} 
\put(75,36){\color{black}\circle{1.7}}  
\end{picture}
\caption{The graphical representation of the irreducible tensorial product of second-quantization operators $\text{A}_{10}$ corresponding to Feynman diagram $\text{CV}_{7}$.}
\label{A_10}
\end{figure}
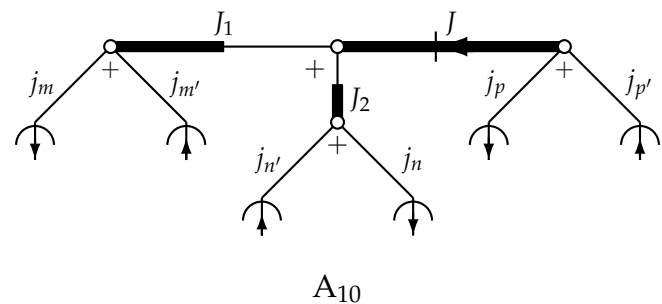

\item[6.2.3] The parts of the original diagram remaining after cutting are closed up according to the graphical technique developed in \cite{JucBan:77a,Gaietal:85a,Gaigalas:85,Gaigalas:89} by the same generalized Clebsch–Gordan coefficient as the irreducible tensorial product. As~a result, we obtain the recouping  matrix, which can be expressed through the 3$nj-$coefficients by using~\cite{JucBan:77a,Gaietal:85a,Gaigalas:89}. For~example, by~applying this rule, we get the recouping matrix $A_{11}$ in graphical representation (see Figure~\ref{A_11}) for Feynman diagram CV$_7$.
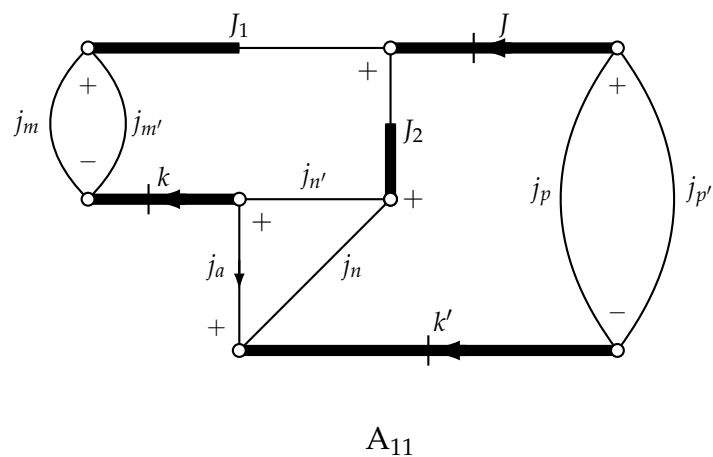
\begin{figure}[H]
\setlength{\unitlength}{1mm}
\begin{picture}(100,59)(4,0)
\thicklines
%
%
\put(7,46){\makebox(0,0){$ j_{m}$}}
\put(35,59){\makebox(0,0){$J_1$}}
\put(23,46){\makebox(0,0){$ j_{m'}$}}
\put(15,51){\makebox(0,0){\small{$+$}}}
\put(15,56.6){\line(1,0){20}}
\put(15,56.5){\line(1,0){20}}
\put(15,56.4){\line(1,0){20}}
\put(15,56.3){\line(1,0){20}}
\put(15,56.2){\line(1,0){20}}
\put(15,56.1){\line(1,0){20}}
\put(15,56){\line(1,0){40}}
\put(15,55.9){\line(1,0){20}}
\put(15,55.8){\line(1,0){20}}
\put(15,55.7){\line(1,0){20}}
\put(15,55.6){\line(1,0){20}}
\put(15,55.5){\line(1,0){20}}
\put(15,55.4){\line(1,0){20}}
\qbezier(15,36)(5,46)(15,56)
\qbezier(15,36)(25,46)(15,56)
\put(15,41){\makebox(0,0){\small{$-$}}}
%
%
\put(52,53){\makebox(0,0){${+}$}}
\put(58,36){\makebox(0,0){${+}$}}
\put(55.6,36){\line(0,1){10}}
\put(55.5,36){\line(0,1){10}}
\put(55.4,36){\line(0,1){10}}
\put(55.3,36){\line(0,1){10}}
\put(55.2,36){\line(0,1){10}}
\put(55.1,36){\line(0,1){10}}
\put(55,36){\line(0,1){20}}
\put(54.9,36){\line(0,1){10}}
\put(54.8,36){\line(0,1){10}}
\put(54.7,36){\line(0,1){10}}
\put(54.6,36){\line(0,1){10}}
\put(54.5,36){\line(0,1){10}}
\put(54.4,36){\line(0,1){10}}
\put(58,45){\makebox(0,0){$J_2$}}
\put(38,33){\makebox(0,0){${+}$}}
\put(15,36.6){\line(1,0){20}}
\put(15,36.5){\line(1,0){20}}
\put(15,36.4){\line(1,0){20}}
\put(15,36.3){\line(1,0){20}}
\put(15,36.2){\line(1,0){20}}
\put(15,36.1){\line(1,0){20}}
\put(15,36){\line(1,0){40}}
\put(15,35.9){\line(1,0){20}}
\put(15,35.8){\line(1,0){20}}
\put(15,35.7){\line(1,0){20}}
\put(15,35.6){\line(1,0){20}}
\put(15,35.5){\line(1,0){20}}
\put(15,35.4){\line(1,0){20}}
\put(23,34){\line(0,4){4}}
\put(28,36.6){\vector(-1,0){3}}
\put(28,35.4){\vector(-1,0){3}}
\put(25,39){\makebox(0,0){$k$}}
\put(45,39){\makebox(0,0){$ j_{n'}$}}
\put(35,36){\line(0,-1){20}}
\put(32,27){\makebox(0,0){$ j_{a}$}}

\put(35,28){\vector(0,-1){4}}
\put(32,19){\makebox(0,0){\small{$+$}}}
\put(55,36){\line(-1,-1){20}}
\put(50,27){\makebox(0,0){$ j_{n}$}}
\put(35,16.6){\line(1,0){50}}
\put(35,16.5){\line(1,0){50}}
\put(35,16.4){\line(1,0){50}}
\put(35,16.3){\line(1,0){50}}
\put(35,16.2){\line(1,0){50}}
\put(35,16.1){\line(1,0){50}}
\put(35,16){\line(1,0){50}}
\put(35,15.9){\line(1,0){50}}
\put(35,15.8){\line(1,0){50}}
\put(35,15.7){\line(1,0){50}}
\put(35,15.6){\line(1,0){50}}
\put(35,15.5){\line(1,0){50}}
\put(35,15.4){\line(1,0){50}}
\put(60,14){\line(0,4){4}}
\put(65,16.6){\vector(-1,0){3}}
\put(65,15.4){\vector(-1,0){3}}
\put(55,4){\makebox(0,0){\large{$\text{A}_{11}$}}}
%
%
\put(75,37){\makebox(0,0){$ j_{p}$}}
\put(70,59){\makebox(0,0){$J$}}
\put(96,37){\makebox(0,0){$ j_{p'}$}}
\put(85,51){\makebox(0,0){\small{$+$}}}
\put(55,56.6){\line(1,0){30}}
\put(55,56.5){\line(1,0){30}}
\put(55,56.4){\line(1,0){30}}
\put(55,56.3){\line(1,0){30}}
\put(55,56.2){\line(1,0){30}}
\put(55,56.1){\line(1,0){30}}
\put(55,56){\line(1,0){30}}
\put(55,55.9){\line(1,0){30}}
\put(55,55.8){\line(1,0){30}}
\put(55,55.7){\line(1,0){30}}
\put(55,55.6){\line(1,0){30}}
\put(55,55.5){\line(1,0){30}}
\put(55,55.4){\line(1,0){30}}
\put(66,54){\line(0,4){4}}
\put(71,56.6){\vector(-1,0){3}}
\put(71,55.4){\vector(-1,0){3}}
\put(62,20){\makebox(0,0){$k'$}}
\put(85,21){\makebox(0,0){\small{$-$}}}
\qbezier(85,16)(70,36)(85,56)
\qbezier(85,16)(100,36)(85,56)
\put(15,56){\color{white}\circle*{1.7}} 
\put(15,56){\color{black}\circle{1.7}}  
\put(55,56){\color{white}\circle*{1.7}} 
\put(55,56){\color{black}\circle{1.7}}  
\put(55,36){\color{white}\circle*{1.7}} 
\put(55,36){\color{black}\circle{1.7}}  
\put(55,36){\color{white}\circle*{1.7}} 
\put(55,36){\color{black}\circle{1.7}}  
\put(15,36){\color{white}\circle*{1.7}} 
\put(15,36){\color{black}\circle{1.7}}  
\put(35,36){\color{white}\circle*{1.7}} 
\put(35,36){\color{black}\circle{1.7}}  
\put(35,16){\color{white}\circle*{1.7}} 
\put(35,16){\color{black}\circle{1.7}}  
\put(85,56){\color{white}\circle*{1.7}} 
\put(85,56){\color{black}\circle{1.7}}  
\put(85,16){\color{white}\circle*{1.7}} 
\put(85,16){\color{black}\circle{1.7}}  
\end{picture}
\caption{The recoupling matrix for Feynman diagram $\text{CV}_{7}$ from Figure~\ref{CV_7}.}
\label{A_11}
\end{figure}
\end{enumerate}
\end{enumerate}
\end{enumerate}

So, using all these rules, we first obtain all the necessary multipliers (including the phase multiplier) and the~tensorial product and the recoupling matrix, both of which are represented by the graphical technique of angular momentum~\cite{JucBan:77a,Gaietal:85a,Gaigalas:89}. Then we already have the tensorial form of the Feynman diagram, which is partly expressed in algebraic terms and partly represented graphically. To~get the full expression in algebraic form, further steps are~required.

So, secondly, the~tensorial product can easily be rewritten from the graphical form into the normal (algebraic) form by using~\cite{JucBan:77a,Gaietal:85a,Gaigalas:89}. For~example, the~tensorial product shown in $A_{10}$ has the following tensorial product:
\begin{equation}
\label{eq:A_10}
   A_{10} = 		\left[\left[\left[\;  a^{\left( j_m \right) }  \times 
  \tilde a^{\left( j_{m^{\prime}} \right) } \; \right] ^{\left( J_1 \right)} \times 
	\left[\; \tilde a^{\left( j_{n^{\prime}} \right) }  \times 
  a^{\left( j_{n} \right) } \; \right] ^{\left( J_2 \right)} \right]^{\left(J \right)}	
  \times
	\left[\; a^{\left( j_{p} \right) }  \times 
  \tilde a^{\left( j_{p^{\prime}} \right) } \; \right] ^{\left( J \right)} \right]^{\left( 0 \right)} .
\end{equation}

Thirdly, the~algebraic expression for the recoupling matrix is obtained by following the necessary steps of the graphical momentum theory~\cite{JucBan:77a,Gaietal:85a,Gaigalas:89}. For~example, after~the transformations of the recoupling matrix $A_{11}$ (such as removing the arrows on the lines (lines $j_a$, $J$, $k$, and~$k'$), removing a swipe from the lines (lines $J$, $k$, and~$k'$),  cutting the diagram in two lines (lines $J_1$, $k$ and lines $J$, $k'$), changing the signs at nodes (vertices), and~rearranging the thick and semi-thick lines), we get the following algebraic expression:
\begin{equation}
\label{eq:A_11}
   A_{11} = - \sqrt{\left[ J_2 \right]} \;
		\left\{
    \begin{array}{ccc}
      j_{n} & j_{n'} & J_2 \\
      k     & k'     & j_{a}  
     \end{array}
		\right\}
		\; \delta\left( J_1, k \right) \; \delta\left( J, k' \right) .
\end{equation}

Thus
, taking into account all these rules (1--
6.2.3) and the above transformations (all three steps), we obtain the final expression of the Feynman diagram CV$_7$ shown in Figure~\ref{CV_7}.

Based on these rules and transformations, it is possible to obtain all the expressions for the Feynman diagrams of Coulomb interaction coming into the second-order expansion of the perturbation theory of the effective operator in irreducible tensorial form. 
At the same time, it is possible to obtain analytical expressions for all the Feynman diagrams that have been dealt with in the previous papers~\cite{Gaigetal:2024CV,Gaigetal:2024C,Gaigetal:2024CC,Gaigetal:2025VV,Gaigetal:2025VVT}.

We also want to point out that the reason we use here the graphical version of the Jucys, Bandzaitis, and~Gaigalas momentum theory~\cite{JucBan:77a,Gaietal:85a,Gaigalas:85,Gaigalas:89} is that it is the most appropriate of all the available graphical methods for representing the secondary quantization operators in the irreducible tensorial form and for all the necessary operations on them (operators)~\cite{Gaietal:85a}. This greatly facilitates finding all the necessary expressions for Feynman diagrams in the irreducible tensorial form in both non-relativistic atomic theory~\cite{Gaigalas:89} and relativistic atomic theory~\cite{Gaigetal:2024CV,Gaigetal:2024C,Gaigetal:2024CC,Gaigetal:2025VV,Gaigetal:2025VVT}.

\section{The Spin-Angular Integration for Second-Order Feynman Diagrams in an Irreducible Tensorial~Form}
\label{Sec:spin_angular}

In the development of this methodology~\cite{Gaigetal:2024CV,Gaigetal:2024C,Gaigetal:2024CC,Gaigetal:2025VV,Gaigetal:2025VVT,Gaigetal:2026CVT}, all the necessary expressions for the Feynman diagrams from Figure~\ref{Feynman_Diagrams} were first obtained in irreducible tensorial form. This makes it possible, as~shown in previous articles, for the~contribution deriving from the CV, C, CC, and~VV correlations of the configurations $K'$ to $E (K \chi J)$ in the second order of the perturbation theory to~be expressed as
\vspace{-14pt}
\begin{adjustwidth}{-\extralength}{0cm}
\begin{eqnarray}
\label{eq:BogEnergy_PT}
  \Delta E_{PT(\text{\text{CV,C,CC,VV}})}
	\nonumber \\
& & \hspace*{-2.5cm}
   =  \Delta \mathcal{E}_{0 \; PT(\text{CV,C,CC,VV})} \left(K J \right) 
	\nonumber \\  [0.2cm]
& & \hspace*{-2.5cm}
	+ \; \sum_{n\ell j} \sum_{k>0} \widetilde{f}_k \left( \ell j^{w}, \; K \chi J  \right) \;
	\Delta \mathcal{F}^{k}_{PT(\text{CV,C,CC,VV})} \left( n \ell j, \; n \ell j \right)
	\nonumber \\
& & \hspace*{-2.5cm}
	+ \; \sum_{n\ell j} \sum_{n'\ell'j' > n\ell j}  \left\{ \sum_{k>0} \widetilde{f}_k \left( \ell j^{w} \; \ell' j'^{w'},
	\; K \chi J  \right) \;
  \Delta \mathcal{F}^{k}_{PT(\text{CV,CC,VV})} \left( n \ell j, \; n' \ell' j' \right)  \right.
	\nonumber \\ [0.2cm]
& & \hspace*{-2.5cm} 
	+ \sum_{k} \widetilde{g}_k \left( \ell j^{w} \; \ell' j'^{w'}, \; K \chi J  \right)	\; 
  \Delta \mathcal{G}^{k}_{PT(\text{CV,CC,VV})} \left( n \ell j, \; n' \ell' j' \right) 
	\nonumber \\ [0.2cm]
& & \hspace*{-2.5cm}
	+ \left. \sum_{k} \widetilde{v}_k \left( \ell j^{w} \; \ell' j'^{w'}, \ell j^{w-2} \; \ell' j'^{w'+2},
	\; K \chi J \; K' \chi' J \right)	\;
  \Delta \mathcal{R}^{k}_{PT(\text{CV,CC,VV})} \left( n \ell j n \ell j, \; n' \ell' j' n '\ell' j' \right) \right\} 
	\nonumber \\
& & \hspace*{-2.5cm}
+ \sum_{\substack{n\ell j \\ n'\ell'j' \, \neq \, n\ell j}}
\; \sum_{\substack{k>0 \\ k',x}} 
\left< \Psi \left\| \biggl[ \bigl[ \tilde{a}^{(j)} \times a^{(j)} \bigr]^{(k)} \times  \Bigl[ \bigl[ a^{(j')} \times \tilde{a}^{(j')} \bigr]^{(x)} \times \bigl[ a^{(j')} \times \tilde{a}^{(j')} \bigr]^{(k')} \Bigr]^{(k)} \biggr]^{(0)} \right\| \Psi \right> 
	\nonumber \\
& & \hspace*{-2.5cm}
\hspace{1.0cm} \times \;
\Delta \widetilde{\mathcal{R}}^{(k,k',x)}_{PT(\text{VV})}
\left( n \ell j \; n '\ell' j' \; n '\ell' j' \right)
	\nonumber \\
& & \hspace*{-2.5cm}
+ \sum_{\substack{n\ell j \\ n'\ell'j' \, \neq \, n\ell j \\ n''\ell''j'' \, \neq \, n\ell j}}
\; \sum_{\substack{k>0 \\ k',x}} 
\left< \Psi \left\| \biggl[ \Bigl[ \bigl[ \tilde{a}^{(j)} \times a^{(j)} \bigr]^{(k)} \times \bigl[ a^{(j'')} \times \tilde{a}^{(j'')} \bigr]^{(x)} \Bigr]^{(k')} \times \bigl[ a^{(j')} \times \tilde{a}^{(j')} \bigr]^{(k')} \biggr]^{(0)} \right\| \Psi \right> 
	\nonumber \\
& & \hspace*{-2.5cm}
\hspace{1.0cm} \times \;
\Delta \widetilde{\mathcal{R}}^{(k,k',x)}_{PT(\text{VV})}
\left( n \ell j \; n '\ell' j' \; n ''\ell'' j'' \right)
	\nonumber \\ [0.2cm]	
& & \hspace*{-2.5cm}
+ \sum_{\substack{n\ell j \\ n'\ell'j' \, \neq \, n\ell j}}
\; \sum_{\substack{k>0 \\ k',x}} 
\left< \Psi \left\| \biggl[ \bigl[ a^{(j)} \times \tilde{a}^{(j)} \bigr]^{(k)} \times  \Bigl[ \bigl[ a^{(j')} \times \tilde{a}^{(j')} \bigr]^{(x)} \times \bigl[ \tilde{a}^{(j')} \times a^{(j')} \bigr]^{(k')} \Bigr]^{(k)} \biggr]^{(0)} \right\| \Psi \right> 
	\nonumber \\
& & \hspace*{-2.5cm} 
\hspace{1.0cm} \times \;
\Delta \widetilde{\widetilde{\mathcal{R}}}^{(k,k',x)}_{PT(\text{CV})}
\left( n \ell j \; n '\ell' j' \; n '\ell' j' \right)
	\nonumber \\
& & \hspace*{-2.5cm} 
+ \sum_{\substack{n\ell j \\ n'\ell'j' \, \neq \, n\ell j \\ n''\ell''j'' \, \neq \, n\ell j}}
\; \sum_{\substack{k>0 \\ k',x}} 
\left< \Psi \left\| \biggl[ \Bigl[ \bigl[ a^{(j)} \times \tilde{a}^{(j)} \bigr]^{(k)} \times \bigl[ \tilde{a}^{(j'')} \times a^{(j'')} \bigr]^{(x)} \Bigr]^{(k')} \times \bigl[ \tilde{a}^{(j')} \times a^{(j')} \bigr]^{(k')} \biggr]^{(0)} \right\| \Psi \right> 
	\nonumber \\
& & \hspace*{-2.5cm} 
\hspace{1.0cm} \times \;
\Delta \widetilde{\widetilde{\mathcal{R}}}^{(k,k',x)}_{PT(\text{CV})}
\left( n \ell j \; n '\ell' j' \; n ''\ell'' j'' \right),
\end{eqnarray}
\end{adjustwidth}
\textls[-15]{where $\left< \; \Psi \; \right\|$ and $\left\| \; \Psi \; \right>$ are configuration state functions, meanwhile
$\widetilde{f}_k \left( ... \right)$, $\widetilde{g}_k \left( \ell j^{w} \; \ell' j'^{w'}, \; K \chi J  \right)$,} and~
$\widetilde{v}_k \left( \ell j^{w} \; \ell' j'^{w'}, \ell j^{w-2} \; \ell' j'^{w'+2},
	\; K \chi J \; K' \chi' J \right)$, respectively, are spin-angular coefficients $f_{abk}$, $g_{abk}$ (see (87) in \cite{Fisetal:16a}), and~$\upsilon_{abcd;\,k}^{\alpha\beta}$ (see (34) in \cite{Gaigalas:2022}) from which submatrix elements $\redmem{\ell j}{\, C^{(k)} \,}{ \ell^{\prime} j^{\prime}}$ are extracted. 
Therefore, summation over $k$ runs over all  possible values instead of the values that satisfy the triangular condition $\left( \ell \ell^{\prime} k\right)$ as it is in the regular case and these coefficients $\redmem{\ell j}{\, C^{(k)} \,}{ \ell^{\prime} j^{\prime}}$ are themselves included in $\Delta \mathcal{F}^{k}_{PT(\text{CV,C,CC,VV})} \left( n \ell j, \; n \ell j \right)$, $\Delta \mathcal{F}^{k}_{PT(\text{CV,CC,VV})} \left( n \ell j, \; n' \ell' j' \right)$, \textls[-23]{$\Delta \mathcal{G}^{k}_{PT(\text{CV,CC,VV})} \left( n \ell j, \; n' \ell' j' \right)$, $\Delta \mathcal{R}^{k}_{PT(\text{CV,CC,VV})} \left( n \ell j n \ell j, \; n' \ell' j' n '\ell' j' \right)$}, $\Delta \widetilde{\mathcal{R}}^{(k,k',x)}_{PT(\text{VV})}
\left( n \ell j \; n '\ell' j' \; n '\ell' j' \right)$, $\Delta \widetilde{\mathcal{R}}^{(k,k',x)}_{PT(\text{VV})}\left( n \ell j \; n '\ell' j' \; n ''\ell'' j'' \right)$, $\Delta \widetilde{\widetilde{\mathcal{R}}}^{(k,k',x)}_{PT(\text{CV})}
\left( n \ell j \; n '\ell' j' \; n '\ell' j' \right)$, and $\Delta \widetilde{\widetilde{\mathcal{R}}}^{(k,k',x)}_{PT(\text{CV})}\left( n \ell j \; n '\ell' j' \; n ''\ell'' j'' \right)$. These latter coefficients are the amplitude of second-order effective operator form Rayleigh–Schr\"odinger many-body perturbation theory~\cite{LinMor:82a,Merk:85} according to terminology~\cite{Gaigalas:89} or effective interaction strength according to terminology~\cite{Gra:2007a}. They are proportional to Slater integrals with some additional coefficients, such as simple multipliers and/or 6$j$-coefficients.

As shown in (\ref{eq:BogEnergy_PT}), the~order in which the ranks are combined into a tensorial product varies. The~tensorial structure of the triple tensor in the first reduced matrix element shows that the rank $x$ (coming from the second pair of operators of second quantization) and the rank $k^{\prime}$ (coming from the third pair of operators of second quantization) are coupled into the rank $k$, and~the rank $k$ (coming from the first pair of operators of second quantization) is coupled with the rank $k$ from the previous coupling into the final rank $0$. The~tensorial structure of the triple tensor in the second reduced matrix element shows that the rank $k$ (coming from the first pair of operators of second quantization) and~rank $x$ (coming from the second pair of operators of second quantization) are coupled into the rank $k^{\prime}$, and~the latter and rank $k^{\prime}$ (coming from the third pair of operators of second quantization) are coupled into the final rank $0$. The~similar structure of triple tensors we have and,~for the rest, reduced matrix elements. The~notations of the ranks $k$, $k^{\prime}$, and~$x$ in tensorial products correspond to the symmetry of the formula, i.e.,~the notation of the rank $k$ indicates that its values are determined by the permissible values of the $\redmem{\ell j}{\, C^{(k)} \,}{ \ell^{\prime} j^{\prime}}$ coefficients, where the coefficients themselves belong to the top interaction line of the Feynman diagram, whereas the notation of the rank $k^{\prime}$ indicates that its values are determined by the $\redmem{\ell j}{\, C^{(k)} \,}{ \ell^{\prime} j^{\prime}}$ coefficients coming from the lower interaction line of the Feynman diagram. Meanwhile, the~notation $x$ indicates that its values are not controlled by the aforementioned coefficients but are determined by the corresponding ranks of the secondary quantization operators. These notations make it easier to understand the formula itself and to implement it programmatically. As~we can see, the~pair of secondary quantization operators with the ranks $j$ is always coupled into the rank $k$, and~the rightmost pair of the operators with the ranks $j^{\prime}$ is coupled into $k^{\prime}$. Since Formula (\ref{eq:BogEnergy_PT}) is applicable to various cases, its tensorial structure varies, as~it is optimized for each specific case under consideration so that the spin-angular method~\cite{Gaietal:97a,Gaigalas:2026a}, which is based on the Racah algebra in the two spaces $j$, $q$, can be fully~utilized.

The contribution of the CV, C, CC, and~VV correlations in the second order of the perturbation theory is expressed over the corresponding spin-angular part (with multiplication, except~for $\Delta \mathcal{E}_{0 \; PT(\text{CV,C,CC,VV})} \left(K J \right)$): 
\begin{itemize}
\item $\Delta \mathcal{E}_{0 \; PT(\text{CV,C,CC,VV})} \left(K J \right)$ for CV~\cite{Gaigetal:2024CV,Gaigetal:2026CVT}, C~\cite{Gaigetal:2024C}, CC~\cite{Gaigetal:2024CC}, and~VV~\cite{Gaigetal:2025VV,Gaigetal:2025VVT} correlations;
 
\item $\Delta \mathcal{F}^{k}_{PT(\text{CV,C,CC,VV})} \left( n \ell j, \; n \ell j \right)$ for CV~\cite{Gaigetal:2024CV,Gaigetal:2026CVT}, C~\cite{Gaigetal:2024C}, CC~\cite{Gaigetal:2024CC}, and~VV~\cite{Gaigetal:2025VV,Gaigetal:2025VVT} correlations;
 
\item $\Delta \mathcal{F}^{k}_{PT(\text{CV,CC,VV})} \left( n \ell j, \; n' \ell' j' \right)$ for CV~\cite{Gaigetal:2024CV,Gaigetal:2026CVT}, CC~\cite{Gaigetal:2024CC}, and~VV~\cite{Gaigetal:2025VV,Gaigetal:2025VVT} correlations;
 
\item $\Delta \mathcal{G}^{k}_{PT(\text{CV,CC,VV})} \left( n \ell j, \; n' \ell' j' \right)$ for CV~\cite{Gaigetal:2024CV}, CC~\cite{Gaigetal:2024CC}, and~VV~\cite{Gaigetal:2025VV} correlations;
 
\item $\Delta \mathcal{R}^{k}_{PT(\text{CV,CC,VV})} \left( n \ell j n \ell j, \; n' \ell' j' n '\ell' j' \right)$ for CV~\cite{Gaigetal:2024CV}, CC~\cite{Gaigetal:2024CC}, and~VV~\cite{Gaigetal:2025VV} correlations;

\item $\Delta \widetilde{\mathcal{R}}^{(k,k',x)}_{PT(\text{VV})} \left( n \ell j \; n '\ell' j' \; n '\ell' j' \right)$ for VV correlations~\cite{Gaigetal:2025VVT};

\item $\Delta \widetilde{\mathcal{R}}^{(k,k',x)}_{PT(\text{VV})}\left( n \ell j \; n '\ell' j' \; n ''\ell'' j'' \right)$ for VV correlations~\cite{Gaigetal:2025VVT};

\item $\Delta \widetilde{\widetilde{\mathcal{R}}}^{(k,k',x)}_{PT(\text{CV})}
\left( n \ell j \; n '\ell' j' \; n '\ell' j' \right)$ for CV correlations~\cite{Gaigetal:2026CVT};
\item $\Delta \widetilde{\widetilde{\mathcal{R}}}^{(k,k',x)}_{PT(\text{CV})}
\left( n \ell j \; n '\ell' j' \; n ''\ell'' j'' \right)$ for CV correlations~\cite{Gaigetal:2026CVT}.
\end{itemize}

 Expressions for the member $\Delta \mathcal{R}^{k}_{PT(\text{CV,CC,VV})} \left( n \ell j n \ell j, \; n' \ell' j' n '\ell' j' \right)$ is coming only from the off-diagonal matrix element for accounting for CV (\ref{eq:CV-off_Diagonal}), CC (\ref{eq:CC-off_Diagonal}), and~VV (\ref{eq:VV-off_Diagonal}) correlations, while the expressions for the remaining parts come only from the diagonal matrix elements for accounting for the rest of the correlations~\cite{Gaigetal:2024CV,Gaigetal:2024C,Gaigetal:2024CC,Gaigetal:2025VV,Gaigetal:2025VVT,Gaigetal:2026CVT}. It should be noted that the $\Delta \mathcal{R}^{k}_{PT(\text{CV,CC,VV})} \left( n \ell j n \ell j, \; n' \ell' j' n '\ell' j' \right)$ factors are not included in the {RCI (RSMBPT)} computations~\cite{Gaigetal:2024CV,Gaigetal:2024C,Gaigetal:2024CC,Gaigetal:2025VV,Gaigetal:2025VVT,Gaigetal:2026CVT}. They are only considered in the {RCI+RSMBPT} calculations~\cite{Gaigetal:2024CV}.

The specificity and advantage of this approach~\cite{Gaigetal:2024CV,Gaigetal:2024C,Gaigetal:2024CC,Gaigetal:2025VV,Gaigetal:2025VVT,Gaigetal:2026CVT} is that the spin-angular parts of vacuum diagrams CC$_1$ and CC$_2$ from Figure~\ref{Feynman_Diagrams} are proportional to the simple multiplier
and the spin-angular parts of one-particle Feynman diagrams (CV$_1$, CV$_2$, CC$_3$, CC$_4$, CC$_5$, and~ VV$_1$ from Figure~\ref{Feynman_Diagrams}) are proportional to the number of subshell occupation, while the spin-angular parts $\widetilde{f}_k \left( \ell j^{w}, \; K \chi J  \right)$, $\widetilde{f}_k \left( \ell j^{w} \; \ell' j'^{w'},
	\; K \chi J  \right)$, $\widetilde{g}_k \left( \ell j^{w} \; \ell' j'^{w'}, \; K \chi J  \right)$, and~$\widetilde{v}_k \left( \ell j^{w} \; \ell' j'^{w'}, \ell j^{w-2} \; \ell' j'^{w'+2}, \; K \chi J \; K' \chi' J \right)$ of two-particle Feynman diagrams (CV$_3$, CV$_4$, CV$_5$, CV$_6$, CC$_6$, and~VV$_2$ from Figure~\ref{Feynman_Diagrams}) can be found using a standard library \texttt{librang}~\cite{Gaigalas:2022} from the {\sc Grasp}2018~\cite{Fro:14a}. 
In addition, the spin-angular part of the vacuum and one-particle Feynman diagrams are independent of the term.
Meanwhile, the~spin-angular coefficients of the three-particle diagrams CV$_7$ and VV$_3$ from Figure~\ref{Feynman_Diagrams} are also partly calculated with the help of these libraries. These would be coefficients $\widetilde{f}_k \left( \ell j^{w}, \; K \chi J  \right))$ and $\widetilde{f}_k \left( \ell j^{w} \; \ell' j'^{w'},
	\; K \chi J  \right)$. Only for the members at the coefficients $\Delta \widetilde{\mathcal{R}}^{(k,k',x)}_{PT(\text{VV})}
\left( n \ell j \; n '\ell' j' \; n '\ell' j' \right)$, $\Delta \widetilde{\mathcal{R}}^{(k,k',x)}_{PT(\text{VV})}\left( n \ell j \; n '\ell' j' \; n ''\ell'' j'' \right)$, $\Delta \widetilde{\widetilde{\mathcal{R}}}^{(k,k',x)}_{PT(\text{CV})}
\left( n \ell j \; n '\ell' j' \; n '\ell' j' \right)$, and~$\Delta \widetilde{\widetilde{\mathcal{R}}}^{(k,k',x)}_{PT(\text{CV})}\left( n \ell j \; n '\ell' j' \; n ''\ell'' j'' \right)$ such an extension of the library \texttt{librang}~\cite{Gaigalas:2022} is needed in order to find the following reduced matrix elements (see (\ref{eq:BogEnergy_PT}) or (24) in \cite{Gaigetal:2025VVT} and (23) in \cite{Gaigetal:2026CVT}):
\begin{equation}
\label{eq:TensorProduct1}
\left< \; \Psi \; \left\| \; \biggl[ \bigl[ \tilde{a}^{(j)} \times a^{(j)} \bigr]^{(k)} \times  \Bigl[ \bigl[ a^{(j')} \times \tilde{a}^{(j')} \bigr]^{(x)} \times \bigl[ a^{(j')} \times \tilde{a}^{(j')} \bigr]^{(k')} \Bigr]^{(k)} \biggr]^{(0)} \; \right\| \; \Psi \; \right> ,
\end{equation}
\begin{equation}
\label{eq:TensorProduct2}
\left< \; \Psi \; \left\| \; \biggl[ \Bigl[ \bigl[ \tilde{a}^{(j)} \times a^{(j)} \bigr]^{(k)} \times \bigl[ a^{(j'')} \times \tilde{a}^{(j'')} \bigr]^{(x)} \Bigr]^{(k')} \times \bigl[ a^{(j')} \times \tilde{a}^{(j')} \bigr]^{(k')} \biggr]^{(0)} \; \right\| \; \Psi \; \right> ,
\end{equation}
\begin{equation}
\label{eq:TensorProduct3}
\left< \; \Psi \; \left\| \; \biggl[ \bigl[ a^{(j)} \times \tilde{a}^{(j)} \bigr]^{(k)} \times  \Bigl[ \bigl[ a^{(j')} \times \tilde{a}^{(j')} \bigr]^{(x)} \times \bigl[ \tilde{a}^{(j')} \times a^{(j')} \bigr]^{(k')} \Bigr]^{(k)} \biggr]^{(0)} \; \right\| \; \Psi \; \right> ,
\end{equation}
\begin{equation}
\label{eq:TensorProduct4}
\left< \; \Psi \; \left\| \; \biggl[ \Bigl[ \bigl[ a^{(j)} \times \tilde{a}^{(j)} \bigr]^{(k)} \times \bigl[ \tilde{a}^{(j'')} \times a^{(j'')} \bigr]^{(x)} \Bigr]^{(k')} \times \bigl[ \tilde{a}^{(j')} \times a^{(j')} \bigr]^{(k')} \biggr]^{(0)} \; \right\| \; \Psi \; \right> .
\end{equation}

Below we discuss how to calculate these reduced matrix elements (\ref{eq:TensorProduct1})$-$(\ref{eq:TensorProduct4}), i.e.,~to calculate spin-angular coefficients of three-particle Feynman diagrams VV$_3$ and CV$_7$ from Figure~\ref{Feynman_Diagrams} for accounting for (\ref{eq:not-CV-a}), (\ref{eq:not-CV-b}), (\ref{eq:not-VV-a}), and (\ref{eq:not-VV-b}) 
 correlations by~extending and using the library \texttt{librang}~\cite{Gaigalas:2022}, in~a way that makes the calculations as efficient as possible, i.e.,~based on the combination of the angular momentum theory as described in Jucys and Bandzaitis~\cite{JucBan:77a}, the~concept of irreducible tensorial sets (Judd~\cite{Jud:67a}, Rudzikas and Kaniauskas~\cite{RudKan:84a}), the~generalized graphical approach (Gaigalas~et~al.~\cite{Gaietal:85a}), the~second quantization in coupled tensorial form (Rudzikas and Kaniauskas~\cite{RudKan:84a}), the~quasispin approach (Rudzikas~\cite{Rud:97a}), and~the use of reduced coefficients of fractional parentage (Gaigalas~et~al.~\cite{Gaietal:98a,Gaietal:2000a}) as was done in papers by Gaigalas~et~al.~\cite{Gaietal:97a,Gaigalas:2022}.
For additional information about the calculation of these reduced matrix elements, see Appendix~\ref{PT_Appendix1}.

\subsection{The Third Type of Valence–Valence and Core–Valence~Correlations}
\label{Sec:spin_angular_third}

For these types of correlations, the~calculation of the spin-angular coefficients at the members of $\Delta \widetilde{\mathcal{R}}^{(J_1,J_2,x)}_{PT(\text{VV})} \left( n \ell j \; n '\ell' j' \; n '\ell' j' \right)$ (see (24) or (26) in \cite{Gaigetal:2025VVT}) and $\Delta \widetilde{\widetilde{\mathcal{R}}}^{(J_1,J_2,x)}_{PT(\text{CV})} \left( n \ell j \; n '\ell' j' \; n '\ell' j' \right)$ (see (21) or (23) in \cite{Gaigetal:2026CVT})
 is more complex and is not fully supported by the library \texttt{librang}~\cite{Gaigalas:2022}. It is essential that these problematic terms have the following tensorial product see ((6) in \cite{Gaigetal:2025VVT}) from (\ref{eq:TensorProduct1})
\begin{equation}
\label{eq:VVT_Tensor12}
\biggl[ \bigl[ \tilde{a}^{(j_m)} \times a^{(j_m)} \bigr]^{(J_1)} \times \Bigl[ \bigl[ a^{(j_n)} \times \tilde{a}^{(j_n)} \bigr]^{(x)} \times \bigl[ a^{(j_n)} \times \tilde{a}^{(j_n)} \bigr]^{(J_2)}  \Bigr]^{(J_1)} \biggr]^{(0)}
\end{equation}
and the following (see (6) in \cite{Gaigetal:2026CVT}) from (\ref{eq:TensorProduct3})
\begin{equation}
\label{eq:CVT_Tensor12}
\biggl[ \bigl[ a^{(j_m)} \times \tilde{a}^{(j_m)} \bigr]^{(J_1)} \times \Bigl[ \bigl[ a^{(j_n)} \times \tilde{a}^{(j_n)} \bigr]^{(x)} \times \bigl[ \tilde{a}^{(j_n)} \times a^{(j_n)} \bigr]^{(J_2)}  \Bigr]^{(J_1)} \biggr]^{(0)}.
\end{equation}

We now discuss below how this problem can be~addressed.

A peculiarity of the methodology~\cite{GaiRud:96a,Gaietal:97a,Gaigalas:2026a} is that in the expression (24) in~\cite{Gaigalas:2022} for the calculation of the reduced matrix element of the two-particle operator
\begin{eqnarray}
\label{eq:mgb}
\redmem{\gamma_{\alpha} J}
       {\, \widehat{G}^{(k_j \ k_j \ 0)} \left( 
			 n_i \ell_i j_{i}, n_j \ell_j j_{j},
			 n_{i^{\prime }} \ell_{i^{\prime }} j_{{i}^{\prime }}, n_{j^{\prime }} \ell_{j^{\prime }} j_{j^{\prime }}
       \right) \,}
			 {\gamma_{\beta} J^{\prime }}
   \nonumber  \\[1ex]
& & \hspace{-7.5cm}
=
   \displaystyle {\sum_{\kappa _{12}}}
   ( -1)^\Delta \;
	 \Theta ^{\prime }\left( 
	n_i \ell_i j_{i}, n_j \ell_j j_{j},
			 n_{i^{\prime }} \ell_{i^{\prime }} j_{{i}^{\prime }}, n_{j^{\prime }} \ell_{j^{\prime }} j_{j^{\prime }},\; \Xi \right) 
\; T\left(j_i, j_j, j_{i^{\prime }}, j_{j^{\prime }}, \Lambda^{bra}, \Lambda^{ket},\Xi,\Gamma \right)
   \nonumber  \\[1ex]
& & \hspace{-7.5cm}
\times \; R\left( j_i, j_j, j_{i^{\prime }}, j_{j^{\prime }}, \Lambda^{bra}, \Lambda^{ket},\Gamma \right),
\end{eqnarray}
the 
 recoupling matrix $R\left( j_i, j_j, j_{i^{\prime }}, j_{j^{\prime }}, \Lambda^{bra}, \Lambda^{ket},\Gamma \right)$, 
the submatrix element $T\bigl(j_i, j_j, j_{i^{\prime }}, j_{j^{\prime }}, \Lambda^{bra}, \break  \Lambda^{ket},\Xi,\Gamma \bigr)$, 
the phase factor $\Delta $, and 
$\Theta ^{\prime }\left(	n_i \ell_i j_{i}, n_j \ell_j j_{j}, n_{i^{\prime }} \ell_{i^{\prime }} j_{{i}^{\prime }}, n_{j^{\prime }} \ell_{j^{\prime }} j_{j^{\prime }},\; \Xi \right)$, which is proportional to the radial part, 
are easily separated from each other and can be treated differently. 
This makes the methodology flexible and allows it to be easily extended to include new class/type operators, such as the three-particle operators (\ref{eq:VVT_Tensor12}) and (\ref{eq:CVT_Tensor12}) we are considering. The~paper~\cite{Gaietal:97a} describes how the expression (11) in ~\cite{Gaietal:97a} is implemented in the library in a regular way. By~the way, the~rank $k_j$ for operator $\widehat{G}^{(k_j \ k_j \ 0)}$ in Equation~(\ref{eq:mgb}) is equal to $J_1$ for the third type of VV and CV~correlations.

In the following, we will discuss only those aspects of the calculation matrix element of three-particle Feynman diagrams $A_5$ from~\cite{Gaigetal:2025VVT,Gaigetal:2026CVT} with the tensorial products (\ref{eq:VVT_Tensor12}) and (\ref{eq:CVT_Tensor12}), which were not covered in the paper~\cite{Gaigalas:2022}.  
The phase factor $\Delta $ and  $\Theta ^{\prime }\left(	n_i \ell_i j_{i}, n_j \ell_j j_{j}, n_{i^{\prime }} \ell_{i^{\prime }} j_{{i}^{\prime }}, n_{j^{\prime }} \ell_{j^{\prime }} j_{j^{\prime }},\; \Xi \right)$ do not bring any additional problems to the calculation of the spin-angular part of this case,
 i.e.,~$\Delta $, according to the methodology~\cite{Gaietal:97a}, is zero, and the $\Theta ^{\prime }\left(	n_i \ell_i j_{i}, n_j \ell_j j_{j}, n_{i^{\prime }} \ell_{i^{\prime }} j_{{i}^{\prime }}, n_{j^{\prime }} \ell_{j^{\prime }} j_{j^{\prime }},\; \Xi \right)$, which is proportional to the radial part, is found in a regular way as it was found in papers~\cite{Gaigetal:2024CV,Gaigetal:2024C,Gaigetal:2024CC,Gaigetal:2025VV}. For~the third type of VV correlations, $\Theta ^{\prime }\left(	n_i \ell_i j_{i}, n_j \ell_j j_{j}, n_{i^{\prime }} \ell_{i^{\prime }} j_{{i}^{\prime }}, n_{j^{\prime }} \ell_{j^{\prime }} j_{j^{\prime }},\; \Xi \right)$ is as follows (see Table 2 in \cite{Gaigetal:2025VVT}) 
 {\small
\begin{equation}
\label{eq:thetaVV3}
\Theta ^{\prime }\left(	n_i \ell_i j_{i}, n_j \ell_j j_{j} , n_{i^{\prime }} \ell_{i^{\prime }} j_{{i}^{\prime }}, n_{j^{\prime }} \ell_{j^{\prime }} j_{j^{\prime }},\; \Xi \right)
= 2 
\sum_{r} 
\left( -1 \right)^{-j_m + j_n + x} \sqrt{\left[ J_1, J_2, x \right]} \; \mathcal{G}\left( J_1 \, J_2 \, x, \, n n, \, m r \right)
\end{equation}}
and for the third type of CV correlations it is as follows (see Table 2 in \cite{Gaigetal:2026CVT}) 
\begin{equation}
\label{eq:thetaCV3}
\Theta ^{\prime }\left(	n_i \ell_i j_{i}, n_j \ell_j j_{j} , n_{i^{\prime }} \ell_{i^{\prime }} j_{{i}^{\prime }}, n_{j^{\prime }} \ell_{j^{\prime }} j_{j^{\prime }},\; \Xi \right)
= 2 
\sum_{a} 
\left( -1 \right)^{j_m + j_n} \sqrt{\left[ J_1, J_2, x \right]} \; \mathcal{G^{\prime}}\left( J_1 \, J_2 \, x, \, n n, \, m a \right),
\end{equation}
where in both cases (\ref{eq:thetaVV3}) and (\ref{eq:thetaCV3}) $n_i \ell_i j_{i} \equiv n$, $ n_j \ell_j j_{j} \equiv m$, $n_{i}^{\prime } \ell_{i}^{\prime } j_{i}^{\prime }  \equiv n$, $n_{j}^{\prime } \ell_{j}^{\prime } j_{j}^{\prime } \equiv  m$, and $\Xi \equiv J_1 \, J_2 \, x$.

Although expressions (\ref{eq:thetaVV3}) and (\ref{eq:thetaCV3}) appear very similar visually, they contain different coefficients $\mathcal{G}\left( J_1 \, J_2 \, x, \, n n, \, m r \right)$ and $\mathcal{G^{\prime}}\left( J_1 \, J_2 \, x, \, n n, \, m a \right)$ and different phases. This is related to the fact that the Feynman diagram expressions themselves differ in phase, multipliers, and~tensorial structure. For~example, in~the expression for the VV$_3$ diagram, we have the phase factor 
$(-1)^{k+k^{\prime}}$, whereas in the case of the CV$_7$ diagram, we have the factor ($-$1). Meanwhile, the~tensorial structure in the first case is
\begin{equation}
\label{eq:tensorVV3}
{\displaystyle{
		\left[\left[\left[\;  a^{\left( j_n \right) }  \times 
  \tilde a^{\left( j_{m} \right) } \; \right] ^{\left( k \right)} \times 
	\left[\; a^{\left( j_{n} \right) }  \times 
  \tilde a^{\left( j_{n} \right) } \; \right] ^{\left( x \right)} \right]^{\left( k^{\prime} \right)}	
  \times
	\left[\; a^{\left( j_{m} \right) }  \times 
  \tilde a^{\left( j_{n} \right) } \; \right] ^{\left( k^{\prime} \right)} \right]^{\left( 0 \right)}
	}},
\end{equation}
and in the second case is:
\begin{equation}
\label{eq:tensorCV3}
{\displaystyle{
		\left[\left[\left[\;  a^{\left( j_m \right) }  \times 
  \tilde a^{\left( j_{n} \right) } \; \right] ^{\left( k \right)} \times 
	\left[\; \tilde a^{\left( j_{n} \right) }  \times 
  a^{\left( j_{n} \right) } \; \right] ^{\left( x \right)} \right]^{\left( k^{\prime} \right)}	
  \times
	\left[\; a^{\left( j_{n} \right) }  \times 
  \tilde a^{\left( j_{m} \right) } \; \right] ^{\left( k^{\prime} \right)} \right]^{\left( 0 \right)}
	}}.
\end{equation}

These expressions are not suitable for calculating the reduced matrix elements of the above operators because~they do not fully exploit the advantages of the Racah algebra. Thus, they must be rearranged accordingly. For~the tensorial product (\ref{eq:tensorVV3}), using \mbox{Expression~(8)} from the paper~\cite{Gaigetal:2025VVT}, and~for the tensorial product (\ref{eq:tensorCV3}), using Expression (11) from the paper~\cite{Gaigetal:2026CVT}, we obtain new expressions for which the~\cite{Gaietal:97a,Gaigalas:2026a} method can already be used when calculating reduced matrix elements. Expressions (\ref{eq:thetaVV3}) and (\ref{eq:thetaCV3}) represent the terms of the triple tensorial product. For~almost all of the multipliers and intermediate sums in these terms, the~corresponding notations $\mathcal{G}\left( J_1 \, J_2 \, x, \, n n, \, m r \right)$ and $\mathcal{G^{\prime}}\left( J_1 \, J_2 \, x, \, n n, \, m a \right)$ have been introduced. Additionally, the~term $\displaystyle{\frac{-1}{~D}}$ and the phase $(-1)^{k^{\prime}}$ are included in the multipliers $\mathcal{G}\left( J_1 \, J_2 \, x, \, n n, \, m r \right)$ and $\mathcal{G^{\prime}}\left( J_1 \, J_2 \, x, \, n n, \, m a \right)$. After~summing all the phase factors defined in the Feynman diagram expression and the phase factors resulting from the tensorial structure transformations listed above, and~including the definitions of the $\mathcal{G}\left( J_1 \, J_2 \, x, \, n n, \, m r \right)$ and $\mathcal{G^{\prime}}\left( J_1 \, J_2 \, x, \, n n, \, m a \right)$ factors, we obtain the following phases, i.e.,~in Expression (\ref{eq:thetaVV3}), they are:
$$k + k^{\prime} + j_m + j_n + k + x - 1 - k^{\prime} =  j_m + j_n + x - 1 = -j_m + j_n+x,$$
and in Expression (\ref{eq:thetaCV3}), it is:
$$1 + j_m + j_n + k^{\prime} - 1 - k^{\prime} = j_m + j_n.$$

As we can see, although~at first glance the structure of Expressions (\ref{eq:thetaVV3}) and (\ref{eq:thetaCV3}) is very similar, their derivation is quite different, and~they differ not only in the phase but also in the values of the multipliers $\mathcal{G}\left( J_1 \, J_2 \, x, \, n n, \, m r \right)$ and $\mathcal{G^{\prime}}\left( J_1 \, J_2 \, x, \, n n, \, m a \right)$.

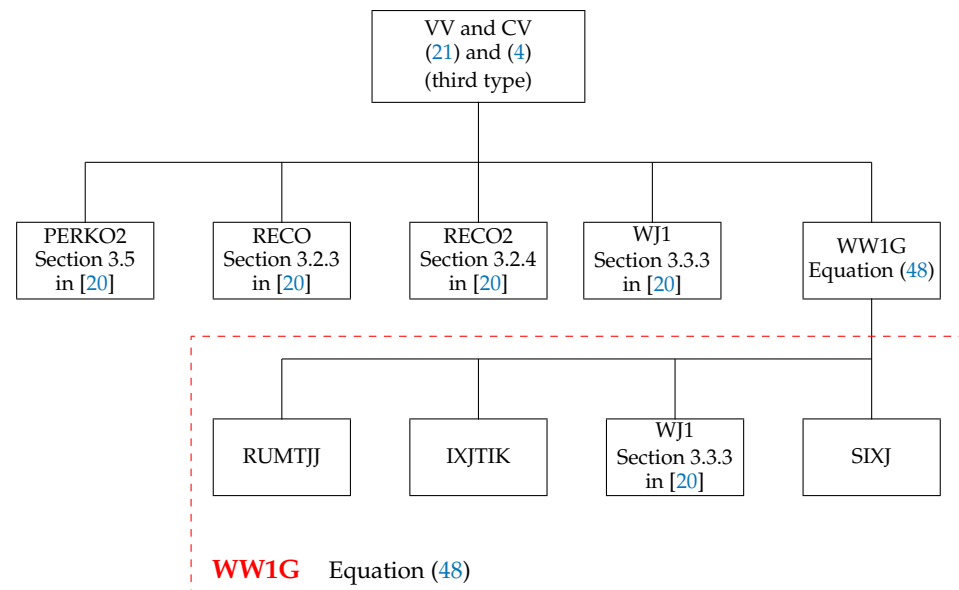
\begin{figure}[H]
\arraybackslash
\setlength{\unitlength}{1mm}
\hspace{5pt}\begin{picture}(123,84)
\put(47,72){\framebox(28,12){\footnotesize {\shortstack{VV and CV \\  (\ref{eq:not-VV-a}) and (\ref{eq:not-CV-a})\\(third type)}}}}
\put(61,64){\line(0,8){8}}
\put(9,64){\line(104,0){104}}
\put(9,56){\line(0,8){8}}
\put(0,46){\framebox(18,10){\footnotesize {\shortstack{PERKO2\\ Section 3.5\\in \cite{Gaigalas:2022}}}}}
\put(35,56){\line(0,8){8}}
\put(26,46){\framebox(18,10){\footnotesize {\shortstack{RECO\\ Section 3.2.3\\in \cite{Gaigalas:2022}}}}}
\put(61,56){\line(0,8){8}}
\put(52,46){\framebox(18,10){\footnotesize {\shortstack{RECO2\\ Section 3.2.4\\in \cite{Gaigalas:2022}}}}}
\put(84,56){\line(0,8){8}}
\put(75,46){\framebox(18,10){\footnotesize {\shortstack{WJ1\\ Section 3.3.3\\in \cite{Gaigalas:2022}}}}}
\put(113,56){\line(0,8){8}}
\put(104,46){\framebox(18,10){\footnotesize {\shortstack{WW1G\\ Equation~(\ref{eq:tg})}}}}
\put(35,38){\line(78,0){78}}
\put(35,30){\line(0,8){8}}
\put(26,20){\framebox(18,10){\footnotesize {\shortstack{RUMTJJ}}}}
\put(61,30){\line(0,8){8}}
\put(52,20){\framebox(18,10){\footnotesize {\shortstack{IXJTIK}}}}
\put(87,30){\line(0,8){8}}
\put(78,20){\framebox(18,10){\footnotesize {\shortstack{WJ1\\ Section 3.3.3\\in \cite{Gaigalas:2022}}}}}
\put(113,30){\line(0,16){16}}
\put(104,20){\framebox(18,10){\footnotesize {\shortstack{SIXJ}}}}
{\color{red} \put(23,7){\dashbox(102,34)}}
{\color{red} \put(33,9){\makebox[18mm]{{\bf WW1G}{\color{black} \; \; \small{Equation (\ref{eq:tg})}}}}}
\end{picture}
\caption{The scheme for the calculation of reduced matrix elements of operators (\ref{eq:VVT_Tensor12}) and (\ref{eq:CVT_Tensor12}), i.e.,~the scheme for the spin-angular coefficients calculation of Feynman diagrams VV$_3$ from~\cite{Gaigetal:2025VVT} and CV$_7$ from~\cite{Gaigetal:2026CVT} for correlations (\ref{eq:not-VV-a}) and (\ref{eq:not-CV-a}).}
\label{op-VV3}
\end{figure}

So it remains to discuss finding  the values of the recoupling matrix $R\bigl( j_i, j_j, j_{i^{\prime }}, j_{j^{\prime }}, \linebreak \Lambda^{bra},   \Lambda^{ket},\Gamma \bigr)$ and the submatrix element $T\left(j_i, j_j, j_{i^{\prime }}, j_{j^{\prime }}, \Lambda^{bra}, \Lambda^{ket},\Xi,\Gamma \right)$, which are related to the calculation of reduced matrix elements of operators (\ref{eq:VVT_Tensor12}) and (\ref{eq:CVT_Tensor12}).
Figure~\ref{op-VV3} shows how  and which subroutines from the program library \texttt{librang} should be used and which expressions should be added for their calculation of~them.

Since we can schematically rewrite the tensorial products (\ref{eq:VVT_Tensor12}) and (\ref{eq:CVT_Tensor12}) as the tensorial product of two operators $A^{(J_{1})}(n_{m}j_{m})$ and $B^{(J_{1})}(n_{n}j_{n})$ acting on different subshells
\begin{equation}
\label{eq:tena}
\left[\; A^{(J_{1})}(n_{m}j_{m}) \times B^{(J_{1})}(n_{n}j_{n}) \; \right]^{(0)},
\end{equation}
the algebraic expression of the recoupling matrix $R\left( j_i, j_j, j_{i^{\prime }}, j_{j^{\prime }}, \Lambda^{bra}, \Lambda^{ket},\Gamma \right)$ can be used from the paper (see (19) in~\cite{Gaietal:97a}); i.e.,~first of all, the~selection rules (see Table 3 in~\cite{Gaigalas:2022}) for the recoupling matrix can be checked by the subroutine \texttt{RECO} (see Section~3.2.3 in~\cite{Gaigalas:2022}), and~after that, it can be computed by the subroutine  \texttt{RECO2} (see Section~3.2.4 in~\cite {Gaigalas:2022}and Figure~\ref{op-VV3}). Therefore, the~library's \texttt{librang} existing capabilities are entirely sufficient for calculating $R\left( j_i, j_j, j_{i^{\prime }}, j_{j^{\prime }}, \Lambda^{bra}, \Lambda^{ket},\Gamma \right)$.

As far as the submatrix element $T\left(j_i, j_j, j_{i^{\prime }}, j_{j^{\prime }}, \Lambda^{bra}, \Lambda^{ket},\Xi,\Gamma \right) $ is concerned, the~situation is different. The~$T\left(j_i, j_j, j_{i^{\prime }}, j_{j^{\prime }}, \Lambda^{bra}, \Lambda^{ket},\Xi,\Gamma \right)$ is equal to the matrix element of the tensorial product (\ref{eq:tena}) between bra and ket functions containing only those subsells to which this tensor operator acts, i.e.,
\begin{eqnarray}
\label{eq:rmatrixa}
T\left(j_i, j_j, j_{i^{\prime }}, j_{j^{\prime }}, \Lambda^{bra}, \Lambda^{ket},\Xi,\Gamma \right)
   \nonumber  \\[1ex]
& & \hspace{-4.5cm}
 =
   \left< (n_m\ell_m)\, j_m^{w_m}\, \alpha_m Q_mJ_m \; (n_n\ell_n) \, j_n^{w_n}\, \alpha_n Q_nJ_n \right.
   \nonumber  \\[1ex]
& & \hspace{-1.8cm}
	        \left\| \, A^{(J_{1})}(n_{m}j_{m}) \, B^{(J_{1})}(n_{n}j_{n}) \, \right\|
   \nonumber  \\[1ex]
& & 
		\left. (n_n\ell_m)\, j_m^{w^{\prime }_{m}} \, \alpha_{m} ^{\prime }Q^{\prime}_{m}J^{\prime }_{m} \;
					(n_n\ell_n)\, j_n^{w^{\prime }_{n}} \, \alpha_n ^{\prime }Q^{\prime}_{n}J^{\prime }_{n} \right>,
\end{eqnarray}
where the binding of the ranks $J_1$, $J_1$, to~$0$ of the tensorial structure (\ref{eq:tena}) is already included in the recoupling matrix $R\left( j_i, j_j, j_{i^{\prime }}, j_{j^{\prime }}, \Lambda^{bra}, \Lambda^{ket},\Gamma \right)$, $n_i \ell_i j_{i} \equiv n$, $ n_j \ell_j j_{j} \equiv m$, $n_{i}^{\prime } \ell_{i}^{\prime } j_{i}^{\prime }  \equiv n$, $n_{j}^{\prime } \ell_{j}^{\prime } j_{j}^{\prime } \equiv  m$, $\Xi \equiv J_1 \, J_2 \, x$, $\Lambda^{bra} \equiv Q_mJ_m Q_nJ_n$, $\Lambda^{ket} \equiv Q^{\prime}_{m}J^{\prime }_{m} Q^{\prime}_{n}J^{\prime }_{n}$, and~$\Gamma$ is an empty array.
In this case, the~tensorial product can be simply split into two parts, i.e.,
\begin{equation}
\label{eq:tenb}
A^{(J_{1})}(n_{m}j_{m}) \; \equiv \bigl[ a^{(q \, j_m)}_{m_{q1}} \times a^{(q \, j_m)}_{m_{q2}} \bigr]^{(J_1)}
\end{equation}
and
\begin{equation}
\label{eq:tenc}
B^{(J_{1})}(n_{n}j_{n}) \;  \equiv \Bigl[ \bigl[ a^{(q \, j_n)}_{m_{q3}} \times a^{(q \, j_n)}_{m_{q4}} \bigr]^{(x)}  \times \bigl[ a^{(q \, j_n)}_{m_{q5}} \times a^{(q \, j_n)}_{m_{q6}} \bigr]^{(J_2)} \Bigr]^{(J_1)}.
\end{equation}

Here, we use the quasispin formalism~\cite{Rud:97a,GaiRud:96a,Gaietal:2000a}, where the operators of second quantization are the components of an
irreducible tensor of rank $q=1/2$ in a quasispin space~\cite{KanRudz:80}.
\begin{eqnarray}
\label{eq:a_tripletensor}
   a^{ \left( q \ j \right)}_{m_q \; m_j} =\left\{
   \begin{array}{ll}
   a^{ \left( j \right)}_{m_j}        & \mbox{ for } m_q = +\frac{1}{2}, \\ \\
   \tilde a^{\left( j \right)}_{m_j} & \mbox{ for } m_q = -\frac{1}{2} .
   \end{array}
   \right.
\end{eqnarray}

In the case where $m_{q2}=m_{q3}=m_{q5}=\frac{1}{2}$ and $m_{q1}=m_{q4}=m_{q6}=-\frac{1}{2}$, then the tensorial product (\ref{eq:tena}) corresponds to the tensorial product (\ref{eq:VVT_Tensor12}) and where $m_{q1}=m_{q3}=m_{q6}=\frac{1}{2}$ and $m_{q2}=m_{q4}=m_{q5}=-\frac{1}{2}$, then the tensorial product (\ref{eq:tena}) corresponds to the tensorial product (\ref{eq:CVT_Tensor12}).

Since the tensorial product can be split into two parts (\ref{eq:tenb}) and (\ref{eq:tenc}), the~reduced matrix elements of these two members $A^{(J_{1})}(n_{m}j_{m})$ and $B^{(J_{1})}(n_{n}j_{n})$ should be considered separately (because they act on different subshells, and~the binding of the ranks of the tensorial structure, as~was mentioned above, is already included in the recoupling matrix). Subroutine  \texttt{WJ1} (see Section 3.3.3 in \cite{Gaigalas:2022} and Figure~\ref{op-VV3}) finds the reduced matrix element of operator~(\ref{eq:tenb}), while the library \texttt{librang} has no suitable subroutine for computing the reduced matrix element of the operator (\ref{eq:tenc}). To~find the latter, it is required to use an expression such as
\begin{eqnarray}
\label{eq:tg}
   \redmem{(n_n\ell_n)\, j_n^w\, \alpha QJ}
	        {\, \Bigl[ \bigl[ a^{(q \, j_n)}_{m_{q3}} \times a^{(q \, j_n)}_{m_{q4}} \bigr]^{(x)}  \times \bigl[ a^{(q \, j_n)}_{m_{q5}} \times a^{(q \, j_n)}_{m_{q6}} \bigr]^{(J_2)} \Bigr]^{(J_1)} \,}
					{(n_n\ell_n)\, j_n^{w^{\prime }} \, \alpha ^{\prime }Q^{\prime}J^{\prime }}
   \nonumber  \\[1ex]
\hspace{-1.5cm}
   =\left( -1\right) ^{J + J^{\prime } + J_1} \ \sqrt{\left[ J_1 \right]} \ 
   \displaystyle {\sum_{\alpha ^{\prime \prime }Q^{\prime \prime }J^{\prime \prime }}}
   \ \left\{
   \begin{array}{ccc}
      x         & J_2 & J_1 \\
      J^{\prime } & J   & J^{\prime \prime }
   \end{array}
   \right\}
   \nonumber  \\[1ex]
\hspace{-1cm}
   \times
   \redmem{(n_n\ell_n)\, j_n^w \,\alpha QJ}{\, \bigl[ a^{(q \, j_n)}_{m_{q3}} \times a^{(q \, j_n)}_{m_{q4}} \bigr]^{(x)} \,}
	        {(n_n\ell_n)\, j_n^{w^{\prime \prime }}\, \alpha ^{\prime \prime }Q^{\prime \prime }J^{\prime \prime }} \
   \nonumber  \\[1ex]
\hspace{-1cm}
   \times
   \redmem{(n_n\ell_n)\, j_n^{w^{\prime \prime }}\, \alpha ^{\prime \prime }Q^{\prime \prime }J^{\prime \prime }}
	        {\, \bigl[ a^{(q \, j_n)}_{m_{q5}} \times a^{(q \, j_n)}_{m_{q6}} \bigr]^{(J_2)} \,}{(n_n\ell_n)\, j_n^{w^{\prime }}\, \alpha ^{\prime }Q^{\prime }J^{\prime }}.
\end{eqnarray}

Therefore, the~library \texttt{librang} has been extended by subroutine  \texttt{WW1G} (see Figure~\ref{op-VV3}), which determines the value of the reduced matrix elements of the operator (\ref{eq:tenc}) according to Expression (\ref{eq:tg}). The~red-framed {WW1G} in Figure~\ref{op-VV3} shows the structure of this new routine, \texttt{WW1G}. There subroutine \texttt{IXJTIK} checks the triangular delta functions of the 6$j$-coefficients from Equation~(\ref{eq:tg}), subroutine \texttt{SIXJ} calculates the values of this coefficient if the triangular delta functions are not zero, and~subroutine \texttt{WJ1} (see Section~3.3.3 in~\cite{Gaigalas:2022}) calculates the reduced matrix elements of the pairs $\bigl[ a^{(q \, j)}_{m_{q}} \times a^{(q \, j)}_{m_{q'}} \bigr]^{(k)}$
of operators of the second quantization from (\ref{eq:tg}). In~this case, the~subroutine \texttt{WJ1} is called twice. Meanwhile, subroutine \texttt{RUMTJJ} prepares the corresponding input arrays for these calculations.
Meanwhile, the~subroutine \texttt{PERKO2} outside the red frame {WW1G} in Figure~\ref{op-VV3} is the interface between {\sc Grasp}2018~\cite{Fro:14a} and the SQ routines group~(see Section~3.3 in \cite{Gaigalas:2022}) from the library \texttt{librang}. For~additional information about the subroutine, see Appendix~\ref{PT_Appendix2}.

All of the above implementations allow simple/easy calculation of the third type of VV and CV correlations with minimal program library \texttt{librang} expansion.

\subsection{The Fourth Type of Valence–Valence and Core–Valence~Correlations}
\label{Sec:spin_angular_fourth}

\textls[-15]{For these types of correlations, the~calculation of the spin-angular coefficients at the members of $\Delta \widetilde{\mathcal{R}}^{(J_1,J_2,x)}_{PT(\text{VV})}
\left( n \ell j \; n '\ell' j' \; n ''\ell'' j'' \right)$ (see (24) or (26) in \cite{Gaigetal:2025VVT}) and $\Delta \widetilde{\widetilde{\mathcal{R}}}^{(J_1,J_2,x)}_{PT(\text{CV})}
\left( n \ell j \; n '\ell' j' \; n ''\ell'' j'' \right)$} (see (21) or (23) in \cite{Gaigetal:2026CVT})
 is complex, and we need to discuss how to find them. These problematic terms have the following tensorial products (see (14) and (21) in \cite{Gaigetal:2025VVT}) from (\ref{eq:TensorProduct2})
\begin{equation}
\label{eq:VVT_Tensor212}
\biggl[ \Bigl[ \bigl[ \tilde{a}^{(j_m)} \times a^{(j_m)} \bigr]^{(J_1)} \times  \bigl[ a^{(j_p)} \times \tilde{a}^{(j_p)} \bigr]^{(x)} \Bigr]^{(J_2)} \times \bigl[ a^{(j_n)} \times \tilde{a}^{(j_n)} \bigr]^{(J_2)}  \biggr]^{(0)}
\end{equation}
and the following (see (14) and (18) in \cite{Gaigetal:2026CVT}) from (\ref{eq:TensorProduct4})
\begin{equation}
\label{eq:CVT_Tensor212}
\biggl[ \Bigl[ \bigl[ a^{(j_m)} \times \tilde{a}^{(j_m)} \bigr]^{(J_1)} \times  \bigl[ \tilde{a}^{(j_p)} \times a^{(j_p)} \bigr]^{(x)} \Bigr]^{(J_2)} \times \bigl[ \tilde{a}^{(j_n)} \times a^{(j_n)} \bigr]^{(J_2)}  \biggr]^{(0)}.
\end{equation}

\begin{figure}[H]
\begin{adjustwidth}{-\extralength}{0cm}
\centering\arraybackslash
\setlength{\unitlength}{1mm}
\begin{picture}(148,45)
\put(60,33){\framebox(28,12){\footnotesize {\shortstack{VV and CV \\ (\ref{eq:not-VV-b}) and (\ref{eq:not-CV-b}) \\(fourth type)}}}}
\put(74,25){\line(0,8){8}}
\put(9,25){\line(130,0){130}}
\put(9,17){\line(0,8){8}}
\put(0,7){\framebox(18,10){\footnotesize {\shortstack{EILE}}}}
\put(35,17){\line(0,8){8}}
\put(26,7){\framebox(18,10){\footnotesize {\shortstack{RECO\\ Section~3.2.3\\ in \cite{Gaigalas:2022}}}}}
\put(61,17){\line(0,8){8}}
\put(52,7){\framebox(18,10){\footnotesize {\shortstack{PERKO2\\ Section~3.5\\ in \cite{Gaigalas:2022}}}}}
\put(87,17){\line(0,8){8}}
\put(78,7){\framebox(18,10){\footnotesize {\shortstack{WJ1\\ Section~3.3.3\\ in \cite{Gaigalas:2022}}}}}
\put(113,17){\line(0,8){8}}
\put(104,7){\framebox(18,10){\footnotesize {\shortstack{RECO2\\ Section~3.2.4\\ in \cite{Gaigalas:2022}}}}}
\put(139,17){\line(0,8){8}}
\put(130,7){\framebox(18,10){\footnotesize {\shortstack{REC3\\ Section~3.2.5\\ in \cite{Gaigalas:2022}}}}}
\end{picture}
\end{adjustwidth}
\caption{The 
 scheme for the calculation of reduced matrix elements of operators (\ref{eq:VVT_Tensor212}) and (\ref{eq:CVT_Tensor212}), i.e.,~the scheme for the spin-angular coefficients calculation of Feynman diagrams $A_5$ and $A_6$ from~\cite{Gaigetal:2025VVT,Gaigetal:2026CVT} for correlations (\ref{eq:not-VV-b}) and (\ref{eq:not-CV-b}).}
\label{op-VV4}

\end{figure}
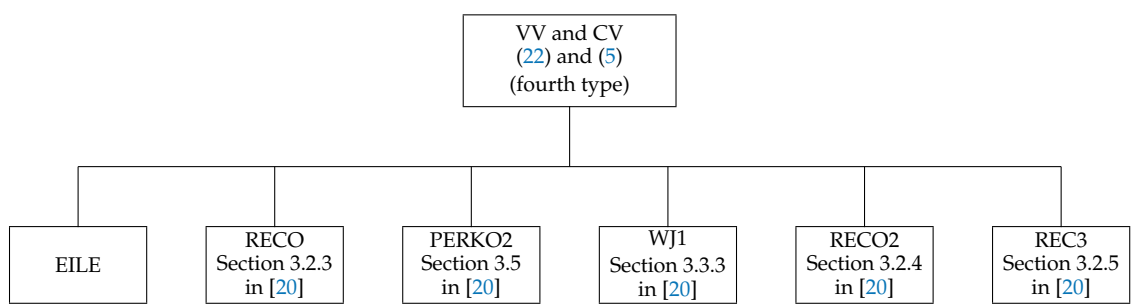

Since we can schematically rewrite the tensorial products (\ref{eq:VVT_Tensor212}) and (\ref{eq:CVT_Tensor212}) as the tensorial product of three operators, $A^{(J_{1})}(n_{m}j_{m})$, $B^{(x)}(n_{p}j_{p})$, $C^{(J_{2})}(n_{n}j_{n})$ acting on different subshells
\begin{equation}
\label{eq:ten2a}
\Bigl[ \bigl[ A^{(J_{1})}(n_{m}j_{m}) \times B^{(x)}(n_{p}j_{p}) \bigr]^{(J_{2})} \times C^{(J_{2})}(n_{n}j_{n}) \Bigr]^{(0)},
\end{equation}
\textls[15]{the algebraic expression of the recoupling matrix $R\left( j_i, j_j, j_{i^{\prime }}, j_{j^{\prime }}, \Lambda^{bra}, \Lambda^{ket},\Gamma \right)$ (see} \linebreak  Equation~(\ref{eq:mgb})) can be used from the paper~(see (19) in \cite{Gaietal:97a}), i.e.,~first of all, the~selection rules for the recoupling matrix can be checked by the subroutine \texttt{RECO} (see Section~3.2.3 in~\cite{Gaigalas:2022}), and~after that, it can be computed by the subroutine  \texttt{RECO2} (see Section~3.2.4 in~\cite{Gaigalas:2022}) in case $J_1 = 0$ or by the subroutine \texttt{REC3} (see Section~3.2.5 in~\cite{Gaigalas:2022}) in case $J_1 \neq 0$ (see Figure~\ref{op-VV4}). Therefore, the~library's \texttt{librang} existing capabilities are entirely sufficient for calculating $R\left( j_i, j_j, j_{i^{\prime }}, j_{j^{\prime }}, \Lambda^{bra}, \Lambda^{ket},\Gamma \right)$.

The submatrix element $T\left(j_i, j_j, j_{i^{\prime }}, j_{j^{\prime }}, \Lambda^{bra}, \Lambda^{ket},\Xi,\Gamma \right)$ (see Equation~(\ref{eq:mgb})) has the following form for the fourth type of VV and CV correlations
\begin{eqnarray}
\label{eq:rmatrixb}
\hspace{-0.5cm}
T\left(j_i, j_j, j_{i^{\prime }}, j_{j^{\prime }}, \Lambda^{bra}, \Lambda^{ket},\Xi,\Gamma \right)
   \nonumber  \\
& & \hspace{-4.5cm}
 =
   \left< (n_m\ell_m)\, j_m^{w_m}\, \alpha_m Q_mJ_m \; (n_n\ell_n) \, j_n^{w_n}\, \alpha_n Q_nJ_n \,  j_p^{w_p}\, \alpha_p Q_pJ_p \right.
	   \nonumber  \\[1ex]
& & \hspace{-1.8cm}
	        \left\| \, A^{(J_{1})}(n_{m}j_{m}) \, B^{(x)}(n_{p}j_{p}) \, C^{(J_{2})}(n_{n}j_{n}) \, \right\|
	   \nonumber  \\[1ex]
& & \hspace{-0.5cm}
					\left. (n_n\ell_m)\, j_m^{w^{\prime }_{m}} \, \alpha_{m} ^{\prime }Q^{\prime}_{m}J^{\prime }_{m} \;
					(n_n\ell_n)\, j_n^{w^{\prime }_{n}} \, \alpha_n ^{\prime }Q^{\prime}_{n}J^{\prime }_{n} \, j_p^{w^{\prime }_p}\, \alpha_p ^{\prime }Q^{\prime}_{p}J^{\prime }_{p} \right> .
\end{eqnarray}

In this case, the~tensorial product can be split into three parts, i.e.,
\begin{equation}
\label{eq:ten2b}
A^{(J_{1})}(n_{m}j_{m}) \; \equiv \bigl[ a^{(q \, j_m)}_{m_{q1}} \times a^{(q \, j_m)}_{m_{q2}} \bigr]^{(J_1)},
\end{equation}
\begin{equation}
\label{eq:ten2c}
B^{(x)}(n_{p}j_{p}) \;  \equiv \bigl[ a^{(q \, j_p)}_{m_{q3}} \times a^{(q \, j_p)}_{m_{q4}} \bigr]^{(x)},
\end{equation}
and
\begin{equation}
\label{eq:ten2d}
C^{(J_{2})}(n_{n}j_{n}) \;  \equiv \bigl[ a^{(q \, j_n)}_{m_{q5}} \times a^{(q \, j_n)}_{m_{q6}} \bigr]^{(J_2)}.
\end{equation}

In the case where $m_{q2}=m_{q3}=m_{q5}=\frac{1}{2}$ and $m_{q1}=m_{q4}=m_{q6}=-\frac{1}{2}$, then the tensorial product (\ref{eq:ten2a}) corresponds to the tensorial product (\ref{eq:VVT_Tensor212}), and when $m_{q1}=m_{q4}=m_{q6}=\frac{1}{2}$ and $m_{q2}=m_{q3}=m_{q5}=-\frac{1}{2}$, then the tensorial product (\ref{eq:ten2a}) corresponds to the tensorial product (\ref{eq:CVT_Tensor212}).

These three members $A^{(J_{1})}(n_{m}j_{m})$, $B^{(x)}(n_{p}j_{p})$, and~$C^{(J_{2})}(n_{n}j_{n})$ in (\ref{eq:ten2a}) should be considered separately (because they act on different subshells, and~the binding of the ranks $J_1$, $x$, to~$J_2$ and $J_2$, $J_2$, to~$0$ of the tensorial structure (\ref{eq:ten2a}) is already included in the recoupling matrix). Subroutine  \texttt{WJ1} (see Section~3.2.3 in \cite{Gaigalas:2022} and Figure~\ref{op-VV4}) finds the reduced  matrix element of all these operators (\ref{eq:ten2b})--(\ref{eq:ten2d}). In~this case, this subroutine is called three times. The~subroutine \texttt{EILE} reorders the operators of second quantization (\ref{eq:VVT_Tensor212}) and (\ref{eq:CVT_Tensor212}) in the same order as the subshells in the CSF from the matrix-reduced element. The~rest of the subroutines are the same as in Figure~\ref{op-VV3}.

So in this case, no new subroutines have to be created. The~old ones are enough, but~to do these calculations, it is necessary to choose the right subroutines already existing in the library and to call them in the right order with the right argument values, as~schematically shown in Figure~\ref{op-VV4}.

\textls[15]{Finally, as~regards the members $\Delta $ and $\Theta ^{\prime }\left(	n_i \ell_i j_{i}, n_j \ell_j j_{j}, n_{i^{\prime }} \ell_{i^{\prime }} j_{{i}^{\prime }}, n_{j^{\prime }} \ell_{j^{\prime }} j_{j^{\prime }},\; \Xi \right)$ in} \mbox{Formula (\ref{eq:mgb}}), they do not pose any problem, as~it was in Section~\ref{Sec:spin_angular_third}. The~$\Delta $, according to the methodology~\cite{Gaietal:97a}, is equal to zero. Meanwhile, the expressions for member $\Theta ^{\prime }\left(	n_i \ell_i j_{i}, n_j \ell_j j_{j}, n_{i^{\prime }} \ell_{i^{\prime }} j_{{i}^{\prime }}, n_{j^{\prime }} \ell_{j^{\prime }} j_{j^{\prime }},\; \Xi \right)$ are published in papers (see Table 3 in \cite{Gaigetal:2025VVT} and Table 3 in~\cite{Gaigetal:2026CVT}) and are, respectively, as~follows for the fourth type of VV correlation 
{\small
\begin{eqnarray}
\label{eq:thetaVV4}
\Theta ^{\prime }\left(	n_i \ell_i j_{i}, n_j \ell_j j_{j}, n_{i^{\prime }} \ell_{i^{\prime }} j_{{i}^{\prime }}, n_{j^{\prime }} \ell_{j^{\prime }} j_{j^{\prime }},\; \Xi \right)
   \nonumber  \\[1ex]
& & \hspace{-6.0cm}
= 
\left( -1 \right)^{j_m + j_n} \sqrt{\left[ J_1, J_2, x \right]} \; \sum_{r} \; \Biggl\{ \left( -1 \right)^{x+1}
\mathcal{G}\left( J_1 \, J_2 \, x, \, n p, \, m r \right) \Biggr.
   \nonumber  \\[1ex]
& & \hspace{-6.0cm}
+ \sum_{k_1, k_2, k_3} 
\left( -1 \right)^{J_1 + J_2 + k_1} \; \left[ k_3 \right]
   \;
		  \left\{
    \begin{array}{ccc}
      j_{n} & j_{p} & k_3 \\
      k_{1} & k_{2} & j_r
    \end{array} \right\} \;
 \mathcal{Q} \left( k_1k_2, n p, m r \right)
\; \mathcal{C}_{12j}\left( j_m j_n j_p, \, k_1 \, k_2 \, k_3, \, J_1 \, J_2 \, x \right)\Biggl\} 
   \nonumber  \\[1ex]
& & \hspace{-6.0cm}
 \times \;
\Biggl( 1 + \mbox{P} 
		  \left( \hspace{-0.15cm}
\begin{array}{lcl}
      n    &\hspace{-0.25cm}\rightleftharpoons&\hspace{-0.25cm}p \\
      k_{1}&\hspace{-0.25cm}\rightleftharpoons&\hspace{-0.25cm}k_{2} \\
			J_2  &\hspace{-0.25cm}\rightleftharpoons&\hspace{-0.25cm}x 
 \end{array}  \hspace{-0.15cm} \right)
 \Biggr)
\end{eqnarray}}
and for the fourth type of CV correlation
{\small
\begin{eqnarray}
\label{eq:thetaCV4}
\Theta ^{\prime }\left(	n_i \ell_i j_{i}, n_j \ell_j j_{j}, n_{i^{\prime }} \ell_{i^{\prime }} j_{{i}^{\prime }}, n_{j^{\prime }} \ell_{j^{\prime }} j_{j^{\prime }},\; \Xi \right)
   \nonumber  \\[1ex]
& & \hspace{-6.0cm}
= 
- 
\left( -1 \right)^{j_m + j_n} \sqrt{\left[ J_1, J_2, x \right]} \; \sum_{a} \; \Biggl\{ \left( -1 \right)^{x+1}
\mathcal{G^{\prime}}\left( J_1 \, J_2 \, x, \, m a, \, n p \right) \Biggr.
   \nonumber  \\[1ex]
& & \hspace{-6.0cm}
+ \sum_{k_1, k_2, k_3} \left( -1 \right)^{J_1 + J_2 + k_1} \; \left[ k_3 \right]
  \;
		  \left\{
    \begin{array}{ccc}
      j_{n} & j_{p} & k_3 \\
      k_{1} & k_{2} & j_a
    \end{array} \right\}
	\;
 \mathcal{Q} \left( k_1k_2, m a, n p \right)
\; \mathcal{C}_{12j}\left( j_m j_n j_p, \, k_1 \, k_2 \, k_3, \, J_1 \, J_2 \, x \right)\Biggl\}
   \nonumber  \\[1ex]
& & \hspace{-6.0cm}
 \times \;
\Biggl( 1 + \mbox{P} 
		  \left( \hspace{-0.15cm}
\begin{array}{lcl}
      n    &\hspace{-0.25cm}\rightleftharpoons&\hspace{-0.25cm}p \\
      k_{1}&\hspace{-0.25cm}\rightleftharpoons&\hspace{-0.25cm}k_{2} \\
			J_2 &\hspace{-0.25cm}\rightleftharpoons&\hspace{-0.25cm}x 
    \end{array}  \hspace{-0.15cm} \right)
 \Biggr).
\end{eqnarray}}

In the expressions, (1 + P) represents two members, where the second member is obtained from the first by making replacements marked in P. This simultaneously simplifies the expression and reveals the symmetry between the two members.
Moreover, the~$\Theta ^{\prime }\left(	n_i \ell_i j_{i}, n_j \ell_j j_{j}, n_{i^{\prime }} \ell_{i^{\prime }} j_{{i}^{\prime }}, n_{j^{\prime }} \ell_{j^{\prime }} j_{j^{\prime }},\; \Xi \right)$ member does not belong to the spin-angular part but to the amplitude of the effective operator. It is therefore only needed to find the reduced matrix element of the Feynman diagram according to Formula (\ref{eq:mgb}).

\section{Calculation of Core–Valence, Core, Core–Core, and~Valence–Valence with a New~Approach}
\label{Sec:Calculation}
This section presents computations using the method
based on the Rayleigh–Schr\"odinger perturbation theory 
in an irreducible tensorial form~\cite{Gaigetal:2024CV,Gaigetal:2024C,Gaigetal:2024CC,Gaigetal:2025VV,Gaigetal:2025VVT,Gaigetal:2026CVT}.
The developed RSMBPT method was applied to select the most significant configuration state functions
and further used them for solving the self-consistent field equations.
The radial wavefunctions obtained by the RSMBPT method were used for further RCI investigations in which
the RSMBPT method was also applied to determine the most important correlations of various~types.

The main goal is to demonstrate how the developed method can be used at different stages of the calculation process (MCDHF and RCI). It is important to check how it works, how the results are reproduced, and how they agree compared with regular {\sc Grasp} calculations. This is a newly developed method, and~this is only the first paper demonstrating the estimation of complete (CV, C, CC, and~VV) correlations using RSMBPT. 
Light atoms/ions are the best candidates for such investigations and standard practice for presentations of new methods.
Therefore, Ar~II, which has a~relatively large core, was selected as the test case for such~investigations.

For this purpose, 3 energy levels ($\mathrm{3s^23p^5~^2P^o_{1/2,3/2}}$ and $\mathrm{3s3p^6~^2S_{1/2}}$) and electric dipole (E1) transitions between these states were computed for Ar~II.
{Since the $\mathrm{3s^23p^5~^2P^o_{1/2,3/2}}$ states are pure, and the $\mathrm{3s3p^6~^2S_{1/2}}$ state 
has strong mixing with the $\mathrm{3s^23p^43d}$ configuration, this configuration was included in the MR set.}
{In the present paper, the~MCDHF computations were performed simultaneously for even and odd states 
using the extended optimal level (EOL) scheme \citep{Dyaetal:89a}.
As a first step, MCDHF calculations were performed  
for the radial wave functions of orbitals belonging to the MR set; these were used in further computations.
In the next steps, MCDHF and RCI calculations were performed using 
the RSMBPT method~\cite{Gaigetal:2024CV,Gaigetal:2024C,Gaigetal:2024CC,Gaigetal:2025VV,Gaigetal:2025VVT,Gaigetal:2026CVT}.
The subsequent subsections provide a detailed description of these calculations, 
while the results are presented in Section~\ref{Results}.
It is important to note that that both (MCDHF and RCI) calculations were performed, including only CSFs that have 
a non-zero spin-angular part of matrix elements with at least one CSF in the MR.
The Breit interactions and leading QED effects (the vacuum polarization and self-energy corrections)
were taken into account at the RCI stage.}

\subsection{Computational Scheme Using the RSMBPT~Method}
\label{Computational_schemes}
\unskip
\subsubsection{The CSF Basis Selection Procedure Applying the RSMBPT~Method}
Using the RSMBPT method, the~orbital space is divided into three sets: $F$, $F'$, and~$G$ 
(see Ref.~\cite{Gaigetal:2024CV} for details).
{The $F$ set defines the core, $F'$-valence, and~$G$-virtual subshells (these belong to orbital sets (OS)).
The classification of subshells in the present calculations is presented in Table~\ref{division_space}.}   

\begin{table}[H]
\caption{{The 
 classification of the orbital space applying the RSMBPT method.}}       
\label{division_space} 
\begin{tabularx}\textwidth{cC}
\toprule
\multicolumn{1}{c}{\textbf{Set}} &\multicolumn{1}{c}{\textbf{Subshells}}  \\
\midrule
\noalign{\smallskip}
$F$& 1s, 2s, $\mathrm{2p_-}$, 2p\\ \midrule
$F'$& 3s, $\mathrm{3p_-}$, 3p, $\mathrm{3d_-}$, 3d \\ \midrule
\multirow{5}{*}{$G$ 
} &$OS_1$ = \{4s,$\mathrm{4p_-}$,4p,$\mathrm{4d_-}$,4d,$\mathrm{4f_-}$,4f\} \\
&$OS_2$ = \{5s,$\mathrm{5p_-}$,5p,$\mathrm{5d_-}$,5d,$\mathrm{5f_-}$,5f,$\mathrm{5g_-}$,5g\}\\
&$OS_3$ = \{6s,$\mathrm{6p_-}$,6p,$\mathrm{6d_-}$,6d,$\mathrm{6f_-}$,6f,$\mathrm{6g_-}$,6g,$\mathrm{6h_-}$,6h\} \\
&$OS_4$ = \{7s,$\mathrm{7p_-}$,7p,$\mathrm{7d_-}$,7d,$\mathrm{7f_-}$,7f,$\mathrm{7g_-}$,7g,$\mathrm{7h_-}$,7h\} \\
&$OS_5$ = \{8s,$\mathrm{8p_-}$,8p,$\mathrm{8d_-}$,8d,$\mathrm{8f_-}$,8f,$\mathrm{8g_-}$,8g,$\mathrm{8h_-}$,8h\} \\
\bottomrule
\end{tabularx}
\end{table}

The estimation of the chosen correlation type applying the RSMBPT method is analogous to {the} procedure used in previous studies~\cite{Gaigetal:2024CV,Gaigetal:2024C,Gaigetal:2024CC,Gaigetal:2025VV,Gaigetal:2025VVT,Gaigetal:2026CVT}. 
{The contribution of each $K'$ configuration is computed according to Equation~(\ref{eq:BogEnergy_PT}).
Further, these configurations are arranged in descending order based on the computed impact of the correlations and are selected by the correlations' impact with the specified fraction of the total correlations' contribution.
Only $K'$ configurations with a correlation contribution larger than {\tt 1.0E-11} are included in the computations;
the remaining configurations are neglected.
It is important to note that the assessment of the correlations using the stationary second-order Rayleigh–Schrödinger many-body perturbation theory 
in an irreducible tensorial form is done for the Coulomb interaction.
Since the V (Equation (\ref{eq:not-V-a})) and C (Equation (\ref{eq:not-C-a})) correlations can not be included using the RSMBPT method,
they were incorporated into the calculations in a regular way.}

\subsubsection{MCDHF Computations Applying the RSMBPT~Method}
\begin{figure}[H]
\includegraphics[width=0.99\textwidth]{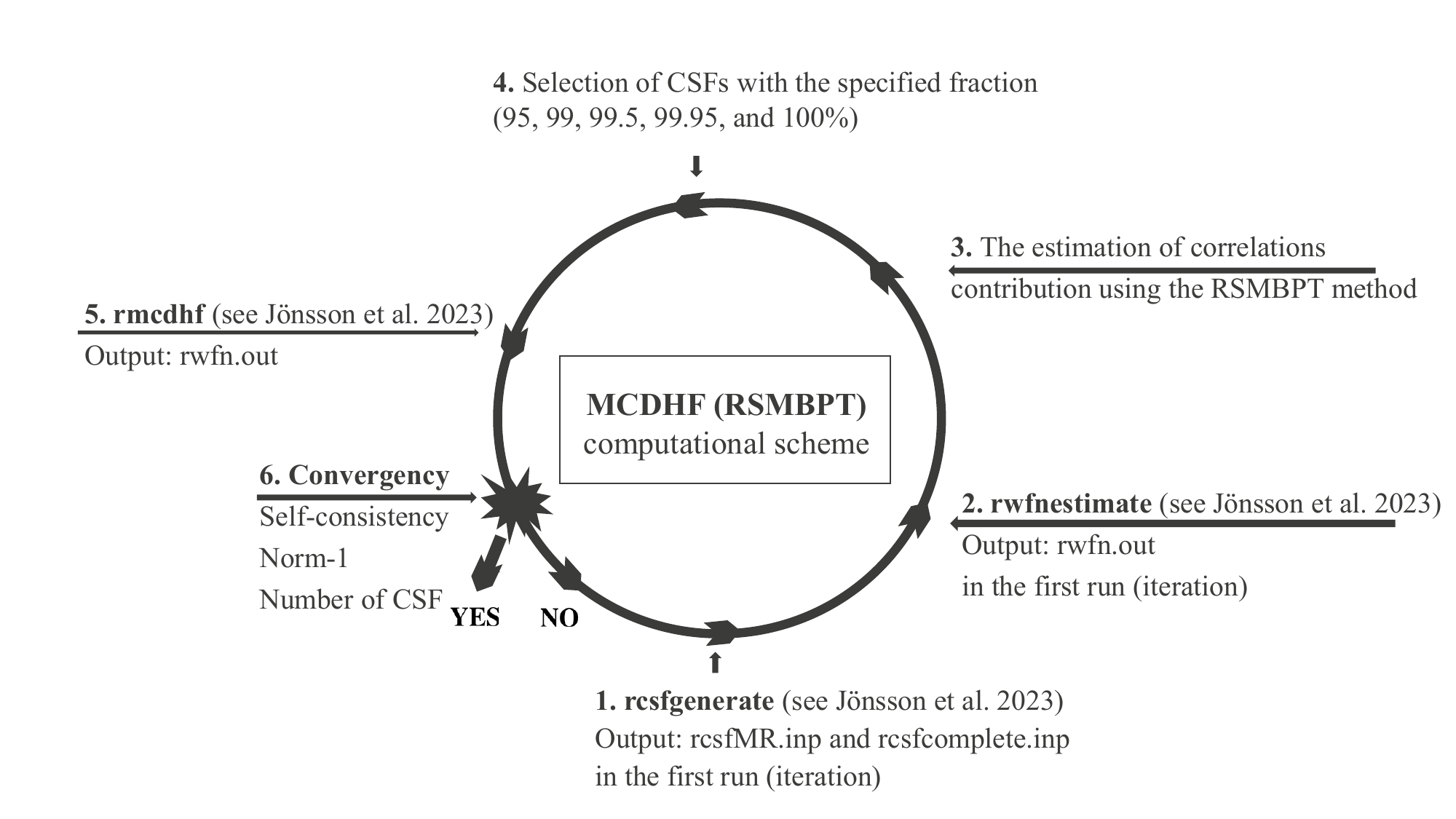}
\caption{\label{PT_MCDHF} Typical 
 sequence of calculation steps for solving the self-consistent field equations using the RSMBPT~method. J\"onsson et~al. 2023 in the figure marks Ref. \cite{Manual_GRASP}.}
\end{figure}

The radial wavefunctions were computed, including different types of correlations. 
There were four types of MCDHF computations performed: (i) including only VV correlations; (ii) including VV and C correlations;
(iii) including VV, C, and~CV correlations; \linebreak  (iv) including VV, C, CV, and~CC correlations.
The procedure for applying the RSMBPT method to construct the CSF basis and~then using it
to solve the self-consistent field equations is described in~\cite{Gaigetal:2025VVT}. 
In this work, the~procedure is repeated, focusing on the main steps.
Figure~\ref{PT_MCDHF} shows the typical sequence of calculation steps 
for solving the self-consistent field equations using the RSMBPT method.
Firstly, the~CSF basis of the $OS_1$ is generated in the regular way (step 1 in Figure~\ref{PT_MCDHF}).
The next step in any MCDHF calculation with a chosen type of correlation is to estimate 
the {initial} radial wavefunctions of the $OS_1$ using the Thomas–Fermi potential (step 2 in Figure~\ref{PT_MCDHF}).
Next, the~contributions of correlations are estimated applying the RSMBPT method, and
the CSF basis is constructed by selecting the most significant correlations 
with a specified fraction of 99.95\% (steps 3 and 4 in Figure~\ref{PT_MCDHF}).
Then, the~self-consistent field equations are solved using a constructed CSF basis (step 5 in Figure~\ref{PT_MCDHF}). 
In step 6, the~convergence of the {\sc self-consistency} and {\sc norm-1} parameters is studied and estimated.
Once convergence has been achieved, the~calculations finish and the radial wavefunctions of the next $OS$ 
 can be computed. 
Otherwise, the~obtained radial wavefunctions are used as the initial ones. 
The selection procedure of the most significant correlations, applying the RSMBPT method with revised radial wavefunctions, 
and MCDHF computation is repeated until convergence is achieved (steps 3--5 are~repeated).

Further, the radial wavefunctions of the new $OS$ are computed.
The same sequence of calculation steps is applied when the radial wavefunctions of the $OS_2$
and further new $OS$--$OS_3$, $OS_4$, and~$OS_5$ are computed.
It is important to highlight a few points described below.
{Computing radial wavefunctions of the $OS_2$,  
the initial radial wavefunctions of the $OS_1$ and $OS_2$ are used in the Thomas–Fermi potential approximation.
Then the contributions of correlations from both ($OS_1$ and $OS_2$) sets are estimated, applying the RSMBPT method,
and the most significant correlations are selected. This is performed in order to ensure that the contributions of 
correlations from both ($OS_1$ and $OS_2$) sets are estimated with the same level of accuracy as radial wavefunctions.
The constructed CSF basis is used for solving the self-consistent field equations of $OS_2$.
In this step, the~$OS_1$ radial wave functions are taken from the final $OS_1$ computations and are fixed.
The obtained radial wavefunctions are taken as initial, and the selection procedure of CSFs is repeated, 
along with a solution of the self-consistent field equations, until~convergence is achieved (as was done in the case of $OS_1$).}
Results from MCDHF computations, including chosen correlations according to the RSMBPT method, are marked as
VV~MCDHF~(RSMBPT) (when only VV correlations are included), 
VV + C~MCDHF~(RSMBPT) (when VV and C correlations are included),
VV~+~C~+~CV~MCDHF~(RSMBPT) (when VV, C, and~CV correlations are included), and~VV~+~C~+~CV~+~CC~MCDHF~(RSMBPT) (when VV, C, CV, and~CC correlations are~included).

\subsubsection{RCI Computations Applying the RSMBPT~Method}
The radial wavefunctions obtained from the above-described MCDHF calculations are used 
for further RCI computations, which are 
performed for all $OS$ ($OS_1$, ..., $OS_5$) in a few ways.
In the first way, RCI calculations are performed with a CSF basis constructed by applying the RSMBPT method
with the same specified fraction (99.95\%) as in MCDHF and including 
only those correlations that were included in computing radial wavefunctions.
These results are marked as 
VV~MCDHF/RCI~(RSMBPT) (when only VV correlations are included in both MCDHF and RCI),
VV~+~C~MCDHF/RCI~(RSMBPT) (when VV and C correlations are included in both MCDHF and RCI),
VV~+~C~+~CV~MCDHF/RCI~(RSMBPT) (when VV, C, and~CV correlations are included in both MCDHF and RCI), and
VV~+~C~+~CV~+~CC~MCDHF/RCI (RSMBPT) (when VV, C, CV, and~CC correlations are included in both MCDHF and RCI).
In the second way, CSFs' basis is constructed by applying the RSMBPT method
with the specified fraction (99.95\%) to include VV, C, CV, and~CC correlations in the RCI calculations.
In this case, radial wavefunctions with only chosen types of correlations included in the MCDHF are used.
Such results are marked as
VV~MCDHF/VV~+~C~+~CV~+~CC~RCI~(RSMBPT) (when only VV correlations are included in MCDHF and VV, C, CV, and~CC correlations in RCI),
VV~+~C~MCDHF/VV~+~C~+~CV~+~CC~RCI~(RSMBPT) (when VV and C correlations are included in MCDHF and VV, C, CV, and~CC correlations in RCI),
and VV~+~C~+~CV~MCDHF/VV~+~C~+~CV~+~CC~RCI (RSMBPT) (when VV, C, and~CV correlations are included in MCDHF and VV, C, CV, and~CC correlations in RCI).

\subsection{Results}
\label{Results}
In this section, the results for Ar~II obtained using the RSMBPT method are presented.
The influence of various types of correlations was studied.
The computed energy levels were compared 
with data from the Atomic Spectra Database (ASD) of the National Institute of Standards and Technology (NIST ASD) \cite{NIST_ASD}.
The E1 transitions were computed between the studied states using various computational schemes described above.
The importance of correlation effects for transition data was also studied. 
The uncertainties of the line strengths obtained in this work were 
estimated based on the quantitative and qualitative evaluation (QQE) method described in~\cite{Kitoviene:2024QQE,Rynkun:2022QQE,Gaigalas:2022QQE}.

\subsubsection{Energy Level~Results}
\label{Ener_results}

\begin{figure}[H]
\includegraphics[width=0.95\hsize]{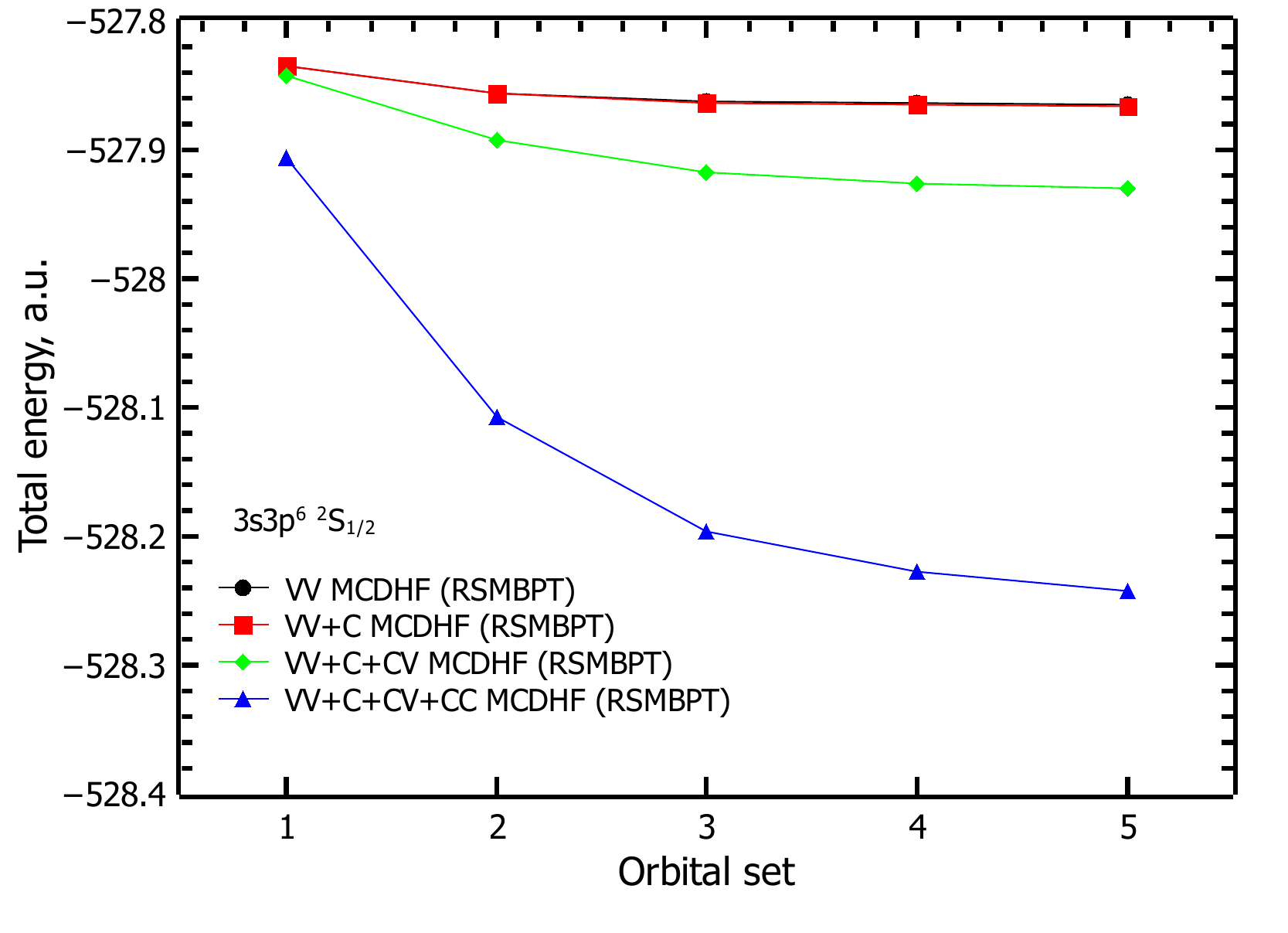}
\caption{\label{convergence_TE_even_1_2} The convergence of the total energy of the $\mathrm{3s3p^6~^2S_{1/2}}$ state 
when various types of correlations are included in the MCDHF~(RSMBPT) computations (non-relativistic interaction) with a specified fraction of 99.95\%. 
The results from the VV~MCDHF~(RSMBPT) and VV~+~C~MCDHF~(RSMBPT) computational schemes overlap.}
\end{figure}

\begin{figure}[H]
\includegraphics[width=\hsize]{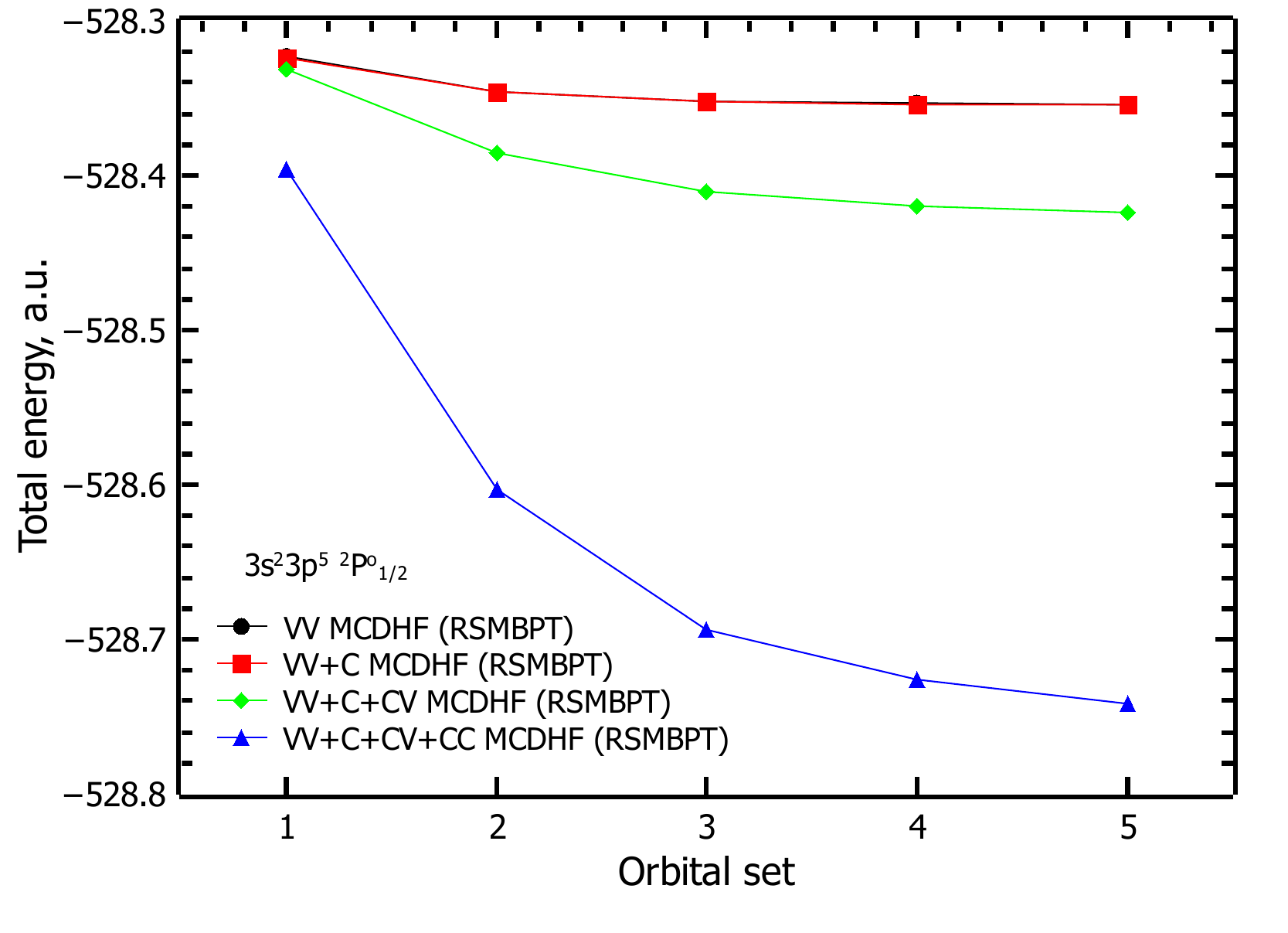}
\caption{\label{convergence_TE_odd_1_2} The convergence of the total energy of the $\mathrm{3s^23p^5~^2P^o_{1/2}}$ state 
when various types of correlations are included in the MCDHF~(RSMBPT) computations (non-relativistic interaction) with a specified fraction of 99.95\%.
The results from the VV~MCDHF~(RSMBPT) and VV~+~C~MCDHF~(RSMBPT) computational schemes overlap.}
\end{figure}

\begin{figure}[H]
\includegraphics[width=\hsize]{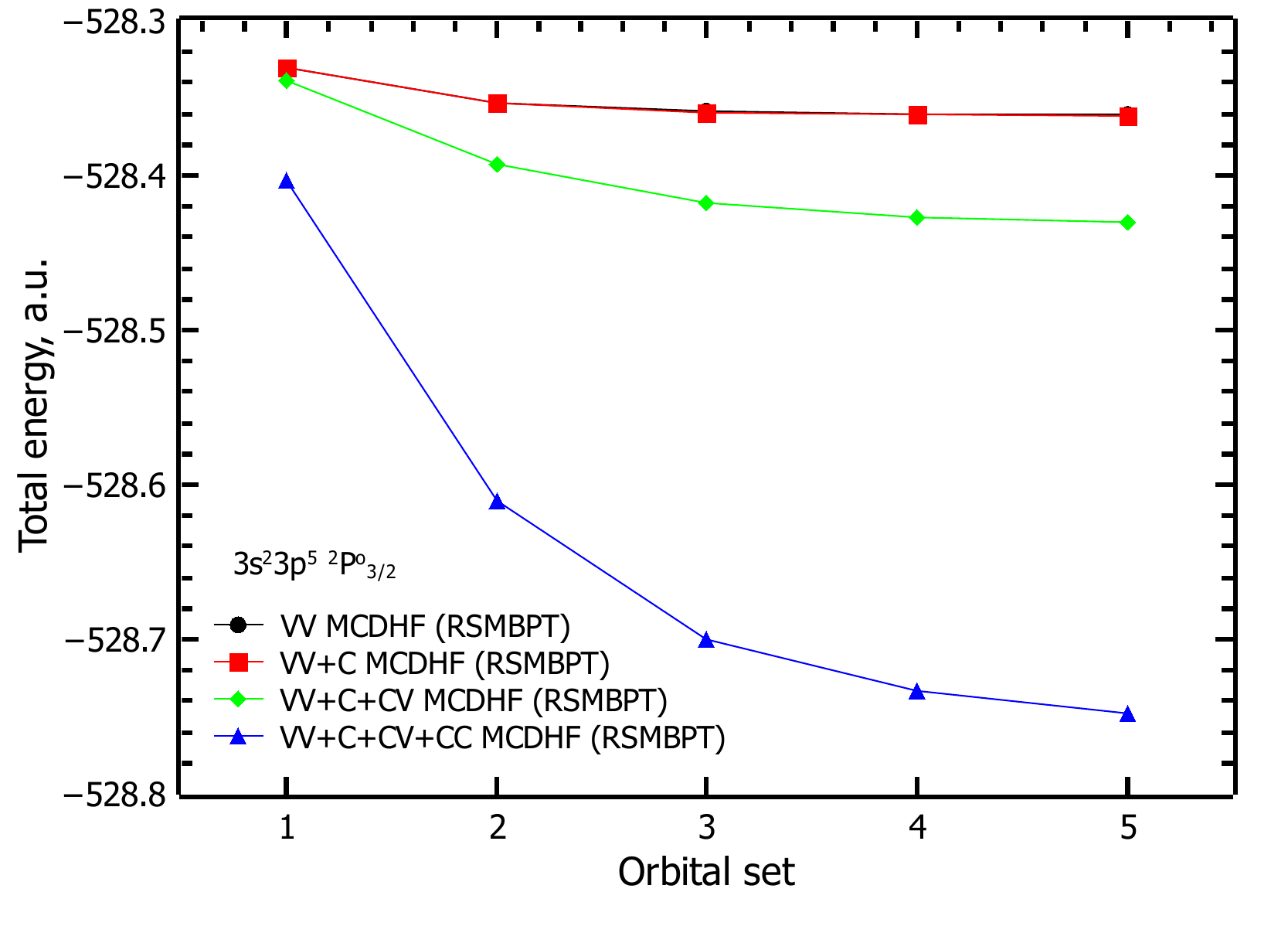}
\caption{\label{convergence_TE_odd_3_2} The convergence of the total energy of the $\mathrm{3s^23p^5~^2P^o_{3/2}}$ state 
when various types of correlations are included in the MCDHF~(RSMBPT) computations (non-relativistic interaction) with a specified fraction of 99.95\%. The~results from the VV~MCDHF~(RSMBPT) and VV~+~C~MCDHF~(RSMBPT) computational schemes overlap.}
\end{figure}

As mentioned above, firstly, using the RSMBPT method, the most significant correlations 
with a specified fraction of 99.95\% were selected for solving self-consistent field equations.
Figures~\ref{convergence_TE_even_1_2}--\ref{convergence_TE_odd_3_2} show the convergence 
of total energies from the MCDHF~(RSMBPT) computations for the three studied states. 
As can be seen in the figures, the~contribution of C correlations does not significantly change 
the total energies compared with the results when only VV correlations are included
(the results from the VV~MCDHF~(RSMBPT) and VV~+~C~MCDHF~(RSMBPT) computational schemes in these figures overlap). 
The energies decrease more when CV correlations are added and change the most
when CC correlations are added. This behavior is independent of the configuration and term for Ar II.

\textls[-15]{Table~\ref{CSF_summary} summarizes the CSF bases used in the RCI computations of $OS_5$ 
when the RSMBPT method with a specified fraction of 99.95\% was applied. 
The number of CSFs is given for each computational scheme 
when different radial wavefunctions are used, and various types of correlations are included in the calculations. 
The CSF bases constructed applying the RSMBPT method, compared with those generated in the regular way,
are smaller.
For example, by~including all correlations with the specified fraction (99.95\%) 
in the MCDHF and RCI calculations (case VV~+~C~+~CV~+~CC~MCDHF/RCI~(RSMBPT)),
the number of CSFs is 377,046 for even $J$ =  1/2, 24,097 for odd $J$ = 1/2, and~55,761 for odd \mbox{$J$ = 3/2}.
These CSF bases are almost twice as small as those constructed in the regular way. 
The number of the CSF, when the VV~+~C~+~CV~+~CC correlations are included at the RCI stage in~the regular way, 
is 603,510 for even $J$ = 1/2, 41,045 for odd $J$ = 1/2, and~95,789 for odd \mbox{$J$ = 3/2}.}

\vspace{-9pt}

\begin{table}[H]
\setlength{\tabcolsep}{3pt}
\caption{Number of CSFs in the RCI computations of $OS_5$ using the RSMBPT method with the specified fraction (99.95\%) and in the regular~way.}            
\label{CSF_summary}

\begin{adjustwidth}{-\extralength}{0cm}
\begin{tabularx}\fulllength{cCCC}
\toprule
\textbf{Computational Scheme} &\textbf{Even} \boldmath{$J$} \textbf{= 1/2} &\textbf{Odd} \boldmath{$J$} \textbf{= 1/2} &\textbf{Odd} \boldmath{$J$} \textbf{= 3/2}  \\
\midrule
\noalign{\smallskip}
VV~MCDHF/RCI~(RSMBPT) & 43,039& 4136& 8367 \\
VV~+~C~MCDHF/RCI~(RSMBPT) & 44,516& 4112& 8380 \\
VV~+~C~+~CV~MCDHF/RCI~(RSMBPT) & 262,109& 18,781& 40,262 \\
VV~MCDHF/VV~+~C~+~CV~+~CC~RCI~(RSMBPT) & 433,323& 27951& 63,217 \\
VV~+~C~MCDHF/VV~+~C~+~CV~+~CC~RCI~(RSMBPT) & 433,352& 27,980& 63,361 \\
VV~+~C~+~CV~MCDHF/VV~+~C~+~CV~+~CC~RCI~(RSMBPT) & 416,468& 26,828& 61,024 \\
VV~+~C~+~CV~+~CC~MCDHF/RCI~(RSMBPT) & 377,046& 24,097& 55,761 \\
VV~+~C~+~CV~+~CC~MCDHF/RCI~regular & 603,510& 41,045& 95,789 \\
\bottomrule
\end{tabularx}
\end{adjustwidth} 
\end{table}

Table~\ref{comp_schem} shows at what stage of the computations the RSMBPT method was applied to select the most significant correlations. 
`RSMBPT' indicates that the CSFs were constructed using the RSMBPT method (with a specified fraction of 99.95\% of chosen correlations), while the mark `regular' indicates that the CSFs were generated using the {\sc Grasp}2018 package in a regular~way.

\begin{table}[H]
\caption{Comparison of computational schemes used in this~work.}            
\label{comp_schem}

\begin{tabularx}\textwidth{CCC}
\toprule
&\multicolumn{1}{c}{\textbf{MCDHF}}  & \multicolumn{1}{c}{ \textbf{RCI}} \\
\midrule
\noalign{\smallskip}
MCDHF~(RSMBPT) & RSMBPT& - \\
RCI~(RSMBPT) & - & RSMBPT \\ 
MCDHF/RCI~(RSMBPT) & RSMBPT & RSMBPT \\
MCDHF/RCI~regular & RSMBPT & regular \\
\bottomrule
\end{tabularx}

\end{table}

Table~\ref{Correlations_contr} shows the contributions of the VV, C, CV, and~CC correlations, as~well as the 
sum of these correlations in the RCI computations
using different radial wavefunctions 
(when different correlations applying the RSMBPT method are included to solve the self-consistent field equations). 
These contributions are presented for three states studied in this work: 
$\mathrm{3s3p^6~^2S_{1/2}}$, $\mathrm{3s^23p^5~^2P^o_{1/2}}$, and~$\mathrm{3s^23p^5~^2P^o_{3/2}}$.
C correlations have the smallest contribution, while VV and CC correlations have the largest.
As seen in the table, the~influence of VV correlations decreases 
when additionally CV and CC correlations are added to MCDHF calculations.
Meanwhile, the~influence of CC correlations increases.
As can be seen, when all correlations are included in MCDHF calculations, 
the total contribution of these correlations to RCI computations decreases compared with VV~+~C~+~CC~cases.

\begin{figure}[H]
\includegraphics[width=0.98\hsize]{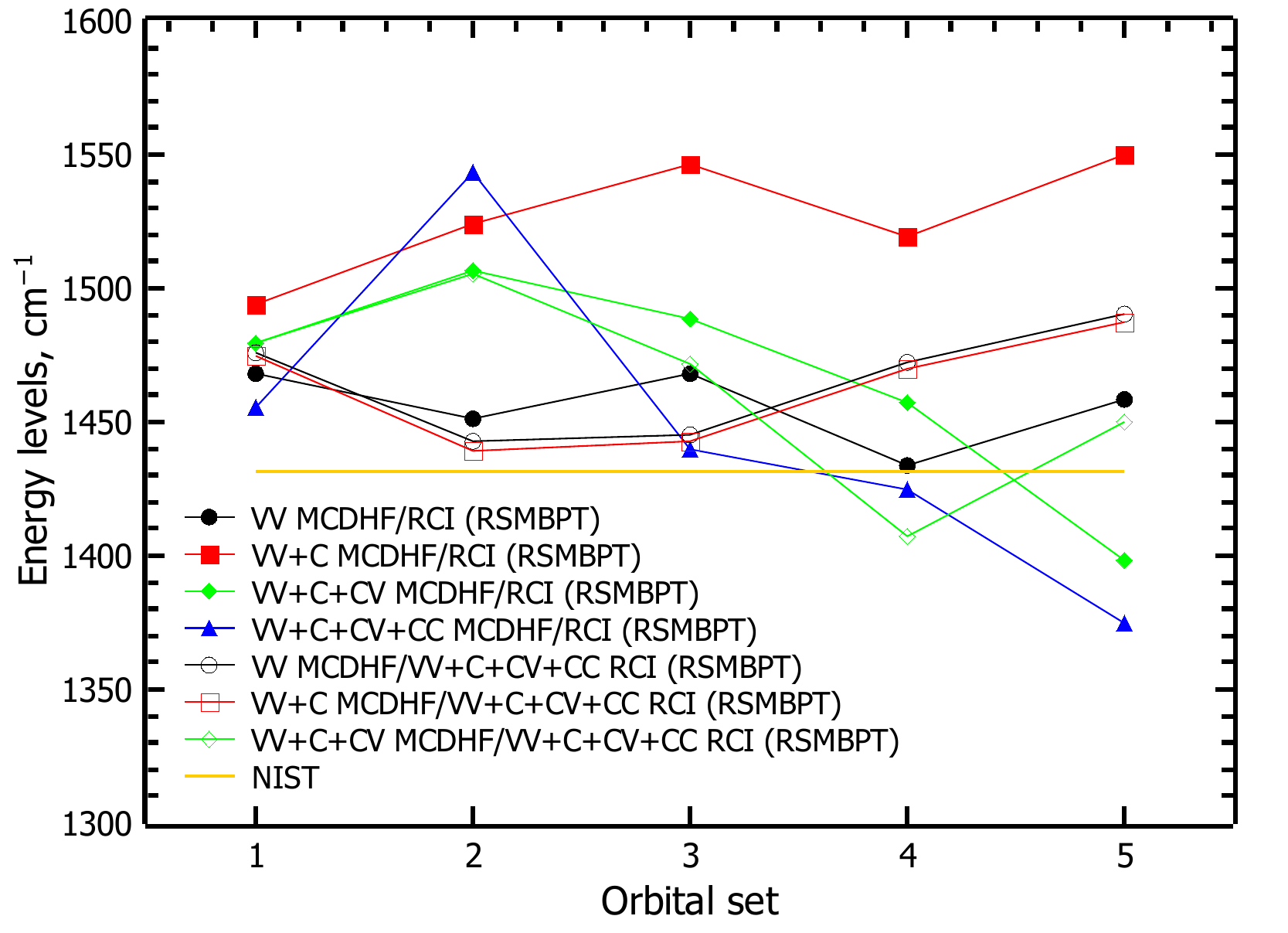}
\caption{\label{convergence_odd} The 
 convergence of the energy level of the $\mathrm{3s^23p^5~^2P^o_{1/2}}$  
when various types of correlations are included in the RCI computations (non-relativistic, Breit, and QED).
}
\end{figure}

\begin{figure}[H]
\includegraphics[width=0.98\hsize]{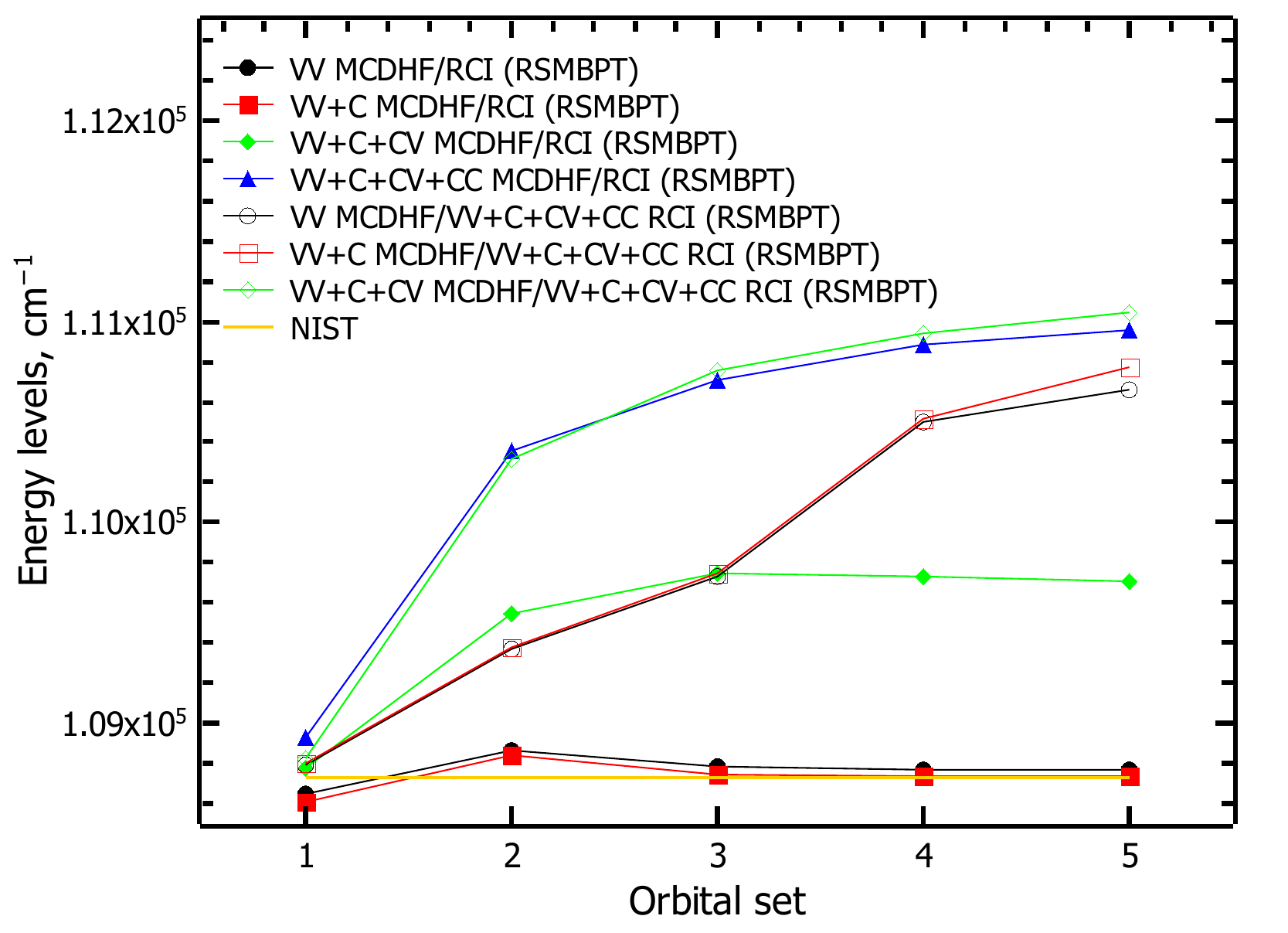}
\caption{\label{convergence_even} The 
 convergence 
 of the energy level of the $\mathrm{3s3p^6~^2S_{1/2}}$  
when various types of correlations are included in the RCI computations (non-relativistic, Breit, and QED). }
\end{figure}

Figures~\ref{convergence_odd} and \ref{convergence_even} show the convergence of energy levels using different
computational schemes from RCI calculations. In~these schemes, the~CSFs bases are constructed by applying the RSMBPT method
with the specified fraction (99.95\%). 
The energy levels obtained using different radial wavefunctions and different types of included correlations are compared with each other
and with data from the NIST ASD~\cite{NIST_ASD}.
Filled symbols mark the results when both (MCDHF and RCI) computations include the same types of correlations.
Meanwhile, empty symbols mark the results when, in the MCDHF, only some correlation types are included, and in the RCI calculations, VV, C, CV, and CC correlations are involved.

The differences between computed energy levels and the NIST results are in the range from 18 to 118 cm$^{-1}$ 
for the $\mathrm{3s^23p^5~^2P^o_{1/2}}$ state and from 14 to 2324 cm$^{-1}$ for the $\mathrm{3s3p^6~^2S_{1/2}}$ state.
The root–mean–square (rms) deviations obtained for the computed energy levels from the NIST data are 38, 84, and~693 cm$^{-1}$
using the following computational schemes: VV~MCDHF/RCI~(RSMBPT), 
VV~+~C~MCDHF/RCI~(RSMBPT), and~
VV~+~C~+~CV~MCDHF/RCI~(RSMBPT), respectively.
By analyzing the energy levels after MCDHF calculations, the~convergence trends are similar to those shown in the figures
when Breit and QED corrections are added to the RCI. 
Depending on the correlations included in the computations,
the energy levels after MCDHF are about \mbox{50--60 cm$^{-1}$} larger 
for the $\mathrm{3s^23p^5~^2P^o_{1/2}}$ state and
about 90--100 cm$^{-1}$ larger for the $\mathrm{3s3p^6~^2S_{1/2}}$ compared with those after RCI.
When all correlations are included in RCI computations, the~rms deviations obtained for the computed energy levels 
from the NIST data are 1372, 1450, 1644, and 1582 cm$^{-1}$, respectively, 
for VV~MCDHF/VV~+~C~+~CV~+~CC~RCI (RSMBPT), 
VV~+~C~MCDHF/VV~+~C~+~CV~+~CC~RCI (RSMBPT), 
VV~+~C~+~CV~MCDHF/VV~+~C~+~CV~+~CC~RCI (RSMBPT), and  
VV~+~C~+~CV~+~CC MCDHF/RCI (RSMBPT).

\begin{table}[H]
\setlength{\tabcolsep}{3pt}
\caption{The contributions of correlations in the RCI computations of $OS_1$--$OS_5$ using the RSMBPT~method.}            
\label{Correlations_contr} 
\footnotesize
\begin{adjustwidth}{-\extralength}{0cm}

\begin{tabularx}\fulllength{Cccccc}
\toprule
\textbf{Computational Scheme} &\multicolumn{1}{c}{\textbf{VV}} &\multicolumn{1}{c}{\textbf{C}} &\multicolumn{1}{c}{\textbf{CV}} &\multicolumn{1}{c}{\textbf{CC}} & \multicolumn{1}{c}{\textbf{Total}} \\
\midrule
\noalign{\smallskip}
&\multicolumn{5}{c}{$\mathrm{3s3p^6~^2S_{1/2}}$} \\
VV~MCDHF/VV~+~C~+~CV~+~CC~RCI~(RSMBPT) & $-3.692 \times 10^{-1}$ 
& $-7.400 \times 10^{-3}$& $-9.316 \times 10^{-2}$& $-2.600 \times 10^{-1}$ & $-7.298 \times 10^{-1}$ \\
VV~+~C~MCDHF/VV~+~C~+~CV~+~CC~RCI~(RSMBPT) & $-3.689 \times 10^{-1}$& $-7.409 \times 10^{-3}$& $-9.512 \times 10^{-2}$& $-2.808 \times 10^{-1}$ & $-7.523 \times 10^{-1}$ \\
VV~+~C~+~CV~MCDHF/VV~+~C~+~CV~+~CC~RCI~(RSMBPT) & $-3.439 \times 10^{-1}$& $-6.985 \times 10^{-3}$& $-1.122 \times 10^{-1}$& $-4.472 \times 10^{-1}$ & $-9.103 \times 10^{-1}$ \\ 
VV~+~C~+~CV~+~CC~MCDHF/RCI~(RSMBPT) & $-2.992 \times 10^{-1}$& $-4.888 \times 10^{-3}$& $-8.144 \times 10^{-2}$& $-4.200 \times 10^{-1}$ & $-8.055 \times 10^{-1}$ \\ \midrule
&\multicolumn{5}{c}{$\mathrm{3s^23p^5~^2P^o_{1/2}}$} \\
VV~MCDHF/VV~+~C~+~CV~+~CC~RCI~(RSMBPT) & $-2.340 \times 10^{-1}$& $-7.713 \times 10^{-3}$& $-8.675 \times 10^{-2}$& $-2.599 \times 10^{-1}$ & $-5.884 \times 10^{-1}$ \\
VV~+~C~MCDHF/VV~+~C~+~CV~+~CC~RCI~(RSMBPT) & $-2.337 \times 10^{-1}$& $-7.722 \times 10^{-3}$& $-8.888 \times 10^{-2}$& $-2.808 \times 10^{-1}$ & $-6.111 \times 10^{-1}$ \\
VV~+~C~+~CV~MCDHF/VV~+~C~+~CV~+~CC~RCI~(RSMBPT) & $-2.094 \times 10^{-1}$& $-7.306 \times 10^{-3}$& $-1.048 \times 10^{-1}$& $-4.466 \times 10^{-1}$ & $-7.681 \times 10^{-1}$ \\ 
VV~+~C~+~CV~+~CC~MCDHF/RCI~(RSMBPT) & $-1.620 \times 10^{-1}$& $-5.076 \times 10^{-3}$& $-7.516 \times 10^{-2}$& $-4.194 \times 10^{-1}$ & $-6.616 \times 10^{-1}$\\\midrule
&\multicolumn{5}{c}{$\mathrm{3s^23p^5~^2P^o_{3/2}}$} \\
VV~MCDHF/VV~+~C~+~CV~+~CC~RCI~(RSMBPT) & $-2.232 \times 10^{-1}$& $-7.173 \times 10^{-3}$& $-8.644 \times 10^{-2}$& $-2.599 \times 10^{-1}$ & $-5.767 \times 10^{-1}$ \\ 
VV~+~C~MCDHF/VV~+~C~+~CV~+~CC~RCI~(RSMBPT) & $-2.229 \times 10^{-1}$& $-7.196 \times 10^{-3}$& $-8.853 \times 10^{-2}$& $-2.808 \times 10^{-1}$ & $-5.994 \times 10^{-1}$ \\
VV~+~C~+~CV~MCDHF/VV~+~C~+~CV~+~CC~RCI~(RSMBPT) & $-1.988 \times 10^{-1}$& $-6.815 \times 10^{-3}$& $-1.043 \times 10^{-1}$& $-4.466 \times 10^{-1}$ & $-7.566 \times 10^{-1}$ \\ 
VV~+~C~+~CV~+~CC~MCDHF/RCI~(RSMBPT) & $-1.516 \times 10^{-1}$& $-4.765 \times 10^{-3}$& $-7.483 \times 10^{-2}$& $-4.193 \times 10^{-1}$ & $-6.506 \times 10^{-1}$ \\
\bottomrule
\end{tabularx}
\end{adjustwidth}

\end{table}

Tables~\ref{comp_TE} and \ref{comp_En_lev} present a comparison of results obtained using the MCDHF/RCI (RSMBPT) 
computational schemes, including different types of correlations with the results from regular computations, 
(when radial wavefunctions were taken from MCDHF (RSMBPT) and a CSF basis was constructed in the regular way at the RCI stage).
For this purpose, 
additional computations at the RCI (RSMBPT) stage using 100\% specified fraction~\cite{Gaigetal:2025VV} and regular computations were performed. 
Table~\ref{comp_TE} shows the comparison of the total energies,
while Table~\ref{comp_En_lev} presents a comparison of the energy levels for the studied states at $OS_5$. 
It is seen that the developed MCDHF/RCI (RSMBPT) method works perfectly when the chosen types of correlations 
are taken into account using the RSMBPT method, and~the results reproduce the regular {\sc Grasp}2018~data.

As was observed in~\cite{Gaigetal:2025VV,Gaigetal:2026CVT}, the~dependence of the contribution of included correlations on the specified fraction 
of the total correlation contribution used in the RCI (RSMBPT) method is linear.
This dependence was also checked in this work. For~this purpose, 
additional computations at the RCI (RSMBPT) stage using other specified fractions: 95, 99, 99.5, 99.95, and~100\% were performed.
Table~\ref{correlations} presents the contribution of the included correlations ($\Delta_{cor}$) using 
the RCI~(RSMBPT) method compared with complete regular RCI computations for the $\mathrm{3s^23p^5~^2P^o_{3/2}}$ state. 
The contribution, $\Delta_{cor}$ (in \%), is computed according to the 
($TE_{RCI~(RSMBPT)}-TE_{MR+})/(TE_{RCI}-TE_{MR+})$ \cite{Gaigetal:2025VV}, where
TE represents the total energy from a certain computational scheme. 
$TE_{RCI~(RSMBPT)}$ is the total energy from the RCI~(RSMBPT) method 
using the specified fraction (in \%) of the total correlations' contribution.
The MR+ results are from computations in which the CSFs' basis consists 
of the MR set, along with the correlations that were not included by applying the RSMBPT method. 
As seen from the table, the~computed $\Delta_{cor}$ is almost identical  
to the specified fraction (in \%) of the total correlation contribution used in the RCI (RSMBPT) method.
There are some larger differences when radial wavefunctions and calculations were included in CV and CC correlations.
The same trends of dependence were obtained for the remaining two states that were studied in this~work.

\begin{table}[H]
\caption{The total energies (in a.u.) for $\mathrm{3s^23p^5~^2P^o_{3/2}}$, $\mathrm{3s^23p^5~^2P^o_{1/2}}$, 
and $\mathrm{3s3p^6~^2S_{1/2}}$ states from VV MCDHF/VV~+~C~+~CV~+~CC~RCI (RSMBPT),
VV~+~C~MCDHF/VV~+~C~+~CV~+~CC~RCI (RSMBPT), VV~+~C~+~CV MCDHF/VV~+~C~+~CV~+~CC~RCI (RSMBPT), and~VV~+~C~+~CV~+~CC~MCDHF/RCI (RSMBPT) 
computational schemes compared with results from regular computations.}            
\label{comp_TE}

\begin{adjustwidth}{-\extralength}{0cm}

\begin{tabularx}\fulllength{cccc}
\toprule
\textbf{Computational Scheme}  & \boldmath{$\mathrm{3s^23p^5~^2P^o_{3/2}}$} & \boldmath{$\mathrm{3s^23p^5~^2P^o_{1/2}}$} & \boldmath{$\mathrm{3s3p^6~^2S_{1/2}}$} \\
\midrule
\noalign{\smallskip}
VV~MCDHF/VV~+~C~+~CV~+~CC~RCI~regular & $-$528.41024839& $-$528.40346015& $-$527.90603684 \\
VV~MCDHF/VV~+~C~+~CV~+~CC~RCI~(RSMBPT)~100\% & $-$528.41024818& $-$528.40345994& $-$527.90603649 \\
VV~MCDHF/VV~+~C~+~CV~+~CC~RCI~(RSMBPT)~99.95\% & $-$528.40990389& $-$528.40311213& $-$527.90569310 \\ \midrule
VV~+~C~MCDHF/VV~+~C~+~CV~+~CC~RCI~regular & $-$528.43013256& $-$528.42337001 & $-$527.92541736 \\
VV~+~C~MCDHF/VV~+~C~+~CV~+~CC~RCI~(RSMBPT)~100\% & $-$528.43013250& $-$528.42336997& $-$527.92541713 \\
VV~+~C~MCDHF/VV~+~C~+~CV~+~CC~RCI~(RSMBPT)~99.95\% & $-$528.42975757& $-$528.42297982& $-$527.92504172 \\ \midrule
VV~+~C~+~CV~MCDHF/VV~+~C~+~CV~+~CC~RCI~regular & $-$528.52044582& $-$528.51373335& $-$528.01392204 \\
VV~+~C~+~CV~MCDHF/VV~+~C~+~CV~+~CC~RCI~(RSMBPT)~100\% & $-$528.52044482& $-$528.51373286& $-$528.01392059 \\
\textls[-15]{VV~+~C~+~CV~MCDHF/VV~+~C~+~CV~+~CC~RCI~(RSMBPT)~99.95\%} & $-$528.51775243& $-$528.51114510& $-$528.01178752 \\ \midrule
VV~+~C~+~CV~+~CC~MCDHF/RCI~regular & $-$528.54094069& $-$528.53432631& $-$528.03480323 \\ 
VV~+~C~+~CV~+~CC~MCDHF/RCI~(RSMBPT)~100\% & $-$528.54091141& $-$528.53430070& $-$528.03479217 \\
VV~+~C~+~CV~+~CC~MCDHF/RCI~(RSMBPT)~99.95\% & $-$528.53577634& $-$528.52951408& $-$528.03021099 \\ 
\bottomrule
\end{tabularx}
\end{adjustwidth}
\end{table}
\vspace{-10pt}

\begin{table}[H]
\caption{Energy levels (in cm$^{-1}$) for $\mathrm{3s^23p^5~^2P^o_{1/2}}$ and $\mathrm{3s3p^6~^2S_{1/2}}$ states 
from VV MCDHF/VV~+~C~+~CV~+~CC~RCI (RSMBPT),
VV~+~C MCDHF/VV~+~C~+~CV~+~CC~RCI (RSMBPT), VV~+~C~+~CV MCDHF/VV~+~C~+~CV~+~CC~RCI (RSMBPT), and~VV~+~C~+~CV~+~CC~MCDHF/RCI (RSMBPT) 
computational schemes compared with results from regular computations.}            
\label{comp_En_lev}
\begin{adjustwidth}{-\extralength}{0cm}
\begin{tabularx}\fulllength{cCC}
\toprule
\multicolumn{1}{c}{\textbf{Computational Scheme}}  & \multicolumn{1}{c}{ \boldmath{$\mathrm{3s^23p^5~^2P^o_{1/2}}$} } & \multicolumn{1}{c}{\boldmath{$\mathrm{3s3p^6~^2S_{1/2}}$}} \\
\midrule
\noalign{\smallskip}
VV~MCDHF/VV~+~C~+~CV~+~CC~RCI~regular & 1489.85& 110,661.65 \\
VV~MCDHF/VV~+~C~+~CV~+~CC~RCI~(RSMBPT)~100\% & 1489.85& 110,661.67 \\
VV~MCDHF/VV~+~C~+~CV~+~CC~RCI~(RSMBPT)~99.95\% & 1490.62& 110,661.48 \\ \midrule
VV~+~C~MCDHF/VV~+~C~+~CV~+~CC~RCI~regular & 1484.21& 110,772.18 \\
VV~+~C~MCDHF/VV~+~C~+~CV~+~CC~RCI~(RSMBPT)~100\% & 1484.20& 110,772.22 \\
VV~+~C~MCDHF/VV~+~C~+~CV~+~CC~RCI~(RSMBPT)~99.95\% & 1487.54& 110,772.33 \\ \midrule
VV~+~C~+~CV~MCDHF/VV~+~C~+~CV~+~CC~RCI~regular & 1473.22& 111,169.12 \\
VV~+~C~+~CV~MCDHF/VV~+~C~+~CV~+~CC~RCI~(RSMBPT)~100\% & 1473.11& 111,169.22 \\
VV+C+CV~MCDHF/VV~+~C~+~CV~+~CC~RCI~(RSMBPT)~99.95\% & 1450.14& 111,046.46 \\ \midrule
VV~+~C~+~CV~+~CC~MCDHF/RCI~regular & 1451.69& 111,084.33 \\ 
VV~+~C~+~CV~+~CC~MCDHF/RCI~(RSMBPT)~100\% & 1450.88& 111,080.33 \\
VV~+~C~+~CV~+~CC~MCDHF/RCI~(RSMBPT)~99.95\% & 1374.41& 110,958.77 \\ 
\bottomrule
\end{tabularx}
\end{adjustwidth}
\end{table}
\vspace{-10pt}

\begin{table}[H]
\caption{The contribution of included correlations ($\Delta_{cor}$ (in \%)) using 
the RCI (RSMBPT) method compared with the regular RCI computations for the $\mathrm{3s^23p^5~^2P^o_{3/2}}$ state.}            
\label{correlations} 
\begin{adjustwidth}{-\extralength}{0cm}
\begin{tabularx}\fulllength{cCCCCC}
\toprule
&\multicolumn{5}{c}{\textbf{Specified Fraction in the RCI (RSMBPT)}} \\ \cmidrule{2-6}
& \multicolumn{1}{c}{\textbf{95\%}} &  \multicolumn{1}{c}{\textbf{99\%}} & \multicolumn{1}{c}{\textbf{99.5\%}} &  \multicolumn{1}{c}{\textbf{99.95\%}} & \multicolumn{1}{c}{\textbf{100\%}} \\
\midrule
\noalign{\smallskip}
VV~MCDHF/VV~+~C~+~CV~+~CC~RCI~(RSMBPT) &95.117& 98.943& 99.410& 99.918& 100.000 \\
VV~+~C~MCDHF/VV~+~C~+~CV~+~CC~RCI~(RSMBPT) & 95.112& 98.940& 99.408& 99.915& 100.000 \\
VV~+~C~+~CV~MCDHF/VV~+~C~+~CV~+~CC~RCI~(RSMBPT) & 93.959& 97.852& 98.517& 99.493& 100.000 \\
VV~+~C~+~CV~+~CC~MCDHF/RCI~(RSMBPT) &  91.028& 95.984& 97.189& 99.065& 99.995 \\
\bottomrule
\end{tabularx}

\end{adjustwidth}
\end{table}

\vspace{-9pt}

The disagreements between the computed $\Delta_{cor}$ and the specified fraction used in the RCI (RSMBPT) method, 
in addition to those mentioned previously in~\cite{Gaigetal:2025VV}, could be related to the following:
(i) the average energy in the denominator of the Feynman diagram expression is used instead of the energy from the theory;
(ii) the contribution of correlation is computed 
for the configuration (occupation of the relativistic subshells without intermediate term), 
but not for the ASF (configuration with intermediate terms);
(iii) the duplication of CSFs that are in both (MR+ and PT) lists.
As mentioned in Section~\ref{Computational_schemes}, the~correlations (Equations (\ref{eq:not-V-a}) and (\ref{eq:not-C-a})),
which cannot be estimated with the RSMBPT method, were added to calculations in a regular way. 
These CSFs were generated using the `rcsfgenarate’ program~\cite{Manual_GRASP}, in~which single substitutions 
were allowed from the valence or core shell. Using the PT theory, 
correlations are estimated by the subshells. Therefore, some CSFs generated in a regular way 
can duplicate these estimates, applying PT theory. This duplication is checked and removed in the atomic calculations. 
However, computing the contribution $\Delta_{cor}$ this duplication remains what could cause the small~discrepancy.

\subsubsection{Transition~Results}
\label{Tr_results}

\begin{figure}[H]
\includegraphics[width=0.92\hsize]{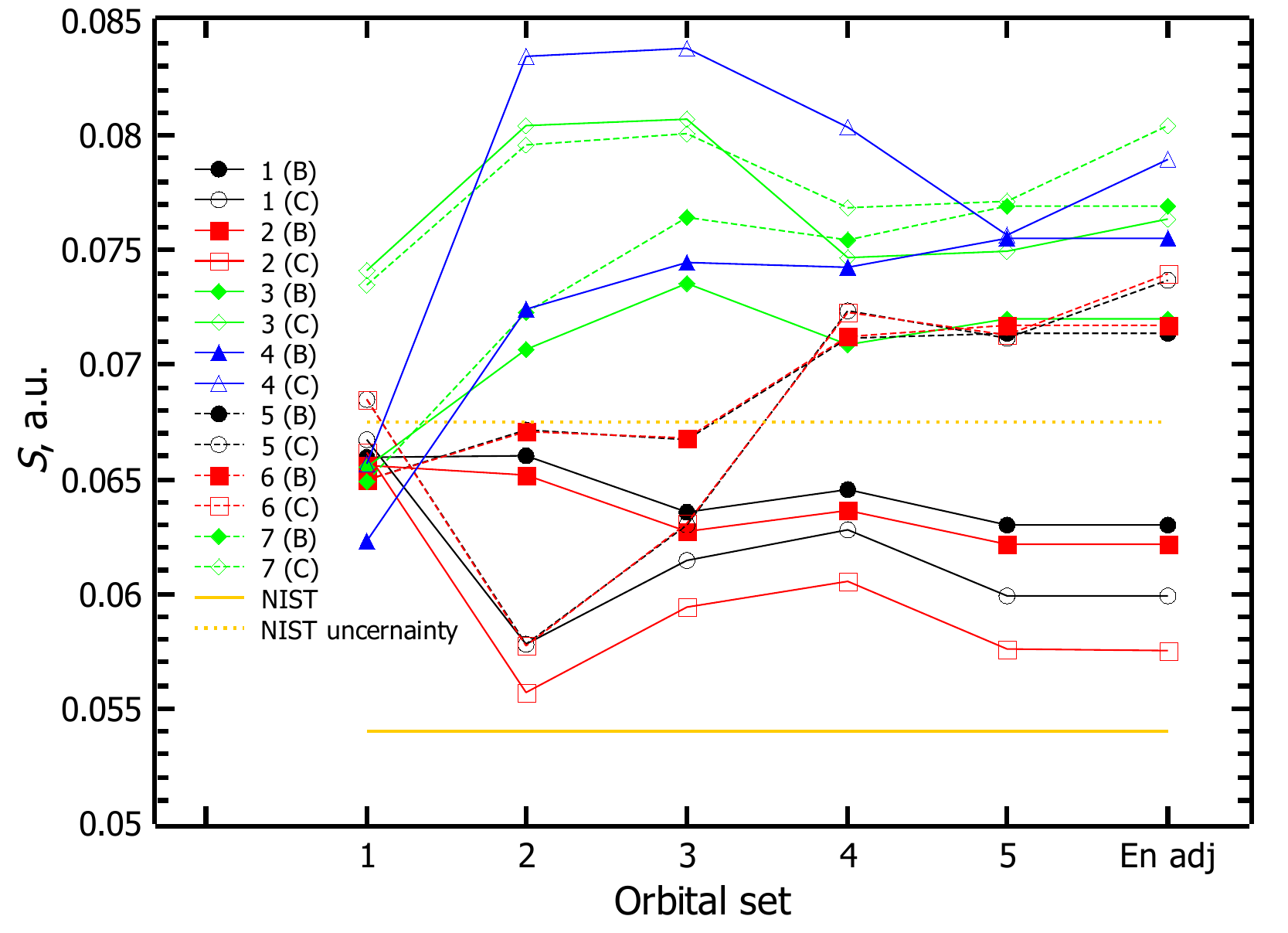}
\caption{\textls[-15]{\label{convergence_trans1_funk_all} The convergence of the line strength of the $\mathrm{3s3p^6~^2S_{1/2}}$--$\mathrm{3s^23p^5~^2P^o_{1/2}}$ transition using different computational schemes. The~computational schemes in the legend are marked: 
\mbox{1---VV} MCDHF/RCI (RSMBPT); 
\mbox{2---VV} +~C~MCDHF/RCI (RSMBPT); 
\mbox{3---VV} +~C~+~CV~MCDHF/RCI (RSMBPT); } 
\mbox{4---VV} +~C~+~CV~+~CC~MCDHF/RCI (RSMBPT);  
\mbox{5---VV} MCDHF/VV~+~C~+~CV~+~CC~RCI (RSMBPT);  
\mbox{6---VV} +~C~MCDHF/VV~+~C~+~CV~+~CC~RCI (RSMBPT); 
\mbox{7---VV} +~C~+~CV~MCDHF/VV~+~C~+~CV~+~CC~RCI (RSMBPT). 
The line strengths are presented in the Babushkin (B) and Coulomb (C) gauges.}
\end{figure}

\begin{figure}[H]
\includegraphics[width=0.92\hsize]{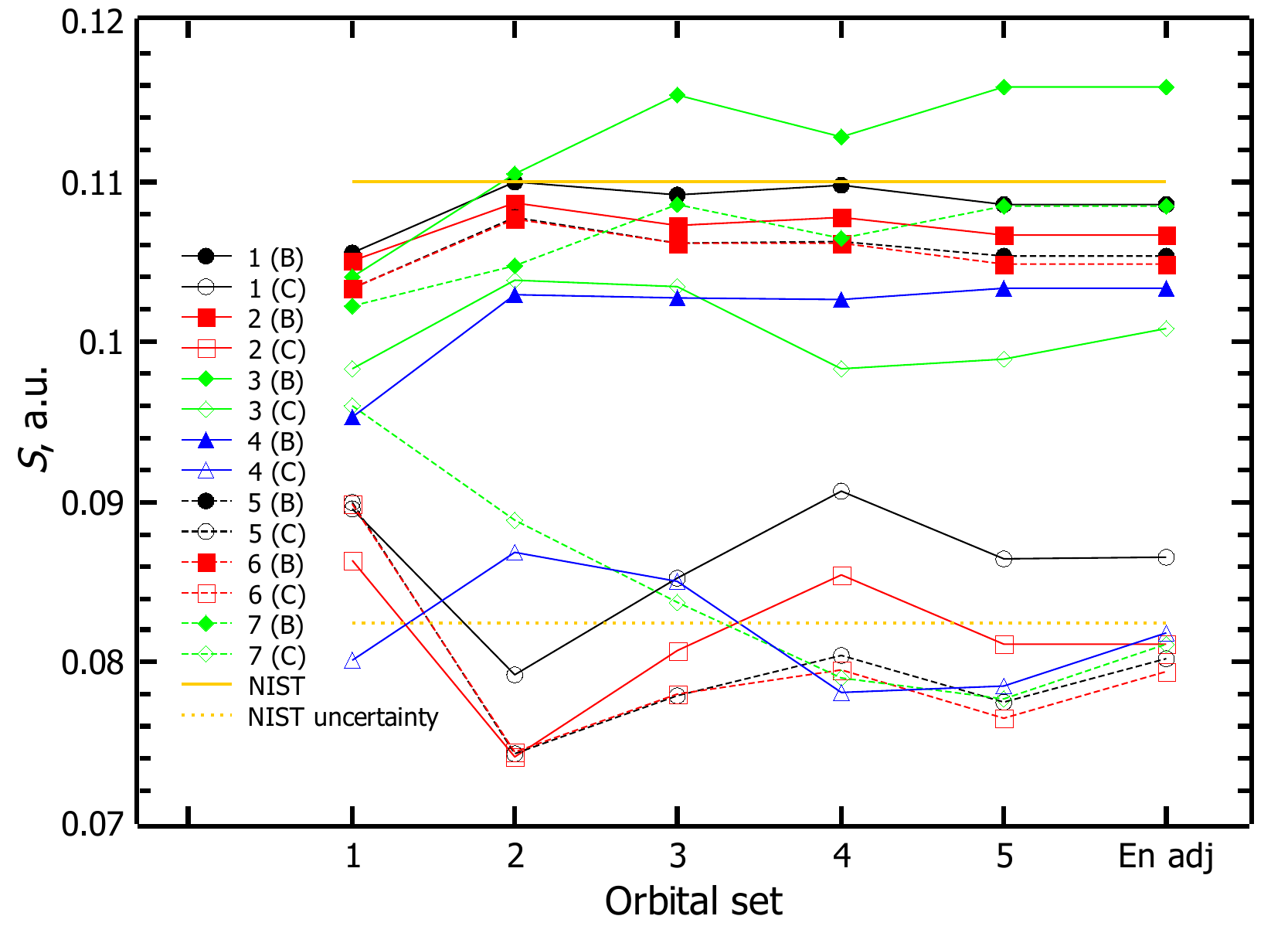}
\caption{\textls[-15]{\label{convergence_trans2_funk_all} The convergence of the line strength of the $\mathrm{3s3p^6~^2S_{1/2}}$--$\mathrm{3s^23p^5~^2P^o_{3/2}}$ transition using different computational schemes. The~computational schemes in the legend are marked: 
\mbox{1---VV} MCDHF/RCI (RSMBPT); 
\mbox{2---VV} +~C MCDHF/RCI (RSMBPT); 
\mbox{3---VV} +~C~+~CV} MCDHF/RCI (RSMBPT);  
\mbox{4---VV} +~C~+~CV~+~CC MCDHF/RCI (RSMBPT);  
\mbox{5---VV} MCDHF/VV~+~C~+~CV~+~CC RCI (RSMBPT);  
\mbox{6---VV} +~C~MCDHF/VV~+~C~+~CV~+~CC RCI (RSMBPT); 
\mbox{7---VV} +~C~+~CV~MCDHF/VV~+~C~+~CV~+~CC RCI (RSMBPT).
The line strengths are presented in the Babushkin (B) and Coulomb (C) gauges.}
\end{figure}

\textls[-15]{Figures~\ref{convergence_trans1_funk_all} and \ref{convergence_trans2_funk_all} 
present the convergence of the line strengths 
of the $\mathrm{3s3p^6~^2S_{1/2}}$--$\mathrm{3s^23p^5~^2P^o_{1/2}}$ 
and $\mathrm{3s3p^6~^2S_{1/2}}$--$\mathrm{3s^23p^5~^2P^o_{3/2}}$ transitions 
using different computational schemes described above.
The results marked as `symbols and straight line' show the results when
the same types of correlations are included in both the MCDHF and RCI computations. 
`Symbols and dashed line' mark the results when only some correlation types are included in the MCDHF and 
VV, C, CV, and~CC correlations are involved in the RCI calculations.
Filled symbols mark the line strengths in the Babushkin gauge, while empty symbols mark results in the Coulomb gauge. 
At the last step in the transition parameters calculations, the experimental transition energy was applied; these results are marked  as `En.~adj.'.}

It can be seen that the values of the line strengths can vary significantly
when the results of two computations with the same radial wavefunctions 
but different types of correlations included at the RCI stage are compared.
This is especially evident when only VV or VV~+~C correlations are included in the computations.
The largest changes occur in the line strengths in the Coulomb gauge.
Line strengths in the Babushkin gauge are less sensitive to the included correlations.

By analyzing the agreement between the two gauges, the~accuracy of the line strength for 
the $\mathrm{3s3p^6~^2S^{}_{1/2}}$--$\mathrm{3s^23p^5~^2P^o_{1/2}}$ transition
at the final $OS_5$ varies between the B and AA accuracy classes, depending on the strategy used in the calculations.
The AA accuracy class is achieved when all (VV, C, CV, and~CC) correlations are included in the RCI calculations.
By applying the experimental transition energy to the transition data calculations, 
the accuracy of the line strength for this transition is evaluated as B+.
The accuracy of the line strength for the $\mathrm{3s3p^6~^2S_{1/2}}$--$\mathrm{3s^23p^5~^2P^o_{3/2}}$ transition
at the final $OS_5$ is estimated as D+. 
The accuracy class of this transition remains D+ when the experimental transition energy 
is used to calculate the transition parameters.
Unlike the line strength in the Coulomb gauge, the~line strength in the Babushkin gauge does not change when the experimental transition energy is applied to the transition data~calculations.

In Figures~\ref{convergence_trans1_funk_all} and \ref{convergence_trans2_funk_all}, 
the line strengths from the NIST database are also shown; 
they are marked by a yellow, straight line.
These line strengths for two transitions are evaluated in the NIST with a C ($\leq$25\%) accuracy class; the~uncertainties
for these lines are marked as yellow, dotted lines.
By comparing these line strengths with the computed ones,
it is seen that they are closer to the line strengths when only VV or VV~+~C correlations are included in the MCDHF and RCI computations. 
The computed line strengths in both gauges for the $\mathrm{3s3p^6~^2S^{}_{1/2}}$--$\mathrm{3s^23p^5~^2P^o_{1/2}}$ transition 
obtained using the VV~MCDHF/RCI~(RSMBPT) and VV~+~C~MCDHF/RCI~(RSMBPT) schemes 
fall within the uncertainty limits of the line strengths given by the NIST. 
For the $\mathrm{3s3p^6~^2S^{}_{1/2}}$--$\mathrm{3s^23p^5~^2P^o_{3/2}}$ transition, the~line strengths in the Babushkin gauge 
from all computational schemes, as~well as the line strengths in the Coulomb gauge from the VV~MCDHF/RCI~(RSMBPT)
and VV~+~C~+~CV~MCDHF/RCI~(RSMBPT) computational schemes, fall within the uncertainty limits of the line strengths given by the~NIST.

\vspace{-9pt}

\unskip

\subsubsection{Lifetime~Results}
\label{Life_results}

The lifetime of the $\mathrm{3s3p^6~^2S_{1/2}}$ state was computed in the Babushkin and Coulomb gauges using the abovementioned computational schemes. 
These results are presented in Table~\ref{Lifetimes} and compared with the experimental results (marked as 'Ex.') 
and other theoretical results (marked as 'Th.'). 
The column labeled as 'En. adj.' in the table presents the lifetimes when the experimental transition energy was applied to the transition parameters calculations. The~lifetime in the NIST row of the table was computed using the transition rates given in the NIST~database.

\begin{table}[H]
\caption{Comparison 
 of the computed lifetimes (in ns) for the $\mathrm{3s3p^6~^2S_{1/2}}$ state with the experimental and theoretical results.
The lifetimes are given in the Babushkin (B) and Coulomb (C) gauges. 'En. adj.' means that in the transition parameters calculations, the experimental transition energy was~applied.}            
\label{Lifetimes} 
\begin{adjustwidth}{-\extralength}{0cm}
\begin{tabularx}\fulllength{ccCCC}
\toprule
\textbf{Computational Scheme} &\multicolumn{2}{c}{\boldmath{$OS_5$}} &\multicolumn{2}{c}{\textbf{En. adj.}} \\
\cmidrule{2-5}
&\multicolumn{1}{c}{\textbf{B}} &\multicolumn{1}{c}{\textbf{C}} & \multicolumn{1}{c}{\textbf{B}} &\multicolumn{1}{c}{\textbf{C}} \\
\midrule
\noalign{\smallskip}
VV~MCDHF/RCI~(RSMBPT) & 4.536&	5.327 & 4.541 & 5.328 \\
VV~+~C~MCDHF/RCI~(RSMBPT) & 4.623& 5.632 & 4.619& 5.630 \\
VV~+~C~+~CV~MCDHF/RCI~(RSMBPT) & 4.038& 4.370 & 4.150& 4.410 \\
VV~MCDHF/VV~+~C~+~CV~+~CC~RCI~(RSMBPT) & 4.190&	4.996& 4.417& 5.084 \\
VV~+~C~MCDHF/VV~+~C~+~CV~+~CC~RCI~(RSMBPT) & 4.181& 5.010& 4.421& 5.104 \\
VV~+~C~+~CV~MCDHF/VV~+~C~+~CV~+~CC~RCI~(RSMBPT) & 3.951& 4.746& 4.211&	4.848 \\ 
VV~+~C~+~CV~+~CC~MCDHF/RCI~(RSMBPT) & 4.104&	4.771& 4.367&	4.871 \\ \midrule
Ex.~\cite{Suzuki_2016} & 4.62 $\pm$ 0.045 &&&\\
Ex.~\cite{SLauer_1999} & 4.684 $\pm$ 0.019 &&&\\
NIST~\cite{NIST_ASD} & 4.83 &&&\\
Th.~\cite{Suzuki_2016} in the B gauge & 4.66 &&&\\
Th.~\cite{Suzuki_2016} in the C gauge & 5.37 &&&\\
Th.~\cite{FROESEFISCHER2006607} &  4.291 &&&\\
\bottomrule
\end{tabularx}
\end{adjustwidth}
\end{table}

The lifetimes in the Babushkin gauge, adjusted by the experimental transition energy, 
agree well with the experimental results~\cite{Suzuki_2016,SLauer_1999} and other theoretical results~\cite{Suzuki_2016,FROESEFISCHER2006607}.
In this case, the~lifetimes from the VV~MCDHF/RCI~(RSMBPT) 
and VV~+~C~MCDHF/RCI (RSMBPT) strategies agree very well with the experimental results.
However, the~agreement between the two gauges is not as good.
For the remaining strategies, the~agreement between the Babushkin and Coulomb gauges improves. 
By comparing the lifetimes adjusted by the experimental transition energy with the experimental ones~\cite{Suzuki_2016,SLauer_1999} and at the same time evaluating the agreement between the two gauges, the VV~+~C~+~CV~+~CC~MCDHF/RCI (RSMBPT) strategy demonstrates good~agreement.

\section{Conclusions}
\label{Sec:Conclusions}

By generalizing the method developed by our group, the~extension of the
approach presented in this paper allows us to achieve the next level of application for more general cases and
essentially completes the methodology.
This approach can also be successfully applied to
the regular Rayleigh–Schrödinger many-body perturbation theory;
however, it should be noted that the radial part of regular perturbation theory is beyond the scope of the present~paper.

The developed method, based on the Rayleigh–Schr\"odinger perturbation theory in an irreducible tensorial form, 
allows estimation of the CV, C, CC, and~VV correlations.
The contribution of these correlations can be investigated for any atom or ion with an arbitrary number of valence and core electrons.
This paper presents all the Feynman diagrams that describe these correlations together with the analytical expressions 
of these diagrams derived in an irreducible tensorial form. 
It also provides rules for obtaining algebraic expressions of any Feynman diagram in this~form.

This newly developed method can be used in three ways: (i) RCI + RSMBPT; (ii) RCI (RSMBPT); and (iii) MCDHF (RSMBPT).
The RSMBPT method can be applied to MCDHF and RCI computations, choosing the same types of correlations 
as well as different types of correlations in both computations.
Selecting the most important correlations with the RSMBPT method 
allows reducing the space of configuration state functions and including correlations in a systematic way. 
This is especially useful and beneficial for calculations involving complex atoms and ions.
Such benefits of the RSMBPT method over the regular method extend the capability of the {\sc Grasp}2018 software package.
The application of the RSMBPT method for atomic calculations in different ways is presented in this~work.


\vspace{6pt}
\authorcontributions{ 
Theory and programming, G.G.; testing program, P.R., and L.K.; calculation, P.R. and L.K.; discussion and theory validation, G.G., P.R., and L.K.; writing---original draft, G.G., P.R., and L.K.; writing---review and editing G.G., P.R., and L.K. All authors have read and agreed to the published version of the manuscript.}

\funding{This research received no external~funding.}

\dataavailability{The data that support the findings of this study are available from the corresponding author upon~reasonable request.}

\conflictsofinterest{The authors declare no conflicts of~interest.}



\abbreviations{Abbreviations}{
The following abbreviations are used in this manuscript:
\\

\noindent 
\begin{tabular}{@{}ll}
{\sc Grasp}     & General Relativistic Atomic Structure package\\
{\sc Grasp}2018& Latest {\sc Grasp} program package version\\
\multirow{2}{*}{{\sc Grasp}2018{\sc\_PT} 
} & Extension of the {\sc Grasp}2018 software package in a combination \\
                & of MCDHF, RCI, and~RSMBPT in the irreducible tensorial form \\
ASF    & Atomic state function  \\
CSF    & Configuration state function \\
MCDHF  & Multiconfiguration Dirac–Hartree–Fock\\
RCI    & Relativistic configuration interaction \\
RME    & Reduced matrix element \\
RSMBPT & Rayleigh–Schr\"odinger many-body perturbation theory \\
CV     & Core–valence \\
CC     & Core–core \\
C      & Core \\
VV     & Valence–valence \\
MCDHF~(RSMBPT) & RSMBPT is applied in the MCDHF computations \\
RCI~(RSMBPT) & RSMBPT is applied in the RCI computations \\
MCDHF/RCI~(RSMBPT) & RSMBPT is applied in both MCDHF and RCI computations \\
\multirow{6}{*}{\textls[-15]{<type>MCDHF/RCI (RSMBPT)}} & RSMBPT is applied in both MCDHF and RCI computations, \\
        & including the chosen <type> of correlations, which\\
				& is the same for both computations. For~example,\\
				& VV~+~C MCDHF/RCI (RSMBPT) means that VV and C \\	
				& correlations were included in both MCDHF and RCI\\
				& computations applying RSMBPT method\\
\multirow{3}{*}{\makecell[l]{<type1>MCDHF/\\<type2>RCI~(RSMBPT)}} & RSMBPT is applied in both MCDHF and RCI computations, \\
       & including <type1> of correlations in the MCDHF computations\\
		& and <type2> of correlations in the RCI computations\\
MR     & Multireference \\
QED    & Quantum electrodynamic corrections \\
OS     & Orbital set \\
QQE    & Quantitative and qualitative evaluation method
\end{tabular}
}

\appendixtitles{yes}
\appendixstart
\appendix
\section[\appendixname~\thesection]{The Reduced Matrix Element of Triple Tensors}
\label{PT_Appendix1}

This appendix presents the specifics of calculating the reduced matrix elements (RME) listed below in the stationary second-order Rayleigh–Schrödinger many-body perturbation theory in an irreducible tensorial form~\cite{Gaigetal:2024CV,Gaigetal:2024C,Gaigetal:2024CC,Gaigetal:2025VV,Gaigetal:2025VVT,Gaigetal:2026CVT}.
{\small
\begin{equation}
\label{eq:PT_Appendix1Product1}
\text{RME$_A$} \equiv \left< \; \Psi \; \left\| \; \biggl[ \bigl[ \tilde{a}^{(j)} \times a^{(j)} \bigr]^{(k)} \times  \Bigl[ \bigl[ a^{(j')} \times \tilde{a}^{(j')} \bigr]^{(x)} \times \bigl[ a^{(j')} \times \tilde{a}^{(j')} \bigr]^{(k')} \Bigr]^{(k)} \biggr]^{(0)} \; \right\| \; \Psi \; \right> ,
\end{equation}
\begin{equation}
\label{eq:PT_Appendix1Product2}
\text{RME$_B$} \equiv \left< \; \Psi \; \left\| \; \biggl[ \Bigl[ \bigl[ \tilde{a}^{(j)} \times a^{(j)} \bigr]^{(k)} \times \bigl[ a^{(j'')} \times \tilde{a}^{(j'')} \bigr]^{(x)} \Bigr]^{(k')} \times \bigl[ a^{(j')} \times \tilde{a}^{(j')} \bigr]^{(k')} \biggr]^{(0)} \; \right\| \; \Psi \; \right> ,
\end{equation}
\begin{equation}
\label{eq:PT_Appendix1Product3}
\text{RME$_C$} \equiv \left< \; \Psi \; \left\| \; \biggl[ \bigl[ a^{(j)} \times \tilde{a}^{(j)} \bigr]^{(k)} \times  \Bigl[ \bigl[ a^{(j')} \times \tilde{a}^{(j')} \bigr]^{(x)} \times \bigl[ \tilde{a}^{(j')} \times a^{(j')} \bigr]^{(k')} \Bigr]^{(k)} \biggr]^{(0)} \; \right\| \; \Psi \; \right> ,
\end{equation}
\begin{equation}
\label{eq:PT_Appendix1Product4}
\text{RME$_D$} \equiv \left< \; \Psi \; \left\| \; \biggl[ \Bigl[ \bigl[ a^{(j)} \times \tilde{a}^{(j)} \bigr]^{(k)} \times \bigl[ \tilde{a}^{(j'')} \times a^{(j'')} \bigr]^{(x)} \Bigr]^{(k')} \times \bigl[ \tilde{a}^{(j')} \times a^{(j')} \bigr]^{(k')} \biggr]^{(0)} \; \right\| \; \Psi \; \right> ,
\end{equation}}
where
\begin{equation*}
\nonumber
\left\| \; \Psi \; \right> \equiv \left\| \; \text{3s}^2 \, \mathrm{3p_-} \, \text{3p}^{3} \, (J_{23}=2) \; \text{3d} \; J=\frac{1}{2} \; \right> .
\end{equation*}

\text{RME$_A$} and \text{RME$_B$} reduced matrix elements come from the triple VV$_3$ Feynman diagram (see Figure~\ref{Feynman_Diagrams}), and \text{RME$_C$} and \text{RME$_D$} reduced matrix elements come from the triple CV$_7$ Feynman diagram (see Figure~\ref{Feynman_Diagrams}). Table~\ref{Apendex1} presents the values of these reduced matrix elements (see Value column) for specific combinations of second quantization operators acting on subshells (see columns $a^{(j)}$, $a^{(j')}$, and~$a^{(j'')}$) and for certain values of the tensorial product ranks $k$, $k'$, and~$x$ (see columns $k$, $k'$, and~$x$). The~last column of the table lists the second quantization operators that form the energy multiplier $D$ (\ref{eq:rule5}) of the Feynman diagrams under consideration and that belong to the $F'$ space. The~examples in the table illustrate certain peculiarities of calculating reduced matrix elements (\ref{eq:PT_Appendix1Product1}) and (\ref{eq:PT_Appendix1Product2}) in the stationary second-order Rayleigh–Schrödinger many-body perturbation theory in an irreducible tensorial form, which we will now~discuss.

\begin{table}[H]
\caption{Values of RME$_A$, RME$_B$, RME$_C$, and~RME$_D$ reduced matrix elements (a test case).
}
\begin{tabularx}\textwidth{ccccccccC}
\toprule 
\textbf{RME}&\boldmath{$a^{(j)}$}&\boldmath{$a^{(j')}$}&\boldmath{$a^{(j'')}$}&\boldmath{$k$}&\boldmath{$k^{\prime}$}&\boldmath{$x$}&\textbf{Value} & \multicolumn{1}{c}{\textbf{\makecell[c]{The Second Quantization\\Operators for \boldmath{$D$} (\ref{eq:rule5})}}} \\
\midrule
RME$_A$ & $\mathrm{3p_-}$ & 3p$^{3}$ & --   & $1$ & $1$ & $1$ &  0.0866025 & $a^{(j)}$, $\tilde{a}^{(j')}$, $\tilde{a}^{(j')}$ \\ 
RME$_A$ & $\mathrm{3p_-}$ & 3p$^{3}$ & --   & $1$ & $2$ & $2$ & $-$0.1936492 & $a^{(j)}$, $\tilde{a}^{(j')}$, $\tilde{a}^{(j')}$ \\ 
RME$_A$ & $\mathrm{3p_-}$ & 3p$^{3}$ & --   & $1$ & $2$ & $3$ &  0.1620185 & $a^{(j)}$, $\tilde{a}^{(j')}$, $\tilde{a}^{(j')}$ \\ \addlinespace[1.0mm]
\text{RME$_B$} & $\mathrm{3p_-}$ & 3d & 3p$^{3}$ & $1$ & $1$ & $0$ &  0.5123475 & $a^{(j)}$, $\tilde{a}^{(j')}$, $\tilde{a}^{(j'')}$ \\ 
\text{RME$_B$} & 3d            & $\mathrm{3p_-}$ & 3s$^{2}$ & $1$ & $1$ & $0$ & 0.4830459 & $a^{(j)}$, $\tilde{a}^{(j')}$, $\tilde{a}^{(j'')}$ \\ 
\text{RME$_B$} & $\mathrm{3p_-}$ & 3d & 3p$^{3}$ & $1$ & $4$ & $3$ &  0.3162278 & $a^{(j)}$, $\tilde{a}^{(j')}$, $\tilde{a}^{(j'')}$  \\ \addlinespace[1.0mm]
\text{RME$_C$} & $\mathrm{3p_-}$ & 3d &  --      & $1$ & $1$ & $1$ & $-$0.0577350 & $a^{(j')}$, $a^{(j')}$, $\tilde{a}^{(j)}$ \\ 
\text{RME$_C$} & $\mathrm{3p_-}$ & 3d &  --      & $1$ & $2$ & $2$ & $-$0.1290994 & $a^{(j')}$, $a^{(j')}$, $\tilde{a}^{(j)}$ \\ 
\text{RME$_C$} & $\mathrm{3p_-}$ & 3d &  --      & $1$ & $2$ & $3$ & $-$0.2121320 & $a^{(j')}$, $a^{(j')}$, $\tilde{a}^{(j)}$ \\ \addlinespace[1.0mm]
\text{RME$_D$} & $\mathrm{3p_-}$ & 3d & 3p$^{3}$ & $1$ & $1$ & $0$ & $-$0.1707825 & $a^{(j')}$, $a^{(j'')}$, $\tilde{a}^{(j)}$ \\ 
\text{RME$_D$} & 3d            & $\mathrm{3p_-}$ & 3s$^{2}$ & $1$ & $1$ & $0$ & 0.0000000 & $a^{(j')}$, $a^{(j'')}$, $\tilde{a}^{(j)}$ \\ 
\text{RME$_D$} & $\mathrm{3p_-}$ & 3d & 3p$^{3}$ & $1$ & $4$ & $3$ & 0.3162278  & $a^{(j')}$, $a^{(j'')}$, $\tilde{a}^{(j)}$ \\ 
\bottomrule
\end {tabularx}
\label{Apendex1}
\end{table}

\begin{itemize}
\item Table~\ref{Apendex1} presents different sets of second quantization operators acting on the subshells for RME$_A$ and RME$_C$. In~the RME$_A$ case, the~operators $a^{(j)}$ and $a^{(j')}$ act on the $\mathrm{3p_-}$ and 3p$^{3}$ subshells, respectively, while in the RME$_C$ case, they act on the $\mathrm{3p_-}$ and 3d subshells. These cases have been specifically chosen because, as~already mentioned, RME$_A$ and RME$_C$ originate from different Feynman diagrams. Therefore, if~we consider the RME$_C$ case with the operators $a^{(j)}$ and $a^{(j')}$ acting on the $\mathrm{3p_-}$ and 3p$^{3}$ subshells, the~reduced matrix element will be non-zero, even though the Feynman diagram CV$_7$ itself is zero. This zero value results from energy multiplier $D$, which is zero, $D = 0$; i.e.,~when two creation operators $a^{(j')}$ (see the last column of Table~\ref{Apendex1}) act on the $\mathrm{3p_-}$ subshell, three electrons should appear there, but~according to atomic theory, the~maximum number of electrons in this subshell can only be two. Therefore, although~the reduced matrix elements themselves are non-zero, they do not need to be calculated using the current methodology. For~the same reason, the~RME$_A$, which involves the second quantization operators $a^{(j)}$ and $a^{(j')}$ (see the last column of Table~\ref{Apendex1}) acting on the $\mathrm{3p_-}$ and 3d subshells, also does not need to be calculated. In~this case, $D$ from VV$_3$ will be zero as~well, since the two annihilation operators $\tilde{a}^{(j')}$ acting on the 3d subshell with a single electron will yield zero.
\item In some cases, it is possible to determine immediately, without~calculation, whether a reduced matrix element is zero or not. For~example, in~the case of RME$_B$, when the second quantization operators $a^{(j)}$, $a^{(j')}$, and~$a^{(j'')}$ (see the last column of Table~\ref{Apendex1}) act on the 3d, $\mathrm{3p_-}$, and~$3s^{2}$ orbitals and have ranks $k=1$, $k'=1$, and~$x=0$, respectively, the~RME$_B$ is non-zero (RME$_B = 0.4830459$), but~with the same values, the~RME$_D$ is zero. This can be easily determined because, when calculating the reduced matrix element, the~creation operator $a^{(j'')}$ (see the last column of Table~\ref{Apendex1}) acts on a closed shell (in our case 3s$^{2}$); thus, according to the theory of second quantization, it is equal to zero (see the second line of the RME$_D$ in Table~\ref{Apendex1}).
\item Special attention should be drawn to the values of RME$_B$ and RME$_D$ presented in the table, which are equal (RME$_B = 0.3162278$ and RME$_D = 0.3162278$) even though the tensorial structures of their respective operators differ. Such cases, where the values of the reduced matrix elements (\ref{eq:PT_Appendix1Product1})–(\ref{eq:PT_Appendix1Product4}) are the same, can be easily explained using a well-known commutation rule of second quantization (see, for~example, (29) from~\cite{Jud:67a}, where the expression is in $LS$-coupling) 
\begin{equation}
\label{eq:Tensor13}
\bigl[ \tilde{a}^{(j_i)} \times a^{(j_k)} \bigr]^{(x)} =
 -\left( -1 \right)^{j_i+j_k-x} \; \bigl[ a^{(j_k)}  \times \tilde{a}^{(j_i)} \bigr]^{(x)}
+ \sqrt{\left[ j_i\right]} \; \delta \left( n_i l_i j_i, n_k l_k j_k \right) \; \delta \left( x, 0 \right).
\end{equation}
When the rank $x$ of these operators is not zero, this expression will take the following~form:
\begin{equation}
\label{eq:Tensor14}
\bigl[ \tilde{a}^{(j_i)} \times a^{(j_k)} \bigr]^{(x \neq 0)} =
 -\left( -1 \right)^{j_i+j_k-x} \; \bigl[ a^{(j_k)}  \times \tilde{a}^{(j_i)} \bigr]^{(x \neq 0)}.
\end{equation}
The tensorial structure present in the reduced matrix element RME$_B$ $-$ in our case, when $k=1$, $k'=4$, and~$x=3$ $-$ can be transformed into the tensorial structure present in RME$_D$ by applying relation (\ref{eq:Tensor14}) three times to the pairs of operators $\bigl[ \tilde{a}^{(j=\frac{1}{2})} \times a^{(j=\frac{1}{2})} \bigr]^{(k=1)}$, $\bigl[ a^{(j''=\frac{3}{2})} \times \tilde{a}^{(j''=\frac{3}{2})} \bigr]^{(x=3)}$, and~$\bigl[ a^{(j'=\frac{5}{2})} \times \tilde{a}^{(j'=\frac{5}{2})} \bigr]^{(k'=4)}$ from (\ref{eq:PT_Appendix1Product2}). From~this, we can see that the values of these matrix elements are equal, since 
\vspace{-12pt}
\begin{adjustwidth}{-\extralength}{0cm}
\begin{eqnarray}
\label{eq:Tensor15}
\hspace*{-0.5cm}
\biggl[ \Bigl[ \bigl[ \tilde{a}^{(j=\frac{1}{2})} \times a^{(j=\frac{1}{2})} \bigr]^{(k=1)} \times \bigl[ a^{(j''=\frac{3}{2})} \times \tilde{a}^{(j''=\frac{3}{2})} \bigr]^{(x=3)} \Bigr]^{(k'=4)} \times \bigl[ a^{(j'=\frac{5}{2})} \times \tilde{a}^{(j'=\frac{5}{2})} \bigr]^{(k'=4)} \biggr]^{(0)} 
	\nonumber \\
& & \hspace*{-12.1cm}
=
\biggl[ \Bigl[ \bigl[ a^{(j=\frac{1}{2})} \times \tilde{a}^{(j=\frac{1}{2})} \bigr]^{(k=1)} \times \bigl[ \tilde{a}^{(j''=\frac{3}{2})} \times a^{(j''=\frac{3}{2})} \bigr]^{(x=3)} \Bigr]^{(k'=4)} \times \bigl[ \tilde{a}^{(j'=\frac{5}{2})} \times a^{(j'=\frac{5}{2})} \bigr]^{(k'=4)} \biggr]^{(0)}.
\end{eqnarray}

\end{adjustwidth}
\textls[35]{Equation~(\ref{eq:Tensor13}) can also explain why RME$_B$ and RME$_D$ have different values} (RME$_B = 0.5123475$ and RME$_D = -0.1707825$) for the ranks $k = 1$, $k' = 1$, and~$x = 0$. Furthermore, based on (\ref{eq:Tensor14}), it is possible to identify cases where the reduced matrix elements have opposite signs.
\end{itemize}

All of the rules listed above are general and apply to (\ref{eq:PT_Appendix1Product1})--(\ref{eq:PT_Appendix1Product4}).

\section[\appendixname~\thesection]{The Subroutine \texorpdfstring{\texttt{\pmb{WW1}}}{WW1G}}
\label{PT_Appendix2}

The subroutine determines the value of the matrix elements (\ref{eq:tg}).

The subroutine has the following arguments:
\begin{enumerate}

\item \verb+IK+ is the array \verb+I+ (see Section~3.3 in~\cite{Gaigalas:2022}) for the bra function.
\item \verb+BK+ is the array \verb+B+ (see Section~3.3 in~\cite{Gaigalas:2022}) for the bra function.
\item \verb+ID+ is the array \verb+I+ for the ket function.
\item \verb+BD+ is the array \verb+B+ for the ket function.
\item \verb+K1+ is the rank $x$.
\item \verb+K2+ is the rank $J_2$.
\item \verb+K+ is the rank $J_1$.
\item \verb+QM1+, \verb+QM2+, \verb+QM3+, and~\verb+QM4+ are the quasi-spin projections in (\ref{eq:tg}).
\item \verb+WW+ is the value of the reduced matrix element (\ref{eq:tg}), which is returned by the subroutine.
\end{enumerate}

Table~\ref{Apendex2} presents three test cases for the subroutine \texttt{WW1G}. They are taken for estimation of the CV and VV correlations, described by the three-particle Feynman diagram, using 
the Rayleigh–Schr\"odinger perturbation theory in an irreducible tensorial form.
For this purpose, the levels of the $\mathrm{3s^23p^43d}$ configuration of Ar~II are studied.
In the calculations, the~1s, 2s, $\mathrm{2p_-}$, and~2p subshells are defined as core subshells (that correspond to the $F$ set); 
the 3s, $\mathrm{3p_-}$, 3p, $\mathrm{3d_-}$, and~3d subshells are defined as valence subshells (that correspond to the $F'$ set). 
The 4s, $\mathrm{4p_-}$, 4p, $\mathrm{4d_-}$, 4d, $\mathrm{4f_-}$, 4f\ subshells are defined as virtual ones (that correspond to the $G$ set).

\begin{table}[H]
\caption{The input and output argument values for the subroutine \texttt{WW1G} in the test~cases.}
\begin{tabularx}\textwidth{CCCCCCCCCCCC}  
\toprule 
\pmb{No.}&\pmb{\verb+IK+}&\pmb{\verb+ID+}&\pmb{\verb+BK+}&\pmb{\verb+BD+}&\pmb{\verb+K1+}&\pmb{\verb+K2+}&\pmb{\verb+K+}&\pmb{\verb+QM1+}&\pmb{\verb+QM2+}&\pmb{\verb+QM3+}&\pmb{\verb+QM4+} \\  
\midrule
\multicolumn{3}{l}{Test case 1}  &   &  &     &     &   &   &    &   & \\ 
 1 & 6 & 6 &  1.0 &  1.0 & 0 & 1 & 1 &  0.5 & $-$0.5 & $-$0.5 &  0.5  \\ 
 2 & 3 & 3 &  2.5 &  2.5 &   &   &   & & &  &   \\ 
 3 & 5 & 5 & $-$1.0 & $-$1.0 &   &   &   & & &  &   \\ 
 4 & 1 & 1 &      &      &   &   &   & & &  &   \\ 
 5 & 2 & 2 &      &      &   &   &   & & &  &   \\ 
 6 & 5 & 5 &      &      &   &   &   & & &  &   \\ 
 7 & 2 & 2 &      &      &   &   &   & & &  &   \\ 
& \multicolumn{3}{l}{\texttt{WW = $-$0.7071068}}    &  &     &     &   &   &    &   & \\ 
\multicolumn{3}{l}{Test case 2}  &   &  &     &     &   &   &    &   & \\
 1 & 3 & 3 &  0.5 &  0.5 & 2 & 2 & 1 &  0.5 & $-$0.5 &  0.5 & $-$0.5  \\ 
 2 & 3 & 3 &  1.5 &  1.5 &   &   &   & & &  &   \\ 
 3 & 3 & 3 &  0.5 &  0.5 &   &   &   & & &  &   \\ 
 4 & 3 & 3 &      &      &   &   &   & & &  &   \\ 
 5 & 1 & 1 &      &      &   &   &   & & &  &   \\ 
 6 & 3 & 3 &      &      &   &   &   & & &  &   \\ 
 7 & 1 & 1 &      &      &   &   &   & & &  &   \\ 
& \multicolumn{3}{l}{\texttt{WW = 1.2247449}}  &  &     &     &   &   &    &   & \\
\multicolumn{3}{l}{Test case 3}  &   &  &     &     &   &   &    &   & \\
 1 & 5 & 5 &  0.0 &  0.0 & 3 & 3 & 4 &  0.5 & $-$0.5 &  0.5 & $-$0.5  \\ 
 2 & 3 & 3 &  2.0 &  2.0 &   &   &   & & &  &   \\ 
 3 & 3 & 3 &  0.0 &  0.0 &   &   &   & & &  &   \\ 
 4 & 2 & 2 &      &      &   &   &   & & &  &   \\ 
 5 & 1 & 1 &      &      &   &   &   & & &  &   \\ 
 6 & 4 & 4 &      &      &   &   &   & & &  &   \\ 
 7 & 0 & 0 &      &      &   &   &   & & &  &   \\ 
& \multicolumn{3}{l}{\texttt{WW = 4.44971919}}    &  &     &     &   &   &    &   & \\
\bottomrule
\end {tabularx}
\label{Apendex2}
\end{table}

Since the parameters \texttt{IK}, \texttt{ID}, \texttt{BK}, and~\texttt{BD} are arrays, the~first column in Table~\ref{Apendex2} indicates the array index, while the remaining columns indicate the values of the subroutine input parameters. For~example, in~the first example, when No. = 3, then \texttt{IK}(3) = 5, \texttt{ID}(3) = 5, \mbox{\texttt{BK}(3) = $-$1.0}, and~\texttt{BD}(3) = $-$1.0. Elements of these arrays corresponding to quantities such as the quantum numbers $n$, $l$, and~$j$ of the subshell or all other characteristics describing the subshell that are necessary for the calculation can be found in Section~3.3 of~\cite{Gaigalas:2022}. The~other parameters are simple variables; their values in the table are in the row where No. = 1. The~value calculated by the subroutine is given in the row \texttt{WW}. For~example, in~the case of the first example, it is \texttt{WW = $-$0.7071068}. This subroutine is available from the corresponding author, [G.G.], upon~reasonable request.

\begin{adjustwidth}{-\extralength}{0cm}
\reftitle{References}

\PublishersNote{}
\end{adjustwidth}
\end{document}